%% file: sigmetrics2025.tex
\documentclass[3p]{elsarticle}

\input{setup}

\usepackage{thmtools}
\usepackage{thm-restate}
\newtheorem{theorem}{Theorem}

\newtheorem{lemma}{Lemma}

\usepackage{blindtext}
\usepackage[utf8]{inputenc}
\usepackage{epsfig,xspace,url,xcolor}
\usepackage{subcaption}

\usepackage{mathtools}
\usepackage{tikz}
\usepackage{caption}
\usepackage{graphicx}
\usepackage[ruled, lined, linesnumbered, commentsnumbered, longend]{algorithm2e}
\usepackage[noend]{algpseudocode}
\usepackage{url}
\usepackage{colortbl}
\usepackage{xcolor}

\newcounter{Deq}
\renewcommand{\theDeq}{DE\arabic{Deq}}

\newcommand{\Dtag}{%
  \refstepcounter{Deq}% increment counter
  \tag{\theDeq}% apply tag
}

\usepackage{wrapfig}

\begin{document}

\definecolor{brown(web)}{rgb}{0.65, 0.16, 0.16}
\definecolor{brown(web)light}{rgb}{0.85, 0.36, 0.36}
\definecolor{amber}{rgb}{1.0, 0.75, 0.0}
\definecolor{amberlight}{rgb}{1.0, 0.844, 0.376}
\definecolor{asparagus}{rgb}{0.53, 0.66, 0.42}

\title{A Control-Theoretic Perspective on\\BBR/CUBIC Congestion-Control Competition}

\author[inst1]{Simon Scherrer}
\author[inst1]{Adrian Perrig}

\author[inst2]{Stefan Schmid}

\address[inst1]{Department of Computer Science, ETH Z\"urich, Switzerland}
\address[inst2]{Faculty IV - Electrical Engineering and Computer Science, TU Berlin, Germany}

\begin{abstract}
To understand the fairness properties of the BBR congestion-control
algorithm (CCA),
previous research has analyzed BBR behavior with a variety of models.
However, previous model-based work suffers from a trade-off
between accuracy and interpretability:
While dynamic fluid models generate
highly accurate predictions through simulation,
the causes of their predictions cannot
be easily understood. In contrast,
steady-state models predict CCA behavior in
a manner that is intuitively understandable, 
but often less accurate.
This trade-off is especially consequential
when analyzing 
the competition between BBR and traditional loss-based CCAs,
as this competition often suffers from instability,
i.e., sending-rate oscillation.
Steady-state models cannot predict this instability at all,
and fluid-model simulation cannot yield analytical
results regarding preconditions and severity of the oscillation. 

To overcome this trade-off, we extend
the recent dynamic fluid model of BBR
by means of control theory.
Based on this control-theoretic analysis, 
we derive quantitative conditions 
for BBR/CUBIC oscillation, identify network settings 
that are susceptible to instability, and 
find that these conditions are frequently 
satisfied by practical networks.
Our analysis illuminates the fairness
implications of BBR/CUBIC oscillation,
namely by deriving and experimentally validating 
fairness bounds that reflect 
the extreme rate distributions during oscillation.
In summary, our analysis shows that BBR/CUBIC oscillation
is frequent and harms BBR fairness,
but can be remedied by means of our control-theoretic framework.
\end{abstract}

\maketitle

%%
%% The code below is generated by the tool at http://dl.acm.org/ccs.cfm.
%% Please copy and paste the code instead of the example below.
%%

\section{Introduction}
\label{sec:introduction}

During the global deployment of the BBR 
congestion-control algorithm (CCA)~\cite{cardwell2017bbr},
researchers have documented several cases of unexpected
BBR behavior, all with significant implications on performance and 
fairness~\cite{gomez2020performance,hock2017experimental,mishra2022we,scherrer2022model,scholz2018towards,ware2019beyond}.
To improve the safety of CCA deployment in the future,
recent research has aimed at a more fundamental
understanding of BBR based on \emph{modeling}, i.e.,
mathematical descriptions that 
enable insights
valid for a wide range of network settings.

This model-based work has taken two different approaches.
First, the \emph{steady-state models} by Ware et al.\ \cite{ware2019modeling} 
and Mishra et al.\ \cite{mishra2022we}
identify a steady state of the congestion-control dynamics,
and predict BBR behavior with
closed-form expressions for network metrics 
in this steady state.
Second, the \emph{dynamic fluid model}
by Scherrer et al.\ \cite{scherrer2022model}
describes the transient behavior of BBR %and the network
by means of ordinary differential equations (ODEs),
and predicts BBR properties by fluid-model simulation, 
i.e., numerical ODE solving.

These two approaches suffer from a
trade-off between accuracy and 
\emph{interpretability}, i.e.,
transparency regarding the \emph{causes}
of model predictions. 
As we demonstrate in this work, 
the dynamic fluid model is more accurate than 
the steady-state models, mainly because the fluid-model 
can represent behavior outside the steady state.
However, the dynamic fluid model 
has so far mostly been used for simulation, which
arrives at predictions through lengthy numerical computation.
As such, the predictions of dynamic fluid models are less
explainable than the predictions of steady-state models,
because steady-state models only consist of a handful of 
closed-form expressions that can be intuitively understood.

In this work, we aim to bridge this gap between accuracy
and interpretability of model predictions.
In particular, we demonstrate that 
a \emph{control-theoretic analysis} of the dynamic fluid model
allows insights that
are not only accurate and interpretable,
but also rigorously provable.
We illustrate the value of this stability analysis
by focusing on the competition
between BBR and CUBIC~\cite{ha2008cubic},
the most prevalent CCAs
in the Internet~\cite{mishra2019great}.
BBR/CUBIC competition frequently suffers from instability, i.e.,
\emph{rate oscillation}, in which flows using different CCAs
regularly obtain highly uneven bandwidth shares.
While congestion-control oscillation 
(e.g., sawtooth in TCP Reno) 
is sometimes a necessary price to maximize
utilization or achieve per-flow fairness, 
BBR/CUBIC oscillation is unnecessary and even harmful
to fairness: Unfairness between BBR and CUBIC flows
is especially pronounced 
when oscillation occurs (cf.~\cref{sec:bbr-cubic:dynamic}).

Crucially, this instability is invisible
in steady-state models, as these models assume stability.
In contrast, fluid-model simulation reveals this
oscillation, but must be conducted for specific network
settings to indicate the occurrence and
the severity of the oscillation.
More general insights into the oscillation are
possible thanks to our control-theoretic analysis of the
fluid model: This analysis yields (i)~mathematical conditions
on network parameters, indicating when oscillation provably
occurs, (ii)~analytical bounds on the severity of the oscillation,
describing the extreme rate distributions of
BBR/CUBIC competition, 
and (iii)~intuitive explanations of stabilizing adaptations
of the BBR algorithm.

{\setlength{\parskip}{5pt}In summary, we present the following contributions:}
% In summary, we present the following contributions:

\textbf{Documentation of instability.}
    We show that steady-state models predict BBR
    fairness towards CUBIC less accurately 
    than dynamic models (\cref{sec:bbr-cubic:model-evaluation}),
    due to their assumption that BBR/CUBIC competition converges. 
    In reality,
    convergence is prevented by persistent sending-rate oscillation,
    which we systematically explain for 
    the first time (\cref{sec:bbr-cubic:dynamic}).
    While the oscillation is predicted accurately
    by fluid-model simulation, such simulation cannot 
    provide a formal 
    understanding of preconditions
    and severity of the oscillation.

\textbf{Stability analysis of BBR/CUBIC fluid model.}
    We apply a control-theoretic analysis to a joint
    BBR/CUBIC fluid model, thereby describing the BBR/CUBIC oscillation
    analytically and accurately.
    For this analysis, we first derive
    the equilibrium of the BBR/CUBIC fluid model
    (\cref{sec:new-model}).
    Then, we derive an instability condition on 
    network parameters, i.e., a network condition that
    is provably conducive to oscillation
    (\cref{sec:bbr-cubic:dynamic:oscillation:model}).
    % In the analysis, we separate BBR/CUBIC competition
    % into short-term dynamics, which are continuous and 
    % asymptotically stable, and long-term dynamics, which
    % are discrete and unstable under mild conditions. 
    This condition is derived using 
    Lyapunov theory, center-manifold theory, and
    fixed-point iteration.
 
\textbf{Insights into temporal fairness.}
    Besides identifying the preconditions for oscillation,
    the stability analysis also illuminates fairness implications
    of BBR/CUBIC oscillation, yielding 
    worst-case fairness bounds and approximate
    tight fairness bounds 
    that we experimentally validate~(\cref{sec:bbr-cubic:dynamic:oscillation:model:fairness}).
   
\textbf{Countermeasure evaluation.}
    We evaluate multiple adaptations of BBR
    regarding their effectiveness in suppressing oscillation,
    and explain the results by means of our control-theoretic
    framework~(\cref{sec:bbr-cubic:dynamic:oscillation:remedies}).
    These redesigned versions include several ad-hoc
    modifications of BBR, which we design and implement,
    and the official releases of BBRv2 and BBRv3.
    Importantly, these redesigns reduce oscillation,
    but still have drawbacks in terms of fairness or responsiveness.

{\setlength{\parskip}{5pt}
This work does not raise any ethical issues.}

\section{Background}
\label{sec:background}

This section provides an overview of 
BBRv1 and CUBIC, the 
main CCAs in our analysis.
BBRv2/v3 are discussed in~\cref{sec:bbr-cubic:dynamic:oscillation:remedies}.

\subsection{BBRv1}
\label{sec:background:bbr}

Fundamentally, a BBRv1 flow maintains a \emph{bottleneck-band-width estimate}~$x^{\mathrm{btl}}$
and a \emph{minimum-RTT estimate}~$\tau^{\min}$.
These two state variables determine the sending rate and are
continuously adjusted by the following probing processes 
(\cref{fig:background:bbr}).

\begin{figure}
    \begin{minipage}{0.47\linewidth}
        \centering
        \input{figures/bbr_explainer}
        \vspace{-5pt}
        \caption{BBR.}
        \label{fig:background:bbr}
    \end{minipage}\quad\vrule\quad
    \begin{minipage}{0.47\linewidth}
        \centering
        \input{figures/cubic_explainer}
        \vspace{-5pt}
        \caption{CUBIC.}
        \label{fig:background:cubic}
    \end{minipage}
    \vspace{-3.5mm}
\end{figure}

\paragraph{Bandwidth probing} 
To regulate~$x^{\mathrm{btl}}$,
a BBR flow cycles through periods of eight phases,
where each phase lasts for the duration of the
minimum-RTT estimate~$\tau^{\min}$.
In six of these phases, the pacing rate of the
BBR flow is simply~$x^{\mathrm{btl}}$.
However, in one phase (\texttt{ProbeBW\_UP}),
the BBR flow raises the pacing rate to~$\nicefrac{5}{4}\cdot x^{\mathrm{btl}}$ 
to discover whether more bandwidth is available.
In the subsequent phase, the pacing rate is decreased to~$\nicefrac{3}{4}\cdot x^{\mathrm{btl}}$
to eliminate potential queues.
The bottleneck-bandwidth estimate $x^{\mathrm{btl}}$
is continuously updated to the maximum measured delivery rate 
(ACK rate) from the last 10 RTTs.

\paragraph{RTT probing} Since the minimum-RTT estimate~$\tau^{\min}$
should match the path propagation delay, 
$\tau^{\min}$ tracks the minimum measured RTT.
If~$\tau^{\min}$ cannot be adjusted for 10 seconds,
BBR performs an~\emph{RTT-probing step}:
Namely, the flow resets~$\tau^{\min}$ and tries 
to drain the buffer by drastically reducing 
its sending rate for 200 milliseconds.
This rate reduction is achieved by
limiting the congestion window to 4 segments.
At any other time, the BBR congestion-window size~$v$ amounts to
twice the estimated BDP, i.e., $v = 2\tau^{\min}x^{\mathrm{btl}}$.
This congestion window may keep the sending rate
below the pacing rate if the 
data volume `in flight', i.e., sent but not acknowledged yet,
grows due to congestion.

\subsection{CUBIC}
\label{sec:background:cubic}

A CUBIC flow likewise maintains two state variables (\cref{fig:background:cubic}),
namely the \emph{maximum congestion-window size}~$w^{\max}$,
which is recorded at the time of the last loss,
and the \emph{window-growth duration}~$s$,
which corresponds to the time since the last loss.
These two variables determine 
the CUBIC congestion-window size~$w$ via the CUBIC
window-growth function~$W$:
\begin{equation}
    w = W(w^{\max}, s) = w^{\max} + c\cdot\left(s - \sqrt[3]{\frac{b}{c}  w^{\max}}\right)^3.
    \label{eq:bbr-cubic:model:basic-model:cubic-window-growth}
\end{equation}
In this function, the recommended parameters are 
$b = 0.3$ and $c = 0.4$~\cite{rhee2018rfc}.
Hence, a packet loss ($s := 0$, $w^{>} := w$) instantly reduces
the congestion-window size~$w$ by 30\%.

\section{Motivation}
\label{sec:motivation}

In this section, we show that
dynamic models predict the outcome of BBR/CUBIC competition 
more accurately than steady-state models  
(\cref{sec:bbr-cubic:model-evaluation}).
Steady-state models are less accurate in
this regard because these models, while 
refreshingly simple, cannot 
represent transient phenomena (\cref{sec:bbr-cubic:dynamic}).
% Hence, we emphasize the need for analytical models
% that describe congestion-control behavior over 
% time (\cref{sec:motivation:conclusion}).

% \begin{figure*}
%     \begin{minipage}{0.2\linewidth}
%         \input{figures/bbr_explainer}
%         \caption{BBR.}
%         \label{fig:background:bbr}
%     \end{minipage}\quad\vrule\quad
%     \begin{minipage}{0.2\linewidth}
%         \input{figures/cubic_explainer}
%         \vspace{-15pt}
%         \caption{CUBIC.}
%         \label{fig:background:cubic}
%     \end{minipage}\quad\vrule\quad
%     \begin{minipage}{0.4\linewidth}
%         \includegraphics[width=\linewidth,trim=0 8 0 0]{figures/BBR_Cubic__Composition__B.pdf}
%         \vspace{-20pt}
%         \caption{Model evaluation for 10~flows.}
%         \label{fig:static:bbr-cubic}
%     \end{minipage}
%     % \centering
%     %     \includegraphics[width=\linewidth,trim=0 8 0 0]{figures/BBR_Cubic__Composition__B.pdf}
%     %     \caption{Model evaluation for 10 flows.}
%     %     \label{fig:static:bbr-cubic}
%     %     \vspace{-10pt}
% \end{figure*}

\subsection{Comparison of Models}
\label{sec:bbr-cubic:model-evaluation}

In previous research, the competition between BBR and CUBIC flows
has been analyzed by means of steady-state models,
namely by Ware et al.~\cite{ware2019modeling} and
Mishra et al.~\cite{mishra2022we}.
These steady-state models have recently been complemented
with a dynamic model by Scherrer et al.~\cite{scherrer2022model},
who analyze BBR/CUBIC competition by means of 
fluid-model simulation, i.e., ODE solving.
Fundamentally, steady-state models have no notion of time,
but allow analytical investigation, whereas fluid-model
simulations describe congestion-control behavior over time,
but rely on numerical computation.

In this section, we evaluate all models
with respect to their prediction accuracy
for BBR/CUBIC competition.
We describe our experiment setting
in~\cref{sec:bbr-cubic:model-evaluation:experiments}, and
present our results in~\cref{sec:bbr-cubic:model-evaluation:results}.

\subsubsection{Experiment Setting}
\label{sec:bbr-cubic:model-evaluation:experiments}

As usual in the literature~\cite{gomez2020performance,hock2017experimental,kfoury2020emulation,nandagiri2020bbrvl,scherrer2022model,scholz2018towards,song2021understanding,ware2019modeling},
we focus on a dumbbell topology 
with $N$ senders sharing a single bottleneck link~$\ell$.
This topology is pervasive in congestion-control studies 
as it captures the core challenge: 
achieving efficient and fair resource allocation among senders
who lack direct communication and 
rely solely on network feedback to adjust their behavior.
Additionally, its simplicity makes it amenable to tractable theoretical analysis, enabling rigorous exploration of key dynamics.
While this fundamental scenario can be extended to include 
more complex setups (e.g., multiple bottlenecks), 
the basic fan-in pattern of traffic remains highly relevant, 
particularly in real-world scenarios such as multiple users 
on a low-capacity residential edge link simultaneously 
downloading content from various sources.

In our experiments, the bottleneck link~$\ell$ 
has capacity~$C_{\ell} = 100$Mbps and
a one-way propagation delay~$\tau_{\ell}^{\mathrm{p}} = 10$ms.
These link properties are in line with the 
capabilities of the network emulator 
Mininet~\cite{mininet_intro}
that we use. 
Each of the~$N$ senders
accesses the bottleneck link~$\ell$ via
an individual non-shared link, also with a 
propagation delay of~10ms. Hence, the overall 
RTT propagation delay
is~$\tau^{\mathrm{p}} = 40$ms for each flow.
The non-shared links and the bottleneck link are
intermediated by a buffer with a size of 1.5 path BDP
($750$KB), as buffer sizes
above 1~BDP are especially interesting for BBR/CUBIC fairness\footnote{In smaller buffers, BBRv1 simply causes starvation of loss-based CCAs such as CUBIC~\cite{scherrer2022model}.}~\cite{scholz2018towards,ware2019modeling}. 
Each experiment lasts 120 seconds, and is
repeated three times. All senders have infinite demand,
generated by the \textit{iperf} utility~\cite{iperf}. 
Unless otherwise stated, these settings are used for the remainder of the paper.

\subsubsection{Results}
\label{sec:bbr-cubic:model-evaluation:results}

\begin{figure}
    \includegraphics[width=0.9\linewidth,trim=0 0 0 0]{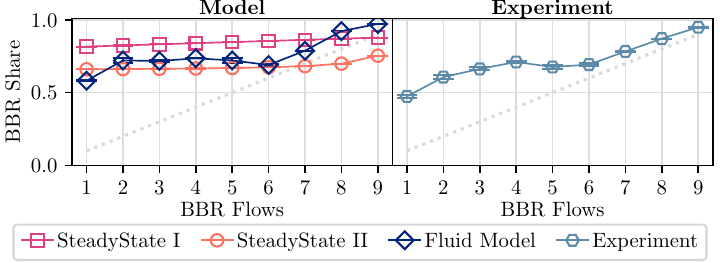}
    \caption{Model evaluation for 10~flows (The dotted gray line indicates the proportional share of BBR flows under perfect per-flow fairness).}
    \label{fig:static:bbr-cubic}
    \vspace{-2mm}
\end{figure}

We consider all BBR/CUBIC combinations in a network with~$N = 10$ flows,
and run all models plus the Mininet experiment for each combination.
Then, we compute the obtained capacity share of all BBR flows.
For the steady-state models,
the capacity share is derived from the predicted steady-state rate distribution.
For the fluid model and the experiments,
the capacity share is the average capacity share over time.
The results in~\cref{fig:static:bbr-cubic}
yield two important insights.

First, the BBR flows consistently obtain a disproportionately large
capacity share, especially if they are in a minority. 
Hence, the fairness from the balanced scenario with 5 flows for each
CCA is not generalizable.

Second and more importantly for our purpose,
the predictive power of the models
grows with their fidelity regarding the time dimension
of the competition: 
The steady-state models (\textit{SteadyState~I}~\cite{ware2019modeling}
and \textit{SteadyState~II}~\cite{mishra2022we}) 
% have no notion of time, but SteadyState~II at least captures that
% the CUBIC flows regularly back off upon loss and is thus more accurate.
% However, both steady-state models %and the fluid-equilibrium model
are less accurate than the fluid-model simulation (\textit{Fluid Model}),
which computes a time series for each network metric.
% The reduced accuracy of the fluid-equilibrium model (\textit{Fluid (Eq)}) is especially
% interesting, given that it is derived from the time-aware fluid model.
This suggests that the steady-state perspective
of BBR/CUBIC competition fails to capture important phenomena
that unfold over time, i.e., the transient behavior of the competition.
% Put differently, the steady-state models predict the outcomes
% of BBR/CUBIC competition from a steady state, and thus implicitly
% assume that the competition dynamics \emph{converge} to this steady
% state. However, this assumption is often questionable, 
% as we will demonstrate in the following sections.

\subsection{The Relevance of Transient Behavior}
\label{sec:bbr-cubic:dynamic}

\begin{figure*}
    \centering
    \includegraphics[width=\linewidth,trim=0 10 0 0]{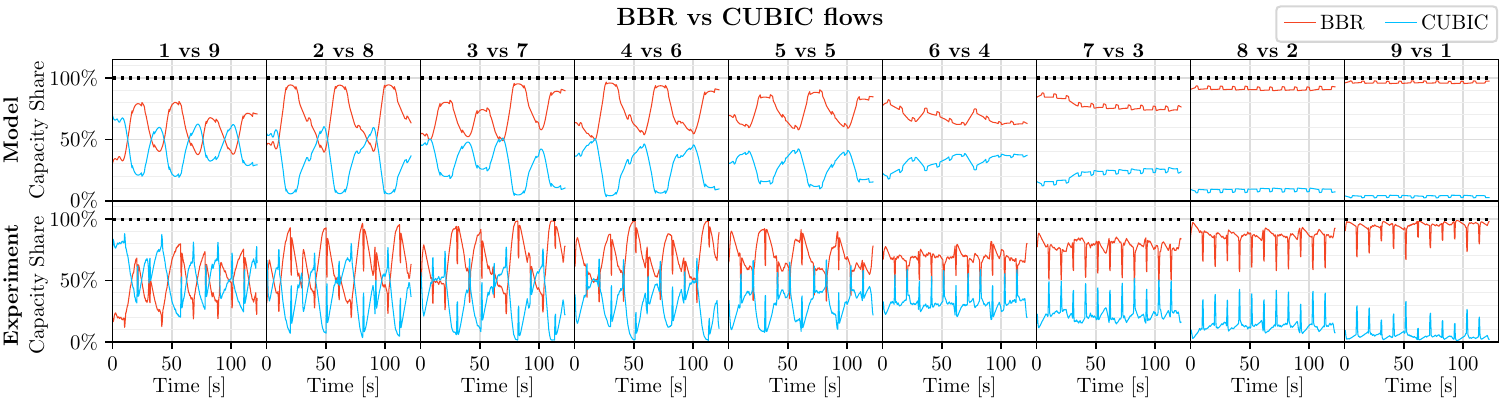}
    \caption{BBR/CUBIC competition over time for different CCA combinations (Aggregate capacity shares).}
    \label{fig:dynamic:bbr-cubic}
\end{figure*}

% The previous section illustrates that BBR/CUBIC
% competition involves important time-varying
% phenomena that reduce the accuracy of predictions 
% from time-agnostic steady-state models.
% In fact, 
The detailed
results from the fluid-model simulations and the
experiments (in~\cref{fig:dynamic:bbr-cubic})
confirm that BBR/CUBIC competition involves
transient phenomena, namely persistent
\emph{sending-rate oscillation}.

This oscillation is relevant for fairness,
as the unfairness in~\cref{fig:static:bbr-cubic}
is particularly high whenever
\cref{fig:dynamic:bbr-cubic} indicates oscillation,
i.e., for 1 to 5 BBR flows.
In particular, the BBR flows obtain a egregiously over-proportional
share of the bandwidth during the recurring peaks 
of the BBR rate oscillation.

While this oscillation has been noted 
in previous work~\cite{scholz2018towards},
it has never been systematically explained.
The following intuitive explanation helps
understand the oscillation, but does not yet
quantify its occurrence or severity; 
such quantitative results
will be derived in \cref{sec:new-model,sec:bbr-cubic:dynamic:oscillation:model,sec:bbr-cubic:dynamic:oscillation:model:fairness}.

Oscillation in BBR/CUBIC competition is an alternation between (i)~the \emph{RTT probing step} of BBR,
and (ii)~the \emph{short-term dynamics}, which are determined by
the preceding RTT-probing step and last until the next
RTT-probing step.
This alternation forms the \emph{long-term dynamics} 
(\cref{fig:bbr-cubic:dynamic:oscillation:why:mechanism,fig:bbr-cubic:dynamic:oscillation:model:experiment}).

\begin{figure*}
    \begin{minipage}{0.36\linewidth}
        \centering
         \input{figures/oscillation-mechanism}
         % \vspace{-6mm}
        \caption{Basic mechanism of BBR/CUBIC oscillation.}
        \label{fig:bbr-cubic:dynamic:oscillation:why:mechanism}
    \end{minipage}\quad\vrule\quad 
    \begin{minipage}{0.58\linewidth}
        \centering
        \input{figures/oscillation-example}
        \input{figures/dynamics_arrows}
        \vspace{-7mm}
        \caption{Oscillation of sending rates for competition between
        one BBR flow and one CUBIC flow (Experiment).
        The green line highlights an inflated MinRTT
        estimate by the BBR flow, leading to subsequent BBR rate growth.}
        \label{fig:bbr-cubic:dynamic:oscillation:model:experiment}
    \end{minipage} %\quad\vrule\quad 
\end{figure*}

In the RTT-probing step, the BBR flow reduces its sending rate to almost zero,
with the goal of emptying the buffer and discovering the propagation delay.
However, the CUBIC flow does not participate in this reduction, and may thus prevent
the complete clearance of the buffer if its congestion-window size~$w$ is
relatively large at the time. The CUBIC congestion-window size~$w$ thus affects the 
\emph{back-off queue length}~$q_{\ell}^-$ during the RTT-probing step, and hence 
also the BBR minimum-RTT estimate~$\tau^{\min}$ computed from probing measurements.
For example, the CUBIC congestion-window size~$w$ in
RTT-Probing Step~\circled{1} in~\cref{fig:bbr-cubic:dynamic:oscillation:model:experiment} 
is relatively large (cf. high CUBIC rate). 
As a result, the buffer is not completely cleared during RTT probing,
and the minimum-RTT estimate~$\tau^{\min}$ after the probing step
(in interval~[\circled{1},\circled{2}] in~\cref{fig:bbr-cubic:dynamic:oscillation:model:experiment}) 
is around 34\% higher than the actual propagation delay.

This inflated minimum-RTT estimate~$\tau^{\min}$ %is relevant because it
co-determines the congestion-window size~$v$ of the BBR flow, and thus 
its sending rate~$x^{\mathrm{B}}$. Since this congestion-window size~$v$
actively constrains the BBR sending rate in scenarios with large buffers, 
the inflated minimum-RTT estimate also co-determines the BBR sending
rate~$x^{\mathrm{B}}$.
In particular, an inflated minimum-RTT estimate~$\tau^{\min}$ 
(such as in time interval~$[\circled{1}, \circled{2}]$
in~\cref{fig:bbr-cubic:dynamic:oscillation:model:experiment}) results
in aggressive bandwidth probing by the BBR flow, which 
increases the BBR sending rate~$x^{\mathrm{B}}$ and reduces
the CUBIC congestion-window size~$w$ by causing packet loss.

Crucially, this inflated~$\tau^{\min}$ is fixed after
RTT probing: Since the BBR flow and the CUBIC flow make
intensive use of the buffer, they prevent a
downward revision of the minimum-RTT estimate~$\tau^{\min}$
during the short-term dynamics.
Hence, these short-term dynamics continue until the 
minimum-RTT estimate~$\tau^{\min}$ times out, 
i.e., for 10 seconds.

At the next RTT probing step, the reduced CUBIC 
congestion-window size~$w$ determines a new minimum-RTT estimate~$\tau^{\min}$: 
For example, the reduced window size $w$ at Step~\circled{2} 
in~\cref{fig:bbr-cubic:dynamic:oscillation:model:experiment}
reduces the minimum-RTT estimate~$\tau^{\min}$ to the propagation delay
and thus induces a relatively low BBR rate~$x^{\mathrm{B}}$.
Hence, the congestion-window size~$w$ increases again until Step~\circled{3}.
At Probing Step~\circled{3}, $w$ is still low enough to allow
a correct estimate of the propagation delay and thus another 
growth phase of~$w$. 
However, at Probing Step~\circled{4},
$\tau^{\min}$ is inflated again and the oscillation pattern
resumes.

This oscillation mechanism
applies to the case with a single flow per CCA.
However, oscillation may also arise for 
multiple flows per CCA, because simultaneous BBR
flows synchronize their RTT-probing behavior 
(cf.~\cref{sec:bbr-cubic:dynamic:oscillation:synchronization}).
% We now explain how and why
% the mechanism generalizes to multiple flows.

\subsection{Conclusion}
\label{sec:motivation:conclusion}

In summary, the fairness of BBR/CUBIC competition is heavily
determined by the transient phenomenon of oscillation.
This oscillation is fundamentally impossible to predict for
the time-agnostic steady-state models, but is predicted with high
accuracy by fluid-model simulations.

However, these simulations must be computed
for a specific network setting of interest, and therefore do
not allow fundamental insights to the same degree 
as analytical investigations would. 
In particular, the numerical simulations do not constitute 
rigorous proofs, e.g., regarding the conditions 
under which oscillation provably occurs,
or regarding worst-case bounds for the unfairness
of BBR/CUBIC competition.

In this paper, we therefore apply
control theory to the dynamic fluid model.
This control-theoretic analysis relies on the
equilibrium of the dynamic fluid model (\cref{sec:new-model}),
which is then leveraged for an analysis of oscillation
conditions~(\cref{sec:bbr-cubic:dynamic:oscillation:model}),
oscillation bounds~(\cref{sec:bbr-cubic:dynamic:oscillation:model:fairness}),
and oscillation suppression~(\cref{sec:bbr-cubic:dynamic:oscillation:remedies}).

\section{Fluid-Model Equilibrium}
\label{sec:new-model}

To enable control-theoretic analysis, we construct a joint fluid model 
of BBR/CUBIC competition by combining our BBR model 
from previous work~\cite{scherrer2022model} with the 
CUBIC model of Vardoyan et al.~\cite{vardoyan2021towards}. 
The resulting system is governed by the differential equations
\cref{eq:bbr-cubic:system-evolution:cubic:wmax,eq:bbr-cubic:system-evolution:cubic:s,eq:bbr-cubic:system-evolution:bbr}. 
From this formulation, we derive the corresponding equilibrium, 
which serves as the foundation for our control-theoretic analysis. 
Our approach remains consistent with prior work, but extends 
these models by providing an integrated framework 
that supports explanatory reasoning and yields additional predictions 
beyond those available from the original formulations.

\paragraph{Relevance} The fluid-model equilibrium is crucial for analyzing
the convergence of BBR and CUBIC flows,
as the fluid-model equilibrium is a fixed point
of the large-scale BBR/CUBIC dynamics:
Once the competing flows adopt the rate distribution
in the fluid-model equilibrium, they maintain this
rate distribution for the entire future of their competition,
apart from small-scale fluctuations inherent in the congestion-control algorithms
(e.g., the $\nicefrac{5}{4}$-$\nicefrac{3}{4}$ pacing-rate variation in BBR).
Therefore, if the BBR/CUBIC dynamics arrived at this equilibrium,
the observed large-scale oscillation would not occur; 
however, it is a separate question 
whether the BBR/CUBIC dynamics arrive at this fixed point 
from a non-equilibrium initial configuration.
This question will be investigated in the next section
(\cref{sec:bbr-cubic:dynamic:oscillation:model}).

\begin{table}[]
    \centering
    \def\arraystretch{1.1}
    \begin{tabular}{ccl}
        \hline
        \multirow{8}{*}{BBR} & $\alpha$ & \makecell[l]{Probing strength\\(Rate coefficient under active bandwidth probing)}\\\cline{2-3}
         & $\beta$ & \makecell[l]{Strength\\(Rate coefficient under inactive bandwidth probing)}\\ \cline{2-3}
         & $\tau^{\min}$ &  Minimum-RTT estimate\\
         & $v = 2x^{\mathrm{btl}}\tau^{\min}$ & Congestion-window size\\\cline{2-3}
         & $x^{\mathrm{B}}$ & BBR sending rate\\\cline{2-3}
         & $x^{\mathrm{btl}}$ & Bottleneck-bandwidth estimate\\\cline{2-3}
         & $x^{\mathrm{dlv}}$ & Measured delivery rate\\\cline{2-3}
         & $\chi > 0$ & Minimum bottleneck-bandwidth estimate\\[2mm]\hline
         \multirow{7}{*}{CUBIC} & $b = 0.3$ & Loss-reduction parameter\\\cline{2-3}
         & $c = 0.4$ & Window-scaling parameter\\\cline{2-3}
         & $s$ & Window-growth duration\\\cline{2-3}
         & $w$ & Congestion-window size\\\cline{2-3}
         & $w^{\max}$ & Maximum congestion-window size\\\cline{2-3}
         & $W(w^{\max}, s)$ & Window-growth function\\\cline{2-3}
         & $x^{\mathrm{C}}$ & CUBIC sending rate\\[2mm]\hline
         \multirow{4}{*}{Network} & $p_{\ell}$ & Loss rate on link~${\ell}$\\\cline{2-3}
         & $q_{\ell}$ & Queue length of link~$\ell$\\\cline{2-3}
         & $q_{\ell}^-$ & Back-off queue length of link~$\ell$ (during BBR RTT probing)\\\cline{2-3}
         & $y_{\ell}$ & Total sending rate on link~$\ell$\\\cline{2-3}
         & $\tau_f$ & Total delay (RTT) experienced by flow~$f$\\\cline{2-3}
         & $\tau^{\mathrm{p}}_f$ & Propagation delay on path used by flow~$f$ \\[2mm]\hline
         \multirow{10}{*}{\makecell{Stability\\analysis}} & $\sigma = (x^{\mathrm{btl}}, w^{\max}, s)$ & System state (Irreducible state variables)\\\cline{2-3}
         & $\tilde{\cdot}$ & Value of state variable in \emph{short-term} equilibrium\\\cline{2-3}
         & $\overline{\cdot}$ & Value of state variable in \emph{long-term} equilibrium\\\cline{2-3}
         & $w^{\leftarrow}(w)$ & \makecell[l]{Window-update function under \emph{incomplete} convergence\\
         (CUBIC congestion-window size at an RTT-probing step\\after a previous RTT-probing step\\where the CUBIC congestion-window size was $w$)}\\\cline{2-3}
         & $\tilde{w}^{\leftarrow}(w)$ & \makecell[l]{Window-update function under \emph{complete} convergence\\(Short-term equilibrium CUBIC congestion-window size\\after an RTT-probing step\\where the CUBIC congestion-window size was $w$)}\\\cline{2-3}
         & $w^{<}$ & Minimum possible CUBIC congestion-window size in equilibrium\\\cline{2-3}
         & $w^{>}$ & Maximum possible CUBIC congestion-window size in equilibrium\\\cline{2-3}
         & $\Omega$ & \makecell[l]{Unstable neighborhood of\\long-term equilibrium CUBIC window size~$\overline{w}$}\\[2mm]\hline
    \end{tabular}
    \caption{Notation used in formal analysis}
    \label{tab:notation}
\end{table}

\subsection{Fluid-Model Basics}
\label{sec:bbr-cubic:model:basic-model}

We consider~$N_{\mathrm{B}}$ BBR flows in set~$F_{\mathrm{B}}$ and
$N_{\mathrm{C}}$ CUBIC flows in set~$F_{\mathrm{C}}$, 
all of which share
a single bottleneck link~$\ell$.
According to Scherrer et al.~\cite{scherrer2022model},
each BBR flow~$i$ generally sends at rate~$x_i^{\mathrm{B}} = \beta_i x^{\mathrm{btl}}_i$,
where~$x^{\mathrm{btl}}_i$ is the bottleneck-bandwidth estimate of flow~$i$,
and~$\beta_i$ is the flow's \emph{strength}.
This strength coefficient~$\beta_i$ enforces the 
congestion-window constraint on the BBR sending rate.
The BBR congestion-window size is~$v_i = 2\tau_i^{\min}x^{\mathrm{btl}}$, where
$\tau^{\min}_i$ is the minimum-RTT estimate of BBR flow~$i$.
\begin{equation}
    \beta_i = \min\left(1,  2 \tau^{\min}_i/ \tau_i\right),
    \label{eq:bbr-cubic:system-evolution:bbr:beta}
\end{equation}
where~$\tau_i$ is the current delay experienced by BBR flow~$i$.

In contrast, each CUBIC flow~$k$ sends at rate~$x_k^{\mathrm{C}} = w_k/\tau_k$,
where~$w_k$ is the current congestion window of CUBIC flow~$k$
and $\tau_k$ is the currently experienced delay of flow~$k$.
The congestion-window size~$w_k$ is
determined by the CUBIC window-growth function~$W(w^{\max}, s)$
in~\cref{eq:bbr-cubic:model:basic-model:cubic-window-growth}.

The total load~$y_{\ell}$ on link~$\ell$ is thus given by:
\begin{equation}
    y_{\ell} = \sum_{i \in F_{\mathrm{B}}} x_i^{\mathrm{B}} +
    \sum_{k \in F_{\mathrm{C}}} x_i^{\mathrm{C}}
    = \sum_{i \in F_{\mathrm{B}}} \beta_i x_i^{\mathrm{btl}} +
    \sum_{k \in F_{\mathrm{C}}} \frac{w_k}{\tau_k}.
    \label{eq:bbr-cubic:model:basic-model:total-load}
\end{equation}
Note that this formulation of~$y_{\ell}$ is an approximation
by Scherrer et al.~\cite{scherrer2022model},
as the BBR flows deviate from the rate~$\beta_i x^{\mathrm{btl}}$
when they are probing for bandwidth (cf.~\cref{sec:background:bbr}),
i.e., when they increase the rate to $\nicefrac{5}{4}\cdot x^{\mathrm{btl}}$ 
in one bandwidth-probing phase and reduce
it to $\nicefrac{3}{4}\cdot x^{\mathrm{btl}}$ in the next phase. 
However, these bandwidth-probing deviations happen in only two
consecutive phases of the 8-phase bandwidth-probing cycle,
and tend to cancel each other when averaged over time.
Hence, we adopt \cref{eq:bbr-cubic:model:basic-model:total-load}
as a suitable
approximation for the large-scale link-load dynamics 
that are relevant for the fluid-model equilibrium.

The loss rate~$p_{\ell}$ is based on the \emph{excess} sending rate: 
\begin{equation}
    p_{\ell} = \begin{cases}
        \frac{y_{\ell} - C_{\ell}}{y_{\ell}} & \text{if } y_{\ell} > C_{\ell} \land q_{\ell} = B_{\ell},\\
        0 & \text{otherwise,}
    \end{cases}
    \label{eq:bbr-cubic:model:basic-model:loss}
\end{equation}
where~$C_{\ell}$, $q_{\ell}$, and~$B_{\ell}$
are the link capacity, queue length and
buffer size at 
the bottleneck link~$\ell$, respectively.

\subsection{CUBIC Equilibrium}
\label{sec:bbr-cubic:model:cubic-equilibrium}

The behavior of a CUBIC flow~$k$ is captured by the two 
variables~$w^{\max}_k$ and~$s_k$, which 
evolve as follows in the model by Vardoyan et al.~\cite{vardoyan2021towards}:
\begin{align}
    \dot{w}^{\max}_k &= \left(w_k - w^{\max}_k\right) \cdot x_k^{\mathrm{C}} \cdot p_{\ell} \label{eq:bbr-cubic:system-evolution:cubic:wmax} \Dtag\\
    \dot{s}_k &= 1 - s_k \cdot x_k^{\mathrm{C}} \cdot p_{\ell} \label{eq:bbr-cubic:system-evolution:cubic:s} \Dtag
\end{align} 

\cref{eq:bbr-cubic:system-evolution:cubic:wmax} adjusts the maximum recorded window~$w^{\max}$
towards the current window for each lost segment.
Similarly, \cref{eq:bbr-cubic:system-evolution:cubic:s} resets the window-growth duration~$s$
to 0 for each lost segment, and lets~$s$ grow linearly in absence of loss. 

Given these CUBIC dynamics,
we are now interested in the \emph{fluid equilibrium} of CUBIC flows, i.e.,
the values $\overline{w}^{\max}_k$ and~$\overline{s}_k$ that are preserved by the dynamics in~\cref{eq:bbr-cubic:system-evolution:cubic:wmax,eq:bbr-cubic:system-evolution:cubic:s}.
This complete stasis is possible in fluid models because fluid models
approximate network metrics in a time-averaged manner; in reality, network metrics fluctuate
around their fluid average and never become fully static.
In the fluid model, the CUBIC state variables are in equilibrium if
\begin{align}
    \forall k \in F_{\mathrm{C}}.  \hspace{15pt}  &\dot{w}^{\max}_k =  \left(\overline{w}_k - \overline{w}^{\max}_k\right) \cdot \overline{x}_k^{\mathrm{C}} \cdot \overline{p}_{\ell} = 0 \label{eq:bbr-cubic:system-evolution:cubic:wmax-eq}\\
    &\dot{s}_k = 1 - \overline{s}_k \cdot \overline{x}_k^{\mathrm{C}} \cdot \overline{p}_{\ell} = 0  \label{eq:bbr-cubic:system-evolution:cubic:s-eq}
\end{align}
\cref{eq:bbr-cubic:system-evolution:cubic:s-eq} implies that 
none of~$\overline{s}_k$, $\overline{x}_k^{\mathrm{C}}$, and~$\overline{p}_{\ell}$
can be zero in equilibrium. We apply this insight on \cref{eq:bbr-cubic:system-evolution:cubic:wmax-eq}:
\begin{equation}
    \begin{split}
    &\dot{w}^{\max}_k = 0 \overset{\text{\cref{eq:bbr-cubic:system-evolution:cubic:wmax-eq}}}{\underset{\overline{x}_k^{\mathrm{C}},\overline{p}_{\ell} \neq 0}{\iff}} \overline{w}_k = \overline{w}^{\max}_k
    \overset{\text{\cref{eq:bbr-cubic:model:basic-model:cubic-window-growth}}}{\iff}\\[-2mm]
    &\overline{w}^{\max}_k = \overline{w}^{\max}_k + c \left(\overline{s}_k - \sqrt[3]{\frac{\overline{w}^{\max}_k b}{c}}\right)^3 \overset{\text{solve}}{\iff} 
    \overline{w}^{\max}_k = \frac{c}{b}\overline{s}_k^{3}. \label{eq:bbr-cubic:system-evolution:cubic:eq:wmax}
    \end{split}
\end{equation}
Hence, in the fluid equilibrium, the congestion-window size~$w_k$
is constant over time. This stasis is enabled by the \emph{CUBIC-stabilizing loss}~$\overline{p}_{\ell}$,
which balances growth and reduction in the fluid-averaged congestion-window size $w_k$ 
(In reality, the CUBIC window size will fluctuate in presence of loss).
This loss~$\overline{p}_{\ell}$ is obtained by 
inserting~$\overline{x}_k^{\mathrm{C}} = \overline{w}^{\max}_k/\overline{\tau}_k$ 
into~\cref{eq:bbr-cubic:system-evolution:cubic:s-eq}:
% Note that fluid models
% average network metrics over time; in reality, the loss actually occurs for
% specific packets rather than continuously, 
% and thereby causes multiplicative-decrease fluctuations in the CUBIC sending rate.
\begin{equation}
     \dot{s}_k \overset{\text{(\ref{eq:bbr-cubic:system-evolution:cubic:s-eq})}}{=} 1 - \overline{s}_k \cdot \overline{x}_k^{\mathrm{C}} \cdot \overline{p}_{\ell} = 0 
     % \underset{\overline{w}_k = \overline{w}_k^{\max}}{\overset{x_k^{\mathrm{C}} = w_k/\tau_k}{\iff}} 
     % 1 - \overline{s}_k \cdot \frac{\overline{w}_k^{\max}}{\overline{\tau}_k} \cdot \overline{p}_{\ell} = 0\\
     \overset{\text{\cref{eq:bbr-cubic:system-evolution:cubic:eq:wmax}}}{\iff}
     \overline{p}_{\ell} = \frac{b\overline{\tau}_k}{c \overline{s}_k^4} > 0.
     \label{eq:bbr-cubic:eq:loss}
\end{equation}
This CUBIC-stabilizing loss~$\overline{p}_{\ell}$ must
correspond to the actually occurring loss from~\cref{eq:bbr-cubic:model:basic-model:loss}, 
which in turn depends on the sending rates
of \emph{both} CUBIC and BBR flows:
\begin{lemma}
\textbf{CUBIC Equilibrium Conditions:}
\begin{equation}
    \forall k \in F_{\mathrm{C}}. \quad \overline{p}_{\ell} = \frac{b \overline{\tau}_k}{c \overline{s}_k^{4}} = \frac{\overline{y}_{\ell} - C_{\ell}}{\overline{y}_{\ell}}
    \label{eq:bbr-cubic:eq:cubic-conditions}
\end{equation}
\label{lmm:bbr-cubic:eq:cubic-conditions}
\end{lemma}

Since $\overline{s}_k > 0$ follows from~\cref{eq:bbr-cubic:system-evolution:cubic:s-eq}, 
\cref{eq:bbr-cubic:eq:loss} implies~$\overline{p}_{\ell} > 0$,
i.e., the equilibrium requires that congestion 
fills the whole buffer ($\overline{q}_{\ell} = B_{\ell}$)
and loss occurs. For that reason, the delays in the fluid equilibrium become:
\begin{equation}
    \forall j \in F_{\mathrm{B}} \cup F_{\mathrm{C}}. \quad \overline{\tau}_j = \tau_j^{\mathrm{p}} + \frac{\overline{q}_{\ell}}{C_{\ell}} = \tau_j^{\mathrm{p}} + \frac{B_{\ell}}{C_{\ell}},
    \label{eq:bbr-cubic:eq:delay}
\end{equation}
where~$\tau_j^{\mathrm{p}}$ is the propagation delay 
experienced by flow~$j$.

\subsection{BBR Equilibrium}

The bottleneck estimate $x_i^{\mathrm{btl}}$ of BBR flow~$i$ tracks the 
maximum delivery rate~$x^{\mathrm{dlv}}_i$~\cite{scherrer2022model},
reflected in a differential equation that
continuously adjusts $x_i^{\mathrm{btl}}$ towards
$x^{\mathrm{dlv}}_i$:
\begin{equation}
    \dot{x}_i^{\mathrm{btl}} = x_i^{\mathrm{dlv}} - x_i^{\mathrm{btl}} \Dtag
    \label{eq:bbr-cubic:system-evolution:bbr}
\end{equation} with
\begin{equation}
    x_i^{\mathrm{dlv}} = \begin{cases}
        \frac{\alpha_i x_i^{\mathrm{btl}} C_{\ell}}{y_{\ell} + (\alpha_i - \beta_i)x^{\mathrm{btl}}_i} & \text{if } y_{\ell} + (\alpha_i - \beta_i)x^{\mathrm{btl}}_i \geq C_{\ell},\\
        \alpha_i x_i^{\mathrm{btl}} & \text{otherwise,}
    \end{cases}
\end{equation} 
where~$\alpha_i$ denotes the \emph{probing strength} of the BBR flow:
In its 8-RTT bandwidth-probing cycle, a BBR flow raises its pacing rate
during a single RTT (\texttt{ProbeBW\_UP} phase) to discover 
whether additional bandwidth is available.
In this phase, the BBR flow sets the pacing rate to~$\nicefrac{5}{4}\cdot x_i^{\mathrm{btl}}$,
while being constrained by the congestion-window size of~$2\tau^{\min}_i x^{\mathrm{btl}}$:
\begin{equation}
     \alpha_i = \min\left(\nicefrac{5}{4},\ 2 \tau^{\min}_i/ \tau_i \right).
      \label{eq:bbr-cubic:system-evolution:bbr:alpha}
\end{equation}

We now extend previous work by identifying the BBR/ CUBIC equilibrium
(outside of the RTT-probing steps).
Since~$\overline{y}_{\ell} > C_{\ell}$ (implied by \cref{lmm:bbr-cubic:eq:cubic-conditions})
and $\alpha_i \geq \beta_i$ (\cref{eq:bbr-cubic:system-evolution:bbr:alpha,eq:bbr-cubic:system-evolution:bbr:beta}),
\cref{eq:bbr-cubic:system-evolution:bbr} implies the BBR equilibrium condition:
\begin{equation}
    \forall i \in F_{\mathrm{B}}. \quad \frac{\overline{\alpha}_i \overline{x}_i^{\mathrm{btl}} C_{\ell}}{\overline{y}_{\ell} + \left(\overline{\alpha}_i - \overline{\beta}_i\right)\overline{x}^{\mathrm{btl}}_i} - \overline{x}^{\mathrm{btl}}_i = 0.
\end{equation}
These conditions admit the equilibrium~$\overline{x}_i^{\mathrm{btl}} = 0\ \forall i$.
However, $\overline{x}_i^{\mathrm{btl}} = 0$ is an artificial equilibrium: 
The BBR implementation never sets~$x^{\mathrm{btl}}$ to zero, 
even under 100\% packet loss,
to allow recovery of the sending rate~\cite{bbrv1-source}.
Effectively, $x^{\mathrm{btl}}$ is thus lower-bounded by some small number~$\chi > 0$.
Hence, we arrive at the following equilibrium conditions:

\begin{lemma}
    \textbf{BBR Equilibrium Conditions:} $\forall i \in F_{\mathrm{B}}.$
    \begin{equation}
        \overline{x}^{\mathrm{btl}}_i =
        \max\left(\chi,\ C_{\ell} - \frac{1}{\overline{\alpha}_i}\left(\sum_{\substack{j\in F_{\mathrm{B}}\setminus\{i\}}} \overline{\beta}_j \overline{x}^{\mathrm{btl}}_j + \sum_{k \in F_{\mathrm{C}}} \overline{x}^{\mathrm{C}}_k\right)\right)
        \label{eq:bbr-cubic:eq:bbr-conditions}
    \end{equation}
    \label{lmm:bbr-cubic:eq:bbr-conditions}
    \vspace{-2mm}
\end{lemma}

To determine the strengths~$\overline{\alpha}_i$ and~$\overline{\beta}_i$ in equilibrium,
% we require the equilibrium minimum-RTT estimate~$\overline{\tau}^{\min}_i$.
% To determine~$\overline{\tau}^{\min}_i$, 
we first note that the CUBIC equilibrium
conditions in~\cref{lmm:bbr-cubic:eq:cubic-conditions} imply a consistently full buffer 
in equilibrium (outside the RTT-probing steps).
Hence, the BBR flows never spontaneously observe decreasing RTT samples in equilibrium,
and must perform an RTT-probing step every 10 seconds. In an RTT-probing step,
a BBR flow~$i$ reduces its congestion window to 4~segments, with the goal of uncovering the 
path propagation delay~$\tau^{\mathrm{p}}_i$.
Intriguingly, BBR flows synchronize their 
RTT-probing steps (cf.~\cref{sec:bbr-cubic:dynamic:oscillation:synchronization})
such that the propagation delay is indeed uncovered if only BBR flows compete.
However, in our scenario, the bottleneck link~$\ell$ is shared with CUBIC flows, which do
not participate in the buffer draining.
Hence, the equilibrium 
min-RTT estimate~$\overline{\tau}^{\min}_i$ 
is inflated by the minimal \emph{back-off queue length}~$\overline{q}^-_{\ell}$ 
remaining in the RTT-probing step:
\begin{equation}
    \overline{q}^-_{\ell} = \left[ 4N_{\mathrm{B}} + (1-b)\sum_{k \in F_{\mathrm{C}}} \overline{w}_k - \tau^{\mathrm{p}}_{\ell} C_{\ell}\right]_0^{B_{\ell}}
    \implies \overline{\tau}^{\min}_i = \tau^{\mathrm{p}}_i + \frac{\overline{q}^-_{\ell}}{C_{\ell}}.
    \label{eq:bbr-cubic:system-evolution:remaining-queue}
\end{equation} 
Since~$\overline{q}^-_{\ell}$ can neither be negative 
nor exceed the bottleneck-buffer capacity~$B_{\ell}$,
we use the notation~$[\cdot]_{0}^{B_{\ell}}$ for projection
to the interval~$[0, B_{\ell}]$.
Intuitively, the remaining queue volume~$\overline{q}_{\ell}^{-}$ contains the inflight volume
of all flows when the BBR flows are in an RTT-probing step
and the CUBIC flows are minimal, i.e., 
back off because of loss (factor $1-b$). 
In case of congestion, this inflight data tends to accumulate on the bottleneck link~$\ell$.
Hence, this inflight data is discounted by the volume
that fits in the pipe, i.e., the BDP $\tau^{\mathrm{p}}_{\ell} C_{\ell}$ ($\tau^{\mathrm{p}}_{\ell}$
being the propagation delay of bottleneck link~$\ell$).

The equilibrium conditions in~\cref{lmm:bbr-cubic:eq:cubic-conditions} (CUBIC flows)
and in~\cref{lmm:bbr-cubic:eq:bbr-conditions} (BBR flows)
form a system of $N = N_{\mathrm{B}} + N_{\mathrm{C}}$ nonlinear equations
with~$N$ variables. This equation system can then be solved to compute an equilibrium
rate distribution. While such solutions are difficult in general,
the starvation of BBR flow~$i$ is easy to derive given~$\overline{\alpha}_i \leq 1$:
\begin{lemma}
    \textbf{Sufficient Condition for BBR Starvation:}
    \begin{equation}
        \overline{\alpha}_i \leq 1 \quad \implies \quad \overline{x}^{\mathrm{btl}}_i = \chi
    \end{equation}
    \label{lmm:bbr-cubic:eq:bbr-starvation}
\end{lemma}
\vspace{-25pt}
\begin{proof}
    From~\cref{lmm:bbr-cubic:eq:cubic-conditions}, we know that~$\overline{y}_{\ell} > C_{\ell}$, and hence:
    \begin{equation}
    \setlength{\jot}{-2pt}
    \begin{split}
        &\overline{y}_{\ell} > C_{\ell} \overset{\overline{\alpha}_i \leq 1}{\iff} \overline{y}_{\ell} > \overline{\alpha}_i C_{\ell} 
        \overset{/\overline{y}_{\ell}}{\underset{-1}{\iff}}
        0 > \frac{\overline{\alpha}_i  C_{\ell}}{\overline{y}_{\ell}} - 1\\
        \overset{\cdot\overline{x}_i^{\mathrm{btl}}}{\underset{\overline{\alpha}_i = \overline{\beta}_i}{\iff}}
        &0 > \frac{\overline{\alpha}_i \overline{x}_i^{\mathrm{btl}} C_{\ell}}{\overline{y}_{\ell}  + (\overline{\alpha}_i - \overline{\beta}_i)\overline{x}^{\mathrm{btl}}_i} - \overline{x}^{\mathrm{btl}}_i = \dot{x}_i^{\mathrm{btl}}
    \end{split}
    \end{equation}
    Therefore, if~$x^{\mathrm{btl}}_i = \chi$, 
    BBR would reduce~$x^{\mathrm{btl}}_i$ further, but cannot. 
    Hence,~$\overline{x}_i^{\mathrm{btl}} = \chi$ is the unique equilibrium.
\end{proof}

\section{Conditions for Oscillation}
\label{sec:bbr-cubic:dynamic:oscillation:model}

% While~\cref{sec:bbr-cubic:dynamic} provides an intuitive explanation why
% oscillation arises in our experiment setting, 
In this section, we derive
% a rigorous
% mathematical characterization of this oscillation.
% With this characterization, we can identify the 
analytical conditions
under which BBR/CUBIC oscillation provably occurs.
% In the next two sections (\S\ref{sec:bbr-cubic:dynamic:oscillation:model:prediction} and \S\ref{sec:bbr-cubic:dynamic:oscillation:model:fairness}), we will
% then apply the model to gain practical insights.

\subsection{Analysis Overview}
\label{sec:bbr-cubic:dynamic:oscillation:model:overview}

We distinguish short-term and long-term dynamics (\cref{fig:bbr-cubic:dynamic:oscillation:why:mechanism}).

The \emph{short-term dynamics}
describe the continuous BBR/CUBIC competition 
\emph{between} the RTT probing steps of BBR. 
Between these probing steps,
the BBR probing strength~$\alpha$ is fixed,
and fully determines the equilibria (\cref{sec:bbr-cubic:dynamic:oscillation:model:eq})
and the convergence behavior (\cref{sec:bbr-cubic:dynamic:oscillation:model:convergence}) 
of the short-term dynamics.

Building on these short-term dynamics, we derive 
the \emph{long-term dynamics}, describing the BBR/CUBIC competition
\emph{across} RTT-probing steps (\cref{sec:bbr-cubic:dynamic:oscillation:model:condition}).
In each RTT-probing step, the CUBIC congestion-window size~$w$
determines the probing strength~$\alpha$ (\cref{eq:bbr-cubic:system-evolution:bbr:alpha}) 
and thus the short-term dynamics for the next 10 seconds.
These short-term dynamics then determine the 
congestion-window size~$w$ at the next RTT-probing step.
Hence, the long-term dynamics form a discrete process.

In the following, we consider the case for one BBR flow and one CUBIC 
flow~($N_{\mathrm{B}} = 1$, $N_{\mathrm{C}} = 1$). 
For simplicity, we thus eliminate the flow-specific subscripts~$i$ and~$k$ where
the association is obvious, e.g., only the BBR flow~$i$ has a probing strength~$\alpha_i = \alpha$.
As discussed in~\cref{sec:bbr-cubic:dynamic:oscillation:synchronization},
the scenarios for multiple flows per CCA are qualitatively similar
for a majority of CCA compositions.

\subsection{Short-Term Equilibria}
\label{sec:bbr-cubic:dynamic:oscillation:model:eq}

In~\cref{sec:bbr-cubic:dynamic}, 
we observed that the minimum-RTT estimate~$\tau^{\min}$ of the BBR flow~$i$ is only periodically
updated in competition with CUBIC.
To investigate the dynamics between these updates, we can thus treat~$\tau^{\min}$
as fixed, and by extension also the probing strength~$\alpha$ (cf.~\cref{eq:bbr-cubic:system-evolution:bbr:alpha}).
This fixed~$\alpha$ then determines the equilibrium~$\tilde{\sigma}(\alpha)$ of the short-term dynamics:

\begin{theorem} \textbf{Unique Short-Term Equilibrium.}
    Given probing strength~$\alpha$, the unique
    short-term equilibrium $\tilde{\sigma}(\alpha) = (\tilde{x}^{\mathrm{btl}}(\alpha),$ $\tilde{w}^{\max}(\alpha), \tilde{s}(\alpha))$
    for competition between BBR flow~$i$ and CUBIC flow~$k$
    on a bottleneck link~$\ell$ can be computed as follows:
    \begin{enumerate}
        \item Find discriminant probing strength~$\hat{\alpha}$ for the considered network by solving:
        \begin{equation}
            \frac{\hat{\alpha}^4(\hat{\alpha}-1)^3}{(\chi+\hat{\alpha}(C_{\ell}-\chi))^3} = \frac{c}{b\overline{\tau}_k(C_{\ell}-\chi)^7}, \label{eq:bbr-cubic:dynamic:oscillation:model:equilibrium:alpha-crit}
        \end{equation}
        \item Find equilibrium CUBIC window-growth duration~$\tilde{s}(\alpha)$ by solving:
        \begin{align}
                \boldsymbol{\alpha \geq \hat{\alpha}:}\quad\quad & \tilde{S}_1(\tilde{s}) = \frac{(\alpha -1)c^2}{\alpha b\overline{\tau}_k} \tilde{s}^7
            - \frac{\left(\alpha-1\right)c}{\alpha} \tilde{s}^3
            - b C_{\ell} \overline{\tau}_k = 0
            \label{eq:bbr-cubic:dynamic:oscillation:model:equilibrium:additional:S1}\\
            \boldsymbol{\alpha < \hat{\alpha}:}\quad\quad &\tilde{S}_2(\tilde{s}) =  \frac{c^{2}}{b\overline{\tau}_k}\tilde{s}^{7}-c\left(C_{\ell}-\alpha\chi\right)\tilde{s}^{4}-c\tilde{s}^{3}-\alpha b\overline{\tau}_k\chi = 0,
            \label{eq:bbr-cubic:dynamic:oscillation:model:equilibrium:additional:S2}
        \end{align}
        \item Compute equilibrium CUBIC congestion window size $\tilde{w}^{\max}(\alpha) = \frac{c}{b} \tilde{s}(\alpha)^{3}$
        \item Compute equilibrium BBR bottleneck estimate 
        $\tilde{x}^{\mathrm{btl}}(\alpha) = \begin{cases}
                C_{\ell} - \frac{\tilde{w}^{\max}(\alpha)}{\alpha \overline{\tau}_k} & \text{if } \alpha \geq \hat{\alpha},\\
                \chi & \text{if } \alpha < \hat{\alpha}
            \end{cases}$
    \end{enumerate}
    \label{thm:bbr-cubic:dynamic:oscillation:model:equilibrium}
\end{theorem}

\begin{proofsketch}

    The full proof of~\cref{thm:bbr-cubic:dynamic:oscillation:model:equilibrium} is provided in~\cref{prf::bbr-cubic:dynamic:oscillation:model:equilibrium}.
    The proof first considers the case~$\alpha \leq 1$, which implies~$\tilde{x}^{\mathrm{btl}}(\alpha) = \chi$
    according to \cref{lmm:bbr-cubic:eq:bbr-starvation}. Inserting this value into the CUBIC equilibrium conditions
    from \cref{lmm:bbr-cubic:eq:cubic-conditions} yields the condition~$\tilde{S}_2(s) = 0$. 
    The septic polynomial~$\tilde{S}_2$
    has a unique root,
    which guarantees a unique equilibrium~$\tilde{s}(\alpha)$. 
    Second, the proof considers the case~$\alpha > 1$,
    and distinguishes the sub-cases~$\hat{x}^{\mathrm{btl}} \geq \chi$ and~$\hat{x}^{\mathrm{btl}} < \chi$,
    where~$\hat{x}^{\mathrm{btl}} = C_{\ell} - \tilde{x}^{\mathrm{C}}/\alpha$ is the equilibrium BBR bottleneck-bandwidth
    estimate without restriction to the domain~$[\chi, \infty)$. If~$\hat{x}^{\mathrm{btl}} \geq \chi$, 
    combining~$\tilde{x}^{\mathrm{btl}} = \hat{x}^{\mathrm{btl}}$ with the CUBIC equilibrium conditions from~\cref{lmm:bbr-cubic:eq:bbr-conditions}
    yields~$\tilde{S}_1(s) = 0$, which again has a unique solution.
    For the other sub-case~$\hat{x}^{\mathrm{btl}} < \chi$, we identify the probing strength~$\hat{\alpha} > 1$ (with \cref{eq:bbr-cubic:dynamic:oscillation:model:equilibrium:alpha-crit})
    such that~$\chi$ is an unrestricted equilibrium, i.e., $\hat{x}^{\mathrm{btl}} = \tilde{x}^{\mathrm{btl}} = \chi$.
    % This critical~$\hat{\alpha}$ is found by solving~\cref{eq:bbr-cubic:dynamic:oscillation:model:equilibrium:alpha-crit}.
    For any~$\alpha < \hat{\alpha}$,  it holds that $\hat{x}^{\mathrm{btl}} < \chi$ 
    and thus~$\tilde{x}^{\mathrm{btl}}(\alpha) = \chi$.
\end{proofsketch}

\subsection{Stability of Short-Term Equilibria}
\label{sec:bbr-cubic:dynamic:oscillation:model:convergence}

\begin{figure*}
    \begin{minipage}{0.45\linewidth}
        \centering
        \includegraphics[width=\linewidth, trim=35 15 16 4,clip]{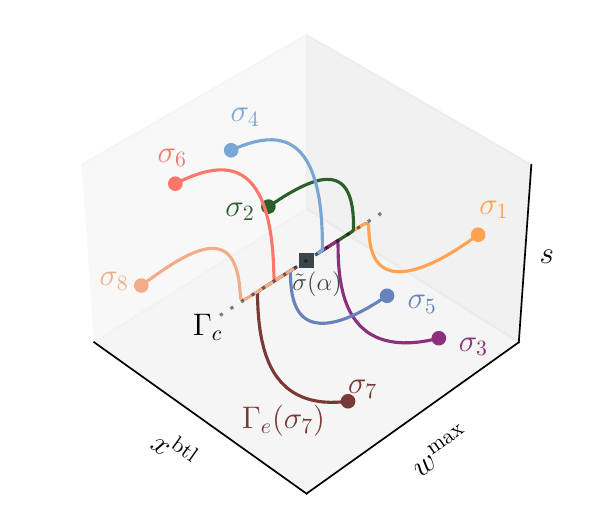}
        % \vspace{-5pt}
        \caption{Two-step convergence
        in short-term dynamics:
        First exponential convergence \emph{towards} the center manifold~$\Gamma_c$
        (a function of~$w^{\max}$), 
        then sub-exponential \emph{along}~$\Gamma_c$ towards the short-term equilibrium~$\tilde{\sigma}(\alpha)$.}
        \label{fig:bbr-cubic:dynamic:oscillation:model:center-manifold}
    \end{minipage}\quad\vrule\quad
    \begin{minipage}{0.45\linewidth}
            \centering
            \includegraphics[width=\linewidth]{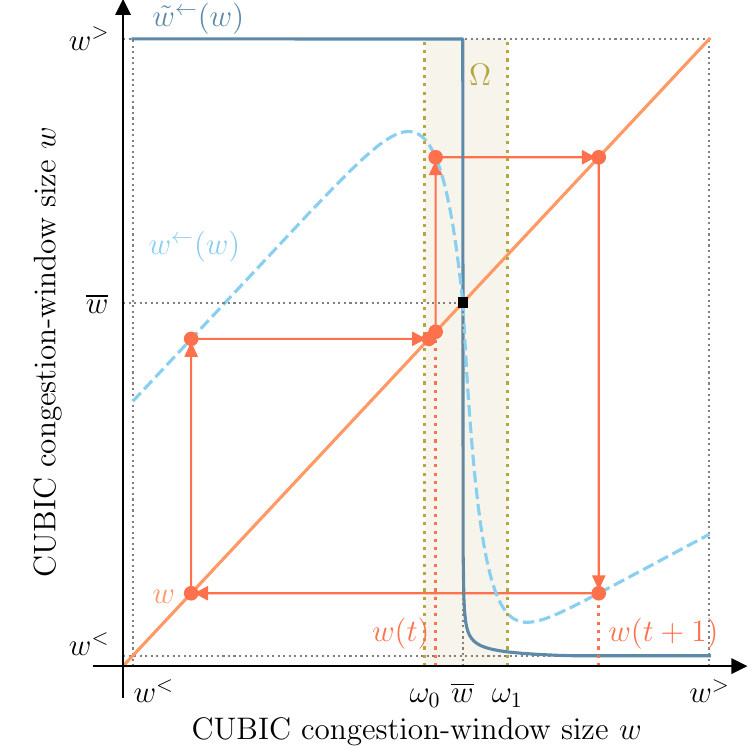}
            \caption{Instability of long-term dynamics:
        Under the condition from~\cref{thm:bbr-cubic:dynamic:oscillation:model:condition}, 
        the dynamics suffer from persistent oscillation as the evolution trajectory moves
        away from the CUBIC long-term equilibrium
        window~$\overline{w}$.}
            \label{fig:bbr-cubic:dynamic:oscillation:model:oscillation}
    \end{minipage}
\end{figure*}

For the short-term equilibria to be relevant,
the short-term dynamics have to converge to
these equilibria. This attractiveness
is necessary for the BBR/CUBIC dynamics to converge
to a new short-term equilibrium rate distribution if the minimum-RTT 
estimate~$\tau^{\min}$ is updated and~$\alpha$ thus 
changes. 
To confirm the attractiveness of the short-term equilibria,
we prove the \emph{asymptotic stability} of these equilibria,
meaning that the competition converges to the short-term equilibrium
if the initial rate distribution is close enough:
\begin{theorem}
    \textbf{Stability of Short-Term Equilibrium.}
    In the competition between one BBR flow and one CUBIC flow,
    the short-term equilibrium $\tilde{\sigma}(\alpha) =$ 
    $(\tilde{x}^{\mathrm{btl}}(\alpha),$
    $\tilde{w}^{\max}(\alpha),\ \tilde{s}(\alpha))$
    of the joint dynamics is asymptotically stable,
    i.e., the BBR/CUBIC dynamics converge to the rate distribution of equilibrium~$\tilde{\sigma}(\alpha)$.
    \label{thm:bbr-cubic:dynamic:oscillation:model:convergence}
\end{theorem}
\begin{proofsketch}
    We provide the full rigorous proof in~\cref{prf:bbr-cubic:dynamic:oscillation:model:convergence},
    and provide a high-level overview here.
    
    Importantly, a straightforward stability proof via linearization
    of the dynamic system fails, for the reasons already 
    noted by Vardoyan et al.~\cite{vardoyan2021towards}
    who considered a CUBIC flow in isolation.
    Namely, when the BBR/CUBIC dynamics approach
    the equilibrium~$\tilde{\sigma}(\alpha)$, the evolution of the CUBIC maximum
    window~$w^{\max}$ is exclusively determined 
    by high-order terms that are not captured by the linearization.
    Formally, the Jacobian matrix of the BBR/CUBIC differential-equation system
    has both zero and negative eigenvalues when evaluated at equilibrium,
    rendering the stability investigation via linearization inconclusive.
    Hence, more advanced methods are required.

    In our case, we can rely on the 
    \emph{center-manifold emergence theorem}~\cite{rard1998topics}:
    Given inconclusive linearization,
    the dynamics converge exponentially fast to the 
    \emph{center manifold} and then approximately follow
    a slow (i.e., sub-exponential) 
    trajectory along the center manifold.
    The center manifold contains the equilibrium, 
    so the behavior of trajectories on this manifold determines convergence to equilibrium. 
    Moreover, because the center manifold is typically low-dimensional, 
    it allows a tractable analysis of the system’s stability.
    
    More precisely, the differential-equation system is first transformed 
    into the eigenbasis of the linearized system (i.e., of the Jacobian matrix), 
    which decouples the state into center and stable subspaces; the center variable
    corresponds to the direction with zero eigenvalue (the center subspace), 
    while the stable variables correspond to directions with negative eigenvalues 
    (the stable subspace). Expressing the stable variables as functions of the center variable, 
    the dynamics are then projected onto the center manifold, 
    giving a one-dimensional equation that governs motion along the center manifold. 
    Stability can then be analyzed directly in this reduced one-dimensional system.
    
    Figure~\ref{fig:bbr-cubic:dynamic:oscillation:model:center-manifold} 
    illustrates this convergence behavior for the dynamics
    between one BBR flow and one CUBIC flow.
    As these dynamics involve the three 
    variables~$x^{\mathrm{btl}}$ (for the BBR flow),
    $w^{\max}$, and~$s$ (both for the CUBIC flow),
    the state space of the competition dynamics is
    three-dimensional. The system evolution
    from a starting point~$\sigma_n$ is a path through the state space. 
    This path is determined 
    by the differential equations governing the three variables,
    and is visualized by a curve~$\Gamma_e(\sigma_n)$ in the 
    three-dimensional space. 
    This evolution curve~$\Gamma_e(\sigma_n)$ approaches the center manifold 
    exponentially quickly. In our proof, this center
    manifold can be expressed by means of the single
    variable~$w^{\max}$, and thus corresponds to another curve~$\Gamma_c$.
    Moreover, the system-evolution path~$\Gamma_e(\sigma_n)$ tracks the
    center-manifold curve~$\Gamma_c$
    towards the equilibrium~$\tilde{\sigma}(\alpha)$, illustrating that the
    short-term equilibrium is asymptotically stable.
\end{proofsketch}

\subsection{Long-Term Dynamics}
\label{sec:bbr-cubic:dynamic:oscillation:model:condition}

So far, our analysis suggests that the BBR/CUBIC dynamics converge to a short-term equilibrium~$\tilde{\sigma}(\alpha)$ for a fixed minimum-RTT estimate~$\tau^{\min}$ and fixed probing 
strength~$\alpha$.
While~$\tau^{\min}$ is indeed fixed for periods of time,
it is also regularly adjusted, at least every 10 seconds:
To readjust~$\tau^{\min}$, the BBR flow measures the 
RTT during RTT probing, meaning the RTT determined by the back-off 
queue length~$q_{\ell}^{-}$ 
from \cref{eq:bbr-cubic:system-evolution:remaining-queue}.
In turn, this back-off queue length is determined 
by the CUBIC congestion-window size~$w$ at the time of RTT probing,
because part of that CUBIC volume resides in the buffer.

During RTT probing, the CUBIC congestion-window size $w$ 
is evolving towards
the window size~$\tilde{w}(\alpha') = \tilde{w}^{\max}(\alpha')$
as defined in~\cref{eq:bbr-cubic:system-evolution:cubic:eq:wmax}).
This target equilibrium depends on the previous probing strength~$\alpha'$.

However, convergence is typically incomplete by the time RTT probing occurs.
The reason is that the state variable~$w^{\max}$, which governs the evolution of the CUBIC window, 
approaches its equilibrium only with sub-exponential speed (see~\cref{thm:bbr-cubic:dynamic:oscillation:model:equilibrium}).
As a consequence, the actual window size at probing, denoted by $w'$, may differ from the short-term equilibrium, 
i.e., $w' \neq \tilde{w}(\alpha')$.
This intermediate size $w'$ is then used to readjust the probing strength~$\alpha$.

Thus, the long-term evolution of the CUBIC window size~$w$
is described by the discrete-time process:
\begin{equation}
% \begin{split}
    \forall t \in \mathbb{N}, t\geq 0. \quad w(t+1) = w^{\leftarrow}(w(t)).
% \end{split}
    \label{eq:bbr-cubic:dynamic:oscillation:model:conditions:discrete}
\end{equation}
Here, $t$ indexes the probing steps of BBR, which are separated by about 10 seconds.  
The function $w^{\leftarrow}$ maps the starting window size $w(t)$ to the resulting window size after one probing interval.  
This function~$w^{\leftarrow}$ reflects the short-term dynamics between $t$ and $t+1$.  
These short-term dynamics are governed by the probing strength~$\alpha$ during the interval, 
which is itself determined by the window size $w(t)$ at the beginning of the interval.  
Within a probing interval, the BBR bandwidth estimate~$x^{\mathrm{btl}}$ converges much faster 
than the CUBIC window (see~\cref{thm:bbr-cubic:dynamic:oscillation:model:equilibrium}).  
Consequently, the initial value~$x^{\mathrm{btl}}(t)$ has negligible influence on $w^{\leftarrow}$.

Hence, we distinguish between two update functions:
\begin{compactitem}
    \item \textbf{Idealized update function} $\tilde{w}^{\leftarrow}(w)$: assumes complete short-term convergence within an interval, 
    producing the short-term equilibrium window size $\tilde{w}$, and
    \item \textbf{Actual update function} $w^{\leftarrow}(w)$: may stop short of equilibrium and is therefore only bounded between~$w$ and $\tilde{w}^{\leftarrow}(w)$.
\end{compactitem}

The process in~\eqref{eq:bbr-cubic:dynamic:oscillation:model:conditions:discrete} 
is guaranteed to have a unique long-term equilibrium window size~$\overline{w}$,
corresponding to the fluid-model equilibrium from \cref{sec:new-model}.  
However, under mild conditions, this equilibrium is \emph{unstable}:

\begin{theorem}
    \textbf{Instability of Long-Term Equilibrium.}
    Consider the dynamics of a BBR flow competing with a CUBIC flow.  
    There exists a unique long-term CUBIC equilibrium window size~$\overline{w}$, defined by the fixed point of $w^{\leftarrow}$.  
    This equilibrium is unstable, meaning that trajectories do not converge to $\overline{w}$ even when starting arbitrarily close,
    if there exists a neighborhood $\Omega$ in which the window-update function~$w^{\leftarrow}(w)$ decreases fast enough in~$w$:
        \begin{equation}
        % \begin{split}
            \exists\ \Omega = [\omega_0, \omega_1],\ \omega_0 < \overline{w} < \omega_1. \quad \forall \omega \in \Omega. 
            \quad \frac{\partial w^{\leftarrow}(\omega)}{\partial \omega} < -1
        % \end{split}
        \end{equation}
    % \end{enumerate}
    \label{thm:bbr-cubic:dynamic:oscillation:model:condition}
\end{theorem}
% \vspace{-10.5mm}
\begin{proofsketch}
    The proof from~\cref{prf::bbr-cubic:dynamic:oscillation:model:condition} 
    can be visualized using the fixed-point diagram in~\cref{fig:bbr-cubic:dynamic:oscillation:model:oscillation}.  

    First, consider the idealized update function $\tilde{w}^{\leftarrow}(w)$.  
    The function~$\tilde{w}^{\leftarrow}$ is fully known, bounded to the finite range $[w^{<}, w^{>}]$
    (derived in~\cref{prf::bbr-cubic:dynamic:oscillation:model:condition}), and monotonic.  
    Thus, $\tilde{w}^{\leftarrow}$ intersects the identity line exactly once, at $\overline{w}$,
    which is thus a unique long-term equilibrium window size~$\overline{w}$.

    In contrast, the actual update function $w^{\leftarrow}(w)$ is not known in closed form.  
    However, we can state that for any starting window size $w$, the resulting value $w^{\leftarrow}(w)$ 
    lies between $w$ and $\tilde{w}^{\leftarrow}(w)$.  
    This property follows from the asymptotic stability of the short-term equilibrium 
    (see~\cref{thm:bbr-cubic:dynamic:oscillation:model:convergence}): 
    the CUBIC window tends toward $\tilde{w}$ while $\alpha$ is fixed, 
    but convergence may be incomplete when $\alpha$ is updated.

    We can now describe the discrete-time evolution:  
    Starting from a point $(w(t), w(t))$ on the identity line, 
    the update step projects vertically to $(w(t), w^{\leftarrow}(w(t)))$ on the curve of $w^{\leftarrow}$.  
    We then obtain the next state $w(t+1)$ by projecting horizontally to the identity line.  
    Repeating this procedure yields the system trajectory.

    The stability of $\overline{w}$ depends on the slope of $w^{\leftarrow}$ near the equilibrium.  
    If the slope satisfies $\frac{\partial w^{\leftarrow}}{\partial w} < -1$, trajectories are pushed away rather than attracted.  
    Formally, in that case, there exists a neighborhood $\Omega$ around $\overline{w}$ 
    such that all trajectories entering $\Omega$ eventually leave it again.  
    Hence, under the stated condition, the long-term equilibrium is unstable.
\end{proofsketch}

This instability has practical implications:
It means that the long-term equilibrium cannot be used to predict aggregate throughput 
as in \cref{fig:static:bbr-cubic}, since the system may never settle at $\overline{w}$.  
Nevertheless, the formal characterization of instability enables new predictions 
that earlier BBR and CUBIC models cannot provide~\cite{mishra2022we,scherrer2022model,vardoyan2021towards,ware2019modeling}.  
In particular, we can now predict the frequency of oscillations (\cref{sec:bbr-cubic:dynamic:oscillation:model:prediction}) 
and the degree of unfairness between BBR and CUBIC during these oscillations (\cref{sec:bbr-cubic:dynamic:oscillation:model:fairness}).

Interestingly, the instability condition in Theorem~\ref{thm:bbr-cubic:dynamic:oscillation:model:condition} does not involve feedback delay.  
Traditionally, delay is seen as the main source of instability in congestion control~\cite{low2002internet,srikant2004mathematics}.  
Here, oscillations arise directly from the idealized dynamics 
of \cref{eq:bbr-cubic:system-evolution:cubic:wmax,eq:bbr-cubic:system-evolution:cubic:s,eq:bbr-cubic:system-evolution:bbr}, even without delay.  
Incorporating feedback delay remains an important direction for future work, as it may reveal additional modes of instability.  
However, \cref{sec:bbr-cubic:dynamic:oscillation:model:prediction} shows 
that our delay-free condition already predicts oscillatory behavior with high accuracy.

\subsection{Incidence of Oscillation}
\label{sec:bbr-cubic:dynamic:oscillation:model:prediction}

\begin{figure*}
    \begin{minipage}{0.45\linewidth}
        \centering
        \includegraphics[width=\linewidth]{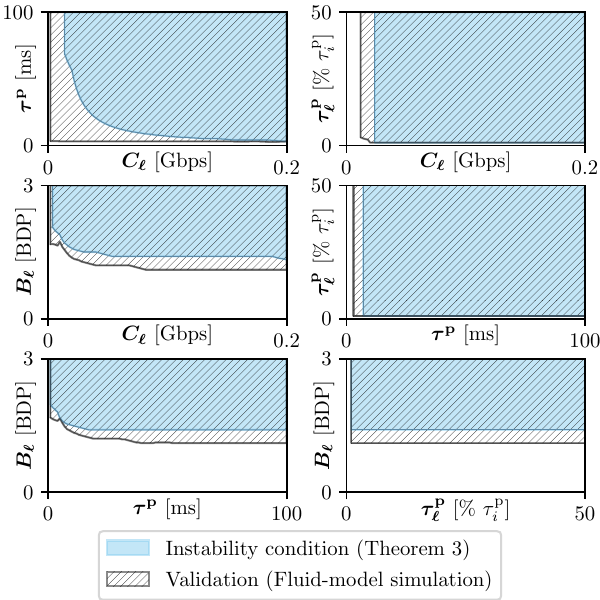}
        \caption{Network-parameter space (blue) that satisfies the oscillation
        condition from~\cref{thm:bbr-cubic:dynamic:oscillation:model:condition}. 
        The shaded area marks the parameter space for which the fluid-model simulation indicates oscillation (Constant parameters listed in~\cref{sec:bbr-cubic:dynamic:oscillation:model:prediction}).}
        \label{fig:bbr-cubic:dynamic:oscillation:model:condition:space}
    \end{minipage}\quad\vrule\quad
    \begin{minipage}{0.45\linewidth}
        \centering
        \includegraphics[width=\linewidth,trim=0 0 0 0]{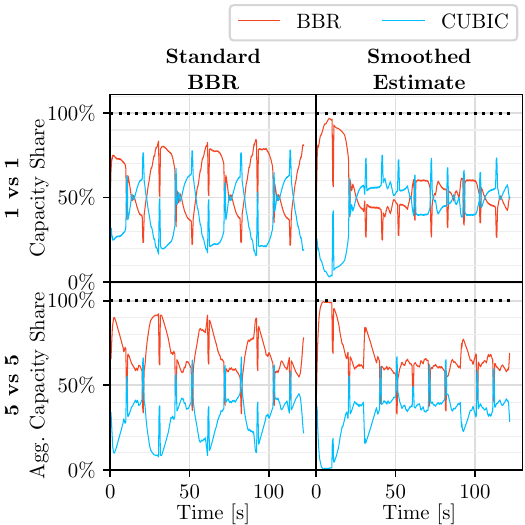}
        \vspace{-6mm}
        \caption{Evaluation of the BBR adaptation using a moving average of the minimum-RTT estimate ($\tau^{\min}$) with $\theta = \nicefrac{1}{6}$).
        As a result of the adaptation, 
        the oscillation is successfully dampened.}
        \label{fig:remedies:smoothed-estimate}
    \end{minipage}
\end{figure*}

After leveraging the oscillation model to elicit abstract conditions
for oscillation, we now investigate how frequently 
these conditions are satisified, i.e., how often
an unstable neighborhood~$\Omega$ around
the long-term equilibrium~$\overline{w}$ exists. 

We first note that the unknown  update function $w^{\leftarrow}$ can be well approximated by
the equilibrium window-update function~$\tilde{w}^{\leftarrow}$ 
around~$\overline{w}$, where the unstable
neighborhood~$\Omega$ should exist:
Around $w \approx \overline{w}$, the difference between the window size~$w$ and 
its associated short-term equilibrium~$\tilde{w}^{\leftarrow}(w)$ shrinks
(because $\overline{w} = \tilde{w}^{\leftarrow}(\overline{w}) = w^{\leftarrow}(\overline{w})$). 
Hence, the congestion window likely manages to converge to the short-term equilibrium,
and thus~$w^{\leftarrow}(w) \approx \tilde{w}^{\leftarrow}(w)$ for~$w \approx \overline{w}$.
Therefore, we substitute the unknown function~$w^{\leftarrow}$ in \cref{thm:bbr-cubic:dynamic:oscillation:model:condition} 
by the known function~$\tilde{w}^{\leftarrow}$ for our analysis.

Given this substitution,
we evaluate the strictness of the condition in~\cref{thm:bbr-cubic:dynamic:oscillation:model:condition} 
by testing a range of parameter combinations similar 
to the experiment configurations from~\cref{fig:bbr-cubic:dynamic:oscillation:model:experiment}.
In particular, we simultaneously vary two critical parameters of the configuration.
Any configuration includes a bottleneck capacity~$C_{\ell} \in [1, 200]\ \text{Mbps}$ 
(default: 100 Mbps),
a path propagation delay~$\tau^{\mathrm{p}} \in [1, 100]\ \text{ms}$ (default: 40 ms),
the bottleneck-link propagation delay~$\tau^{\mathrm{p}}_{\ell}$ in percent of~$\tau^{\mathrm{p}}$ (default: 25\%), 
and the bottleneck buffer capacity~$B_{\ell} \in [0.1, 3]$ as a multiple of path BDPs (default: 1.5).

The parameter exploration in \cref{fig:bbr-cubic:dynamic:oscillation:model:condition:space} 
shows that instability is promoted by large values of $C_{\ell}$, $\tau^{\mathrm{p}}$, and $B_{\ell}$. 
Fluid-model simulations, which are highly accurate, confirm the boundary of the unstable region predicted
by the oscillation condition from \cref{thm:bbr-cubic:dynamic:oscillation:model:condition}. 
These simulations were automatically evaluated for convergence of the minimum-RTT estimate~$\tau^{\min}$. 
Importantly, the instability region predicted by \cref{thm:bbr-cubic:dynamic:oscillation:model:condition} 
is always contained within the simulation-validated region, consistent with the theorem’s status
as a sufficient (but not necessary) condition. 
Thus, while additional factors such as feedback delay (absent from the theorem but present in simulation)
can also trigger oscillations, the predicted instability region closely aligns with the simulation-based one, 
indicating that feedback delay plays only a secondary role.

\section{Fairness under Oscillation}
\label{sec:bbr-cubic:dynamic:oscillation:model:fairness}

% We now consider the \emph{fairness implications} of BBR/CUBIC oscillation.
% In particular, 
% the long-term BBR/CUBIC dynamics from~\cref{eq:bbr-cubic:dynamic:oscillation:model:conditions:discrete}
% involve CUBIC congestion-window sizes $\{w(t)\}_{t \in \mathbb{N}, t \geq 0}$,
% Given this association between any oscillation state~$w(t)$ and sending rates,
Fairness under oscillation depends on the fairness of the \emph{oscillation pattern},
i.e., the evenness of all rate 
distributions during the oscillation. 
Formally, this oscillation pattern can be reduced
to the series~$\{w(t)\}_{t \in \mathbb{N}, t \geq 0}$ of CUBIC window sizes from~\cref{eq:bbr-cubic:dynamic:oscillation:model:conditions:discrete}.

This oscillation pattern has a worst-case form,
which is found in~\cref{sec:bbr-cubic:dynamic:oscillation:model:fairness:bound}.
From this worst-case pattern, fairness bounds are computed and
experimentally validated in~\cref{sec:bbr-cubic:dynamic:oscillation:model:fairness:computing}.

\subsection{Bounding the Oscillation Pattern}
\label{sec:bbr-cubic:dynamic:oscillation:model:fairness:bound}

The fairness of the oscillation pattern can be lower-bounded 
by maximizing the \emph{amplitude} of the oscillation pattern, 
i.e., the variance of~$x^{\mathrm{C}}(t)$ and~$x^{\mathrm{B}}(t)$ over time:

\begin{theorem}
    \textbf{Worst-Case Oscillation Pattern.}
    Given oscillation with maximum amplitude, the CUBIC flow oscillates between the congestion-window
    sizes $\hat{w}_0 = \tilde{w}^{\leftarrow}(w^{>})$ and $\hat{w}_1 = \tilde{w}^{\leftarrow}(w^{<})$
    when competing with a BBR flow ($\hat{w}_0 < \hat{w}_1$).
    \label{thm:bbr-cubic:dynamic:oscillation:model:maximal}
\end{theorem}
\begin{proofsketch}
    The proof in~\cref{prf:bbr-cubic:dynamic:oscillation:model:maximal} rests on two
    observations regarding the oscillation amplitude. First, this oscillation amplitude
    corresponds to the window-size changes in the update intervals, i.e.,
    the difference between the window size~$w$ at the start of the interval
    and the convergence result~$w^{\leftarrow}(w)$ at the end of the interval (cf.~\cref{fig:bbr-cubic:dynamic:oscillation:model:oscillation}).
    This difference~$|w - w^{\leftarrow}(w)|$ is upper-bounded by~$|w - \tilde{w}^{\leftarrow}(w)|$,
    i.e., the difference for complete convergence to the short-term equilibrium.
    Hence, the amplitude of the oscillation is maximized by assuming~$w^{\leftarrow} = \tilde{w}^{\leftarrow}$.
    Second, the oscillation amplitude correlates with the size of the unstable
    neighborhood~$\Omega$
    around the long-term equilibrium~$\overline{w}$, as the oscillation pattern typically involves
    a window-size change across~$\Omega$, e.g., from~$w(t) < \omega_0$ to~$w(t+1) > \omega_1$.
    % Remember that~$\Omega$ must involve a strictly decreasing part of the 
    % update function~$w^{\leftarrow}$ (here:~$\tilde{w}^{\leftarrow}$).
    % Since~$\tilde{w}^{\leftarrow}$ can be shown to be strictly decreasing for arguments
    % in~$[\hat{w}_0, \hat{w}_1]$ and constant otherwise, we assume that the neighborhood~$\Omega$
    % covers the entire decreasing part to maximize the oscillation amplitude, 
    % i.e.,~$\Omega = [\hat{w}_0, \hat{w}_1]$.
    In the feasible maximum, $\Omega$ covers the entire decreasing part of~$\tilde{w}^{\leftarrow}$.

    Under these two assumptions, the oscillation corresponds to
    a \emph{limit cycle}, cyclically revisiting the window sizes~$\hat{w}_0 = \tilde{w}^{\leftarrow}(w^{>})$ 
    and~$\hat{w}_1 = \tilde{w}^{\leftarrow}(w^{<})$,
    where $[w^{<}, w^{>}]$ is the finite value range of~$\tilde{w}^{\leftarrow}$.
    % As explained in~\cref{sec:bbr-cubic:dynamic:oscillation:model:condition},
    % ~$[w^{<}, w^{>}]$ is the value range of the update function~$\tilde{w}^{\leftarrow}$.
    % Note that the window sizes in~$L$ do not necessarily correspond to~$w^{>}$ or~$w^{<}$.
    % Instead the values of~$\tilde{w}^{\leftarrow}(w^{<})$ and~$\tilde{w}^{\leftarrow}(w^{>})$
    % depend on the nature of the intersection~$W_{\cap} = [w^{<}, w^{>}] \cap \Omega$, 
    % which is guaranteed to contain the unstable equilibrium~$\overline{w}$.
    % For any form of the intersection~$W_{\cap}$ (i.e., relative position of~$[w^{<}, w^{>}]$ and~$\Omega$),
    % we confirm the limit cycle~$L$ in two aspects.
    This limit cycle is \emph{sound} for the process in~\cref{eq:bbr-cubic:dynamic:oscillation:model:conditions:discrete}, 
    i.e., $\tilde{w}^{\leftarrow}(\hat{w}_0) = \hat{w}_1$
    and $\tilde{w}^{\leftarrow}(\hat{w}_1) = \hat{w}_0$,
    Moreover, the limit cycle is \emph{attractive},
    i.e., the process in~\cref{eq:bbr-cubic:dynamic:oscillation:model:conditions:discrete}
    eventually enters the limit cycle.
    % Given these two properties,
    % the long-term dynamics in~\cref{eq:bbr-cubic:dynamic:oscillation:model:conditions:discrete}
    % eventually exhibit the oscillation described by limit cycle~$L$
    % for maximum oscillation amplitude.
\end{proofsketch}

While difficult to interpret in this form,
\cref{thm:bbr-cubic:dynamic:oscillation:model:maximal} allows to
compute the worst-case fairness bounds in the following section.

\subsection{Computing Fairness Bounds}
\label{sec:bbr-cubic:dynamic:oscillation:model:fairness:computing}

\begin{figure*}
    \centering
    \includegraphics[width=\linewidth,trim=0 10 0 0]{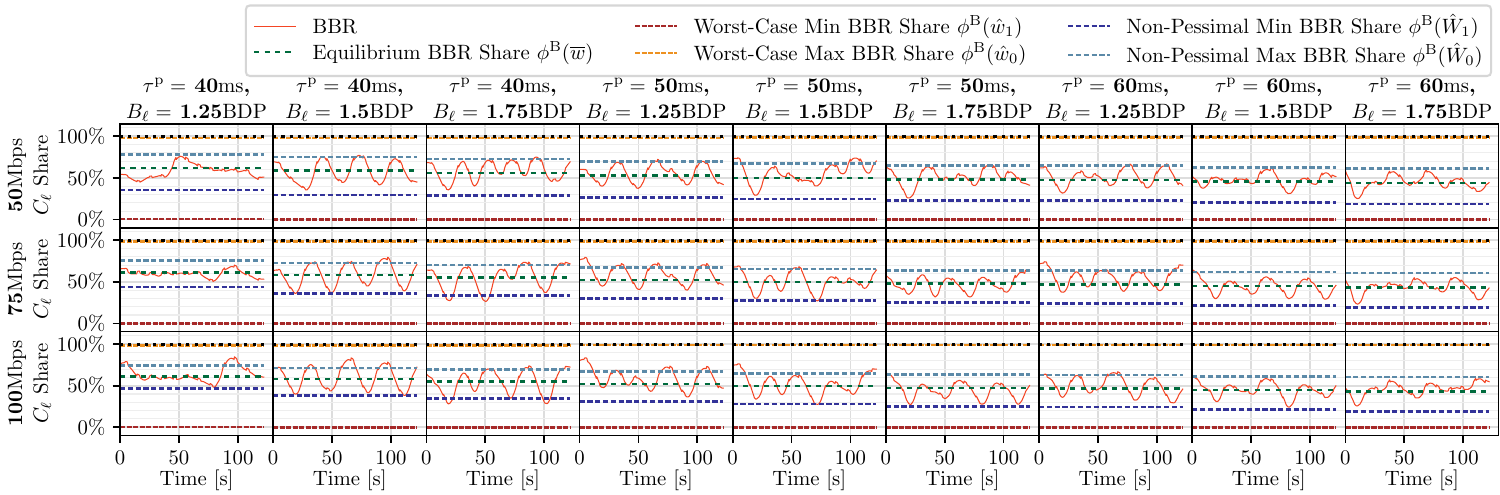}
    \caption{Experimental validation of fairness bounds (cf.~\cref{sec:bbr-cubic:dynamic:oscillation:model:fairness}) (1 flow per CCA).}
    \label{fig:bbr-cubic:dynamic:oscillation:model:prediction:experiments}
\end{figure*}

In the following, we compute and validate fairness bounds for specific network configurations. 
In particular, we test a range of 27 configurations,
which are variations of the experiment settings in~\cref{sec:bbr-cubic:model-evaluation:experiments}
regarding bottleneck-link capacity~$C_{\ell}$, path-propagation delay~$\tau^{\mathrm{p}}$,
and buffer capacity~$B_{\ell}$. 
These Mininet experiments result in 120-second traces (cf.~\cref{fig:bbr-cubic:dynamic:oscillation:model:prediction:experiments}).
% that involve two flows and
% are depicted in~\cref{fig:bbr-cubic:dynamic:oscillation:model:prediction:experiments}.

\subsubsection{Worst-Case Bounds}
For each of these configurations, we find the worst-case oscillation pattern 
from~\cref{thm:bbr-cubic:dynamic:oscillation:model:maximal}.
In this oscillation pattern, the CUBIC congestion-window size alternates between two
values~$\hat{w}_0$ and~$\hat{w}_1$ at every RTT-probing step.
These congestion-window sizes constitute the \emph{extreme points} of the oscillation:
$\hat{w}_0$ is the smallest congestion-window size attained by the CUBIC flow
during the oscillation, whereas~$\hat{w}_1$ is the largest window size.
Hence, $x^{\mathrm{C}}(\hat{w}_0) = \hat{w}_0/\overline{\tau}_k$ is the minimal attained
CUBIC rate, whereas $x^{\mathrm{C}}(\hat{w}_1)$ is maximal. 

For BBR, the maximum sending rate
is (symmetric for~$\hat{w}_1$):
\begin{equation}
    x^{\mathrm{B}}(\hat{w}_0) = \beta(\hat{w}_1) \cdot \max\left(\chi, C_{\ell} - \frac{x^{\mathrm{C}}(\hat{w}_0)}{\alpha(\hat{w}_1)}\right),
\end{equation}
which is based on the BBR equilibrium conditions from \cref{lmm:bbr-cubic:eq:bbr-conditions},
because~$\hat{w}_0$ and~$\hat{w}_1$ represent short-term equilibria.
Note that the minimum-RTT estimate (and thus also the
strengths~$\alpha$ and~$\beta$) of a BBR flow is based on the
CUBIC congestion-window size at the previous RTT-probing step,
which is~$\hat{w}_1$ when the current congestion-window size is~$\hat{w}_0$.

Based on these derived sending rates, we also identify the
maximum BBR capacity share:
\begin{equation}
    \phi^{\mathrm{B}}(\hat{w}_0) = \frac{x^{\mathrm{B}}(\hat{w}_0)}{x^{\mathrm{C}}(\hat{w}_0) + x^{\mathrm{B}}(\hat{w}_0)}.
\end{equation}
The minimum BBR share~$\phi^{\mathrm{B}}(\hat{w}_1)$ is found analogously.

% \begin{figure*}
%     \centering
%     \input{figures/bbr_vs_cubic__fairness_bounds_annotated}
%     \caption{Time CDF of BBR load share, compared to theoretical BBR load-share bounds,
%     computed for the experiments in~\cref{fig:bbr-cubic:dynamic:oscillation:model:prediction:experiments}.}
%     \label{fig:bbr-cubic:dynamic:oscillation:model:fairness}
% \end{figure*}

To validate these bounds, we compare the theoretically
derived bounds to the experimentally found oscillation
patterns in~\cref{fig:bbr-cubic:dynamic:oscillation:model:prediction:experiments}.
For bound correctness, the oscillation pattern of the BBR flow is 
supposed to be contained within these bounds, which is clearly true.
However, the worst-case bounds are relatively loose,
as they are derived with worst-case assumptions,
which are overly pessimistic for the settings 
in~\cref{fig:bbr-cubic:dynamic:oscillation:model:prediction:experiments}.

\subsubsection{Non-Pessimal Bounds}

% Interestingly, we can also leverage our
% model-based analysis to derive approximate bounds 
% that reflect the oscillation outside the worst case.

As described above,  worst-case fairness is realized if the
BBR/CUBIC competition dynamics always converge completely
to the current short-term equilibrium
between RTT probing steps.
In reality, however, convergence often takes more than 10 seconds,
and is thus often incomplete 
at the time of the next RTT-probing step.
Given this insight on incomplete convergence, 
we can find approximate non-pessimal fairness 
bounds~$\phi^{\mathrm{B}}(\hat{W}_0)$ 
and~$\phi^{\mathrm{B}}(\hat{W}_1)$, where:
\begin{equation}
    \hat{W}_0 = W(\overline{w}, 0) \hspace{10mm} \hat{W}_1 = W(\overline{w}, 10) 
\end{equation}
and $W$ is the CUBIC window-growth function 
from~\cref{eq:bbr-cubic:model:basic-model:cubic-window-growth}.
The non-pessimal bounds are informed
by insights on the convergence behavior 
of the CUBIC state variables ($w^{\max}$ and~$s$)
from~\cref{thm:bbr-cubic:dynamic:oscillation:model:convergence,thm:bbr-cubic:dynamic:oscillation:model:condition}:
These insights suggest that (1) the CUBIC state variable $w^{\max}$
is usually close to the CUBIC window size~$\overline{w}$ from
the long-term equilibrium ($w^{\max} \approx \overline{w}$),
and (2) the CUBIC window-growth duration~$s$ is usually 
between 0 seconds (as it cannot be negative) and 10 seconds
(i.e., the duration between RTT-probing steps, where loss typically occurs), 
i.e., $s \in [0, 10]$.
The full derivation is presented in~\cref{sec:non-pessimal}.
% Hence, this incomplete convergence must be characterized
% in order to estimate a realistic oscillation pattern.
% To characterize this convergence behavior, we rely
% on a number of key insights from our analysis.

% First, the proof of \cref{thm:bbr-cubic:dynamic:oscillation:model:convergence}
% suggests that the CUBIC window-growth duration~$s$ converges
% exponentially to its short-term equilibrium~$\tilde{s}$, whereas
% the convergence of the recorded maximum window~$w^{\max}$ is slower
% (also visible in~\cref{fig:bbr-cubic:dynamic:oscillation:model:center-manifold}).
% Hence, it is plausible that the window-growth duration~$s$ really 
% oscillates between its worst-case equilibrium values, i.e.,
% \begin{equation}
%     \hat{s}_0 \overset{\text{(\ref{eq:bbr-cubic:system-evolution:cubic:eq:wmax})}}{=} \sqrt[3]{\frac{b}{c} \hat{w}_0} \hspace{50pt} \text{and} \hspace{50pt} \hat{s}_1 \overset{\text{(\ref{eq:bbr-cubic:system-evolution:cubic:eq:wmax})}}{=} \sqrt[3]{\frac{b}{c} \hat{w}_1}.
% \end{equation}

The non-pessimal bounds $\phi^{\mathrm{B}}(\hat{W}_0)$
and $\phi^{\mathrm{B}}(\hat{W}_1)$ are also shown in
\cref{fig:bbr-cubic:dynamic:oscillation:model:prediction:experiments}.
These bounds are sometimes violated due to their
approximate nature, but predict 
the range of oscillating BBR-share values
better than the worst-case bounds.
% The non-pessimal prediction is more accurate
% not least because the non-pessimal bounds 
% are more sensitive to the parameters of 
% the experiment setting.

\color{black}

\section{Preventing Oscillation}
\label{sec:bbr-cubic:dynamic:oscillation:remedies}

% The preceding sections confirm that rate oscillation in BBR/ CUBIC
% competition is a frequent problem with severe fairness implications.
To prevent unfairness from oscillation, we evaluate different
countermeasures, and explain their effects by means of
our control-theoretic framework.

\subsection{Smoothed Minimum-RTT Estimate}
\label{sec:bbr-cubic:dynamic:oscillation:remedies:stabilizing}

\begin{figure*}
    \begin{minipage}{0.45\linewidth}
        \centering
        \includegraphics[width=\linewidth,trim=10 0 0 0]{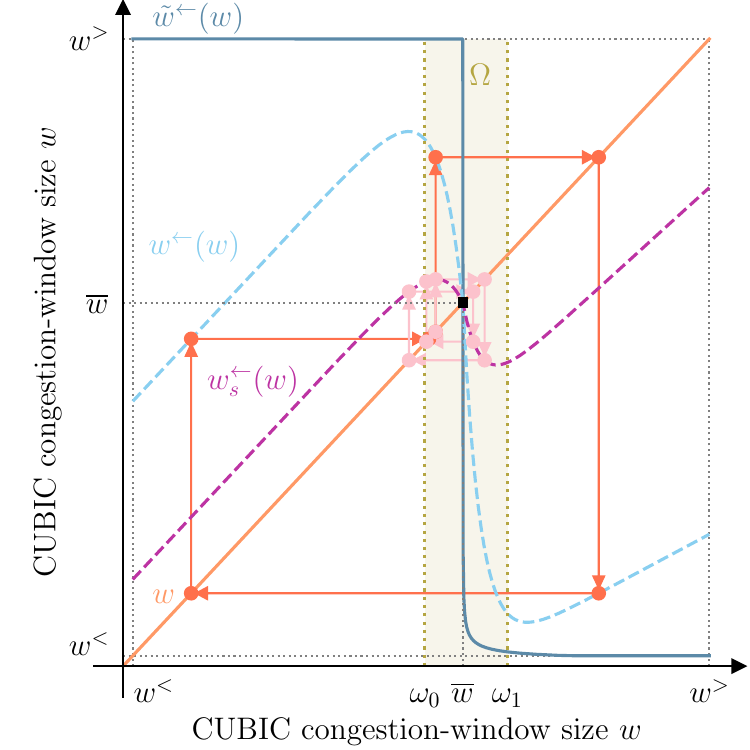}
        \vspace{-6mm}
        \caption{Theoretical explanation of oscillation reduction by smoothed-estimate adaptation.}
        \label{fig:remedies:smoothed-estimate:theo}
    \end{minipage}\quad\vrule\quad
    \begin{minipage}{0.45\linewidth}
        \centering
        \includegraphics[width=\linewidth,trim=10 0 0 0]{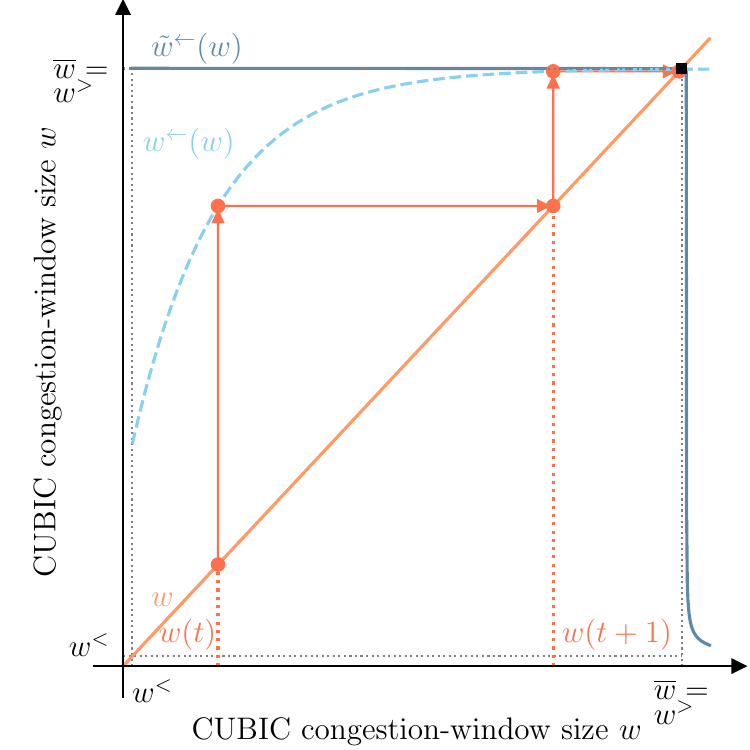}
        \vspace{-6mm}
        \caption{Theoretical explanation of oscillation elimination by BBRv2 and BBRv3.}
        \label{fig:remedies:bbr2:theo}
    \end{minipage}
\end{figure*}

% First, we identify a countermeasure based on our 
% analysis of instability in.
In the analysis from~\cref{sec:bbr-cubic:dynamic:oscillation:model:condition}, 
we find that the long-term
BBR/ CUBIC dynamics oscillate because the update
function~$w^{\leftarrow}$ of the CUBIC congestion-window
size~$w$ is too steep around
the long-term equilibrium~$\overline{w}$.
This steepness is caused by the fact 
that the minimum-RTT estimate~$\tau^{\min}$
(and thus the probing strength~$\alpha$
and the window-update function~$w^{\leftarrow}$)
of the BBR flows is highly sensitive to the 
CUBIC congestion-window size~$w$ at the time
of RTT probing. 
To reduce this sensitivity and thereby reduce oscillation, 
% the adjustment of the minimum-RTT estimates
% might be restricted. 
% Using a classic idea from control theory, 
we use a classic control-theoretic idea:
We slow down the adjustment of the minimum-RTT estimate
to new RTT measurements using a moving average.
Formally, an RTT-probing step at time~$t$ 
updates the minimum-RTT estimate~$\tau^{\min}$
as follows:
\begin{equation}
    \tau^{\min}(t) = \theta \cdot \left(\tau_i^{\mathrm{p}} + \frac{q_{\ell}^-(t)}{C_{\ell}}\right) + \left(1-\theta\right) \cdot \tau^{\min}\left(t-10\right)
\end{equation}
where~$\theta$ denotes the adaptation speed of~$\tau^{\min}$ to the
most recently measured RTT (\cref{eq:bbr-cubic:system-evolution:remaining-queue}).

Indeed, this smoothing of the minimum-RTT estimate can dampen,
but not completely eliminate oscillation (\cref{fig:remedies:smoothed-estimate}).
\Cref{fig:remedies:smoothed-estimate:theo} explains this
effect for the situation 
of~\cref{fig:bbr-cubic:dynamic:oscillation:model:oscillation}: 
As the window-update function~$w^{\leftarrow}$ is transformed into the
update function~$w_s^{\leftarrow}$, the original oscillation trajectory 
(in orange) is reduced to a much more narrow trajectory (in rose),
corresponding to oscillation with lower amplitude.

However, we note that 
this smoothing approach reduces responsiveness
when a path change alters the propagation delay,
which may be acceptable in fixed networks, but
is problematic in mobile networks.
We also implement and evaluate two alternative
BBR adaptations in~\cref{sec:bbr-modification:evaluation},
but these adaptations have less desirable effects.

\subsection{BBRv2 and BBRv3}
\label{sec:bbr-cubic:dynamic:oscillation:remedies:bbr2}

% The previous section demonstrates that suppressing oscillation
% in BBR/CUBIC competition with ad-hoc changes to the BBR algorithm
% is difficult and comes with several drawbacks. 
% In this section, we show that a more fundamental redesign of BBR
% represented by BBRv2~\cite{cardwell2019bbrv2} and BBRv3~\cite{cardwell2023bbrv3}
% eliminates oscillation, but does not yet guarantee fairness.

Similarly to the smoothed-estimate adaptation in the preceding
section, BBRv2 eliminates oscillation by reducing the sensitivity
of the BBR rate 
on the minimum-RTT estimate~$\tau^{\min}$:
While BBRv1 constrains the inflight volume only
with the BBR congestion-window size of~$2 x^{\mathrm{btl}} \tau^{\min}$,
BBRv2 constrains this inflight volume (and thus the sending rate)
more strongly, i.e., by multiplying the minimum-RTT estimate~$\tau^{\min}$ 
with a factor lower than $2 x^{\mathrm{btl}}$.
In particular, BBRv2 has a long-term inflight bound (\textit{inflight\_hi}),
which is at most $\nicefrac{5}{4} \cdot x^{\mathrm{btl}} \tau^{\min}$,
and a short-term inflight bound (\textit{inflight\_lo}), which is at most
$x^{\mathrm{btl}} \tau^{\min}$. Both these bounds are even further reduced
if loss occurs frequently, as in the competition with CUBIC.

\begin{figure*}
    \begin{minipage}{0.7\linewidth}
        \centering
        \includegraphics[width=\linewidth,trim=0 10 0 10]{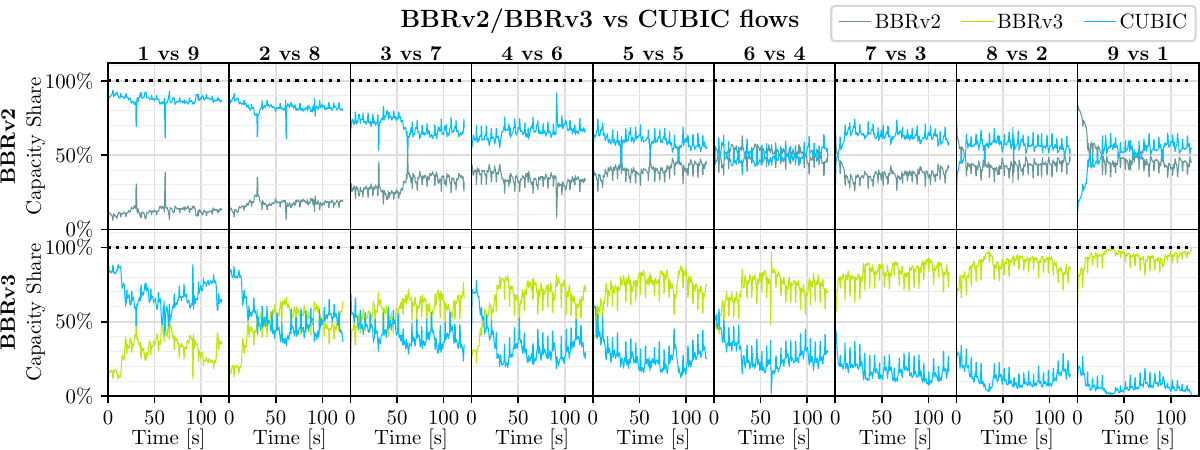}
        \caption{Competition between BBRv2/BBRv3 and CUBIC (Aggregate capacity share by CCA).}
        \label{fig:dynamic:bbr2-cubic}
    \end{minipage}\quad\vrule\quad
    \begin{minipage}{0.24\linewidth}
        \includegraphics[width=\linewidth,trim=0 10 0 0]{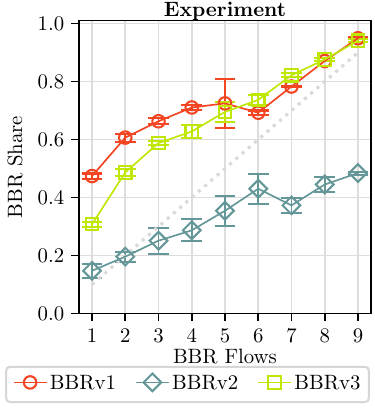}
        \caption{Capacity shares obtained by BBR versions.}
        \label{fig:bbr-cubic:dynamic:oscillation:remedies:versions}
    \end{minipage}
\end{figure*}

These new inflight bounds alter the formulation
of the strengths~$\alpha$ and~$\beta$,
as shown by Scherrer et al.~\cite{scherrer2022model}:
\begin{align}
    \text{Limited by \textit{inflight\_hi}:} \quad &\alpha' \overset{\text{\cite{scherrer2022model}}}{=} \frac{5}{4} \cdot \min\left(1, \tau^{\min}/\tau\right) &\leq 
    \min\left(\frac{5}{4}, 2\tau^{\min}/\tau\right) \overset{\text{(\ref{eq:bbr-cubic:system-evolution:bbr:alpha})}}{=} \alpha\\
    \text{Limited by \textit{inflight\_lo}:} \quad &\beta' \overset{\text{\cite{scherrer2022model}}}{=} \min\left(1, \tau^{\min}/\tau\right) &\leq 
    \min\left(1, 2\tau^{\min}/\tau\right) \overset{\text{(\ref{eq:bbr-cubic:system-evolution:bbr:beta})}}{=} \beta
\end{align}
As $\alpha' \leq \alpha$ and $\beta' \leq \beta$, 
the sending rate in BBRv2 is less strongly affected by varying minimum-RTT estimates 
than in BBRv1, and oscillation is prevented. 
\cref{fig:remedies:bbr2:theo} illustrates this effect 
theoretically: Computing the equilibrium window-update 
function~$\tilde{w}^{\leftarrow}$ with~$\alpha'$ and~$\beta'$
lowers the `maximum plateau' at~$w^{>}$. Hence, a different intersection point 
(long-term equilibrium)~$\overline{w} = w^{>}$
results, crucially located on the plateau, which allows convergence.
The convergence is confirmed by the experiments in~\cref{fig:dynamic:bbr2-cubic}.
% which also includes predictions by the BBRv2 fluid model by Scherrer et al.~\cite{scherrer2022model}.

However, the BBRv2 features that suppress oscillation have two drawbacks.
First, BBRv2 flows might obtain less than the fair bandwidth share.
This unfairness is reflected by (i)~the theoretical observation that the CUBIC window 
size~$\overline{w}$ in the long-term equilibrium of~\cref{fig:remedies:bbr2:theo}
is stable and relatively high, suggesting a low BBR capacity share in equilibrium,
and (ii)~the experimental observation that most cases in~\cref{fig:bbr-cubic:dynamic:oscillation:remedies:versions} 
show sub-proportional throughput for BBRv2. 
% This lack of assertiveness might hamper the adoption of BBRv2~\cite{mishra2022we}.
Second, the relatively static inflight bounds of BBRv2 
slow down the discovery of newly available bandwidth~\cite{yang2022bbrv2+}.

To remedy these issues, Google has recently released BBRv3, which is
largely equal to BBRv2, but more aggressively probes for 
additional bandwidth~\cite{cardwell2023bbrv3}.
Due to this higher aggressiveness,  BBRv3
obtains an over-proportional bandwidth share 
(cf.~\cref{fig:bbr-cubic:dynamic:oscillation:remedies:versions}),
and is thus similarly unfair as BBRv1, albeit stable.

In summary, BBRv 2 and BBRv3 solve the instability
problem of BBRv1, but still do not achieve perfect per-flow
fairness. Designing a BBR version that is stable \emph{and} fair
is an interesting task for future work, and can benefit from 
our control-theoretic methodology.

\section{Related Work}
\label{sec:related-work}

\paragraph{Most relevant related work}
Our work has a close connection to the 
two previous steady-state models of BBR/ CUBIC competition~\cite{ware2019modeling,mishra2022we},
and the fluid models for CUBIC~\cite{vardoyan2021towards}
and BBR~\cite{scherrer2022model}.
Another highly relevant recent work is by 
Arun et al.~\cite{arun2022starvation},
who investigate the delay oscillations caused 
by delay-bounding CCAs (such as BBR) and loss-based CCAs (such as CUBIC) 
in \emph{homogeneous} settings. 
In particular,
they find that the amplitude of these delay oscillations 
should exceed the random, non-congestive jitter
that is expected, in order to avoid starvation
of flows measuring a distorted RTT.
Our work complements this analysis by investigating 
oscillation in \emph{heterogeneous}
settings, i.e., given CCA competition.
Moreover, we identify an additional condition for BBR
starvation involving delay measurements, 
namely \cref{lmm:bbr-cubic:eq:bbr-starvation}.
% but demonstrate that BBR can recover from this starvation
% in the long term when competing with CUBIC.

\paragraph{Congestion-control models}
Congestion-control algorithms have been analyzed with a wide range of modeling techniques~\cite{srikant2004mathematics,olsen2003stochastic}. 

For example, \emph{steady-state models} describe only 
the equilibrium of CCA execution 
in terms of network metrics.
Most influentially, Mathis et al.~\cite{mathis1997macroscopic} and 
Padhye et al.~\cite{padhye1998modeling} provide closed-form
functions for long-term Reno throughput based on RTT and packet loss;
their methodology has been extended to short-lived flows and other 
CCAs~\cite{cardwell2000modeling,kumar1998comparative,bao2010model}.
% As mentioned, also the competition between BBR and CUBIC flows has
% been analyzed with steady-state models~\cite{ware2019modeling,mishra2022we}.

In contrast, \emph{dynamic models} represent the full CCA
behavior over time. This evolution is sometimes represented
in discrete time~\cite{akella2002selfish,arun2021toward,zarchy2019axiomatizing,poojary2011analytical},
but more often in continuous time by fluid models.
Specific fluid models exist for Reno~\cite{low2002internet,misra2000fluid},
Vegas~\cite{bonald1998comparison}, CUBIC~\cite{vardoyan2021towards},
and BBRv1 and v2~\cite{scherrer2022model}.
While these models have been applied to analyze fairness and
stability~\cite{kelly1998rate,kelly2003fairness,johari2001end,liu2003fluid},
our work is the first to theoretically investigate inter-CCA fairness
with a dynamic fluid model.
% and to rigorously demonstrate the oscillation in BBR/CUBIC competition.

\paragraph{BBR} Motivated by the excessive queuing of loss-based
CCAs (e.g., Reno~\cite{fall1996simulation}, CUBIC~\cite{ha2008cubic}) and competitiveness issues of
latency-based CCAs (e.g., Vegas~\cite{brakmo1994tcp}),
Google proposed the BBR CCA in 2017~\cite{cardwell2017bbr}.
BBR was enabled for YouTube soon afterwards~\cite{cardwell2017bbr-ietf},
and was used by around 40\% of Internet traffic in 2019~\cite{mishra2019great}.
Given that BBR competes with other CCAs in the Internet,
its fairness towards other CCAs has received much attention.
In a first independent study, Hock et al.~\cite{hock2017experimental}
demonstrate that BBR is over-aggressive against loss-based CCAs
for settings with small buffers, and under-aggressive given large buffers.
In these large buffers, the BBR congestion window restricts the BBR sending rate, 
which is confirmed by follow-up work~\cite{scholz2018towards,dong2018pcc,turkovic2019fifty,xu2022measurement}. 
This insight about BBR fairness triggered the release of BBRv2~\cite{cardwell2019bbrv2},
which is less aggressive, but also  less responsive than 
BBRv1~\cite{yang2022bbrv2+,kfoury2020emulation,nandagiri2020bbrvl,song2021understanding}.
Finally, the sub-optimal assertiveness and responsiveness of BBRv2
led to the release of BBRv3 in 2023~\cite{cardwell2023bbrv3}.

\paragraph{Congestion-control fairness}
% In evaluating CCAs,
% much attention is traditionally devoted to \emph{fairness}, i.e., 
% the equality of resource sharing under distributed CCA execution.
In congestion-control research, fairness is typically measured by 
some aggregation (e.g., Jain fairness index~\cite{jain1999throughput}) 
of the throughput-share distribution
across flows on a bottleneck link~\cite{dong2018pcc,cardwell2017bbr,ware2019modeling,mishra2022we,scholz2018towards,hock2017experimental}.
Alternative fairness measures
focus on flow-completion times~\cite{dukkipati2006flow},
on quality of experience (QoE)~\cite{hossfeld2016definition},
or on compatibility with CCAs already deployed in the Internet~\cite{ware2019beyond}, 
or avoid flows as entities of the fairness 
definition~\cite{briscoe2007flow}.
% Regarding fairness between different CCAs,
% TCP friendliness (i.e., fairness towards the Reno CCA)
% has been shown to require a fundamental trade-off with
% other goals such as throughput and responsiveness~\cite{zarchy2019axiomatizing,brown2020future}.
% In this work, we also demonstrate that fairness has an important
% time component: If throughput shares are averaged over a long 
% enough time, high transitory inequality might be obscured.

\section{Conclusion}
\label{sec:conclusion}

% The work in this paper is motivated by limitations of
% previous work on the modeling of BBR/CUBIC competition. 
% In particular, the simulation of dynamic fluid models 
% yields predictions that are highly accurate, but lack 
% the explainability of predictions by steady-state models, 
% and certainly do not provide rigorous guarantees.
Dynamic fluid models have an
under-explored potential for generating interpretable
and even provable insights into complex
congestion-control phenomena. To tap into this potential,
we leverage these models for a 
theoretical stability analysis of BBR/CUBIC competition.
Through this control-theoretic analysis, we provide a 
mathematical explanation for the
oscillation that regularly occurs in BBR/CUBIC competition. 
Moreover, we derive quantitative conditions on networks
that are susceptible to oscillation, and quantitative bounds
on the unfairness caused by extreme rate distributions during
the oscillation.
While these extreme rate distributions are transient, 
they are substantially more unfair than the 
long-term average rate distribution, and thus 
also matter for the fairness among short flows.
Finally, our control-theoretic framework also explains the effects
and shortcomings of different stabilizing adaptations of BBR.

By pioneering the use of theoretical stability analysis 
for inter-CCA competition, 
this paper opens numerous avenues for future research. 
While our analysis is extensive, it represents only 
the beginning of what control theory can offer. 
The application of advanced control-theoretic techniques 
to hybrid CCA fluid models 
has the potential to uncover even deeper insights. 
Notably, we see significant promise in 
Lyapunov functions for understanding which initial configurations
allow convergence to the equilibria 
(global asymptotic stability and region of attraction), 
the Laplace transform for modeling time-delay effects, and 
robust-controller design for rigorously eliminating oscillations
in a broad range of network settings.

Although being extensible in these many ways, 
our analysis already now provides practically relevant insights. 
By characterizing when oscillations and unfairness arise, 
our results can inform the design and parameter tuning of congestion-control algorithms, 
and guide practitioners in anticipating performance issues in heterogeneous deployments.

\section*{Acknowledgements}

We gratefully acknowledge funding from ETH Zurich, and
from the German Research Foundation (DFG) 
for project `Algorithms for Flexible Networks (FlexNets)' (Grant 470029389).
Moreover, we thank Andr\'es Ferragut for shepherding, 
and the anonymous reviewers for valuable suggestions.

% \input{sections/pcc-cubic-static}
% \input{sections/pcc-cubic-dynamic}
% \input{sections/fact}

%\label{bodyLastPage}
%{\balance
\bibliographystyle{plainurl}
\bibliography{ref}
%\label{LastPage}
%\bibliographystyle{ACM-Reference-Format}
%\bibliography{ref}
% }

\pagebreak
\newpage
\appendix

\begin{appendix}

\section{Oscillation of multiple flows}
\label[appendix]{sec:bbr-cubic:dynamic:oscillation:synchronization}

The explanation of the oscillation mechanism in~\cref{sec:bbr-cubic:dynamic}
applies to the case with a single flow per CCA.
% However, as~\cref{fig:dynamic:bbr-cubic} illustrates,
% we observe oscillation also for multiple flows per CCA.
We now explain how and why
the mechanism generalizes to multiple flows.

\begin{figure}
    \centering
    \input{figures/oscillation-synchronization}
    \vspace{-5pt}
    \caption{Self-synchronization of RTT probing
    (highlighted with red rectangles)
    among 4 BBR flows.}
    \label{fig:bbr-cubic:dynamic:oscillation:synchronization}
\end{figure}
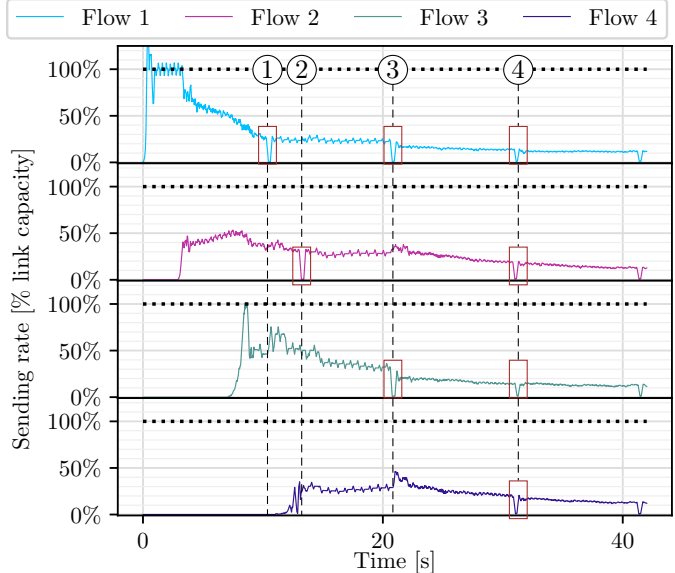

The generalization to multiple CUBIC flows is straightforward:
The explanation in~\cref{sec:bbr-cubic:dynamic} is 
equally valid for multiple CUBIC flows, with the slight adaptation that
the \emph{aggregate} congestion-window size of all CUBIC flows
affects the minimum-RTT estimate~$\tau^{\min}$ of a BBR flow.

Interestingly, multiple BBR flows also behave analogously to a 
single BBR flow for the purpose of oscillation 
because BBR flows synchronize their RTT-probing steps. 
This synchronization has been documented by previous work \cite{cardwell2017bbr,scholz2018towards},
and is experimentally confirmed in~\cref{fig:bbr-cubic:dynamic:oscillation:synchronization}.
In that experiment, 4 BBR flows are initiated
sequentially with 4-second delays, and compete with 6 CUBIC flows (not pictured).
The BBR flow 1 performs the first RTT probing at Time~\circled{1}, which shrinks the 
queue and reduces the RTT. In fact, the RTT is reduced enough such that Flow~$3$
measures a new minimum RTT. Hence, Flow~1 and Flow~3 start the reset timers of
their minimum-RTT estimates at the same time, and therefore simultaneously probe the RTT
at Time~\circled{3} = \circled{1} $+ 10$ seconds. At this Time~\circled{3},
also the remaining two flows 2 and 4 become synchronized with flows 1 and 3
such that all flows perform a simultaneous RTT probing at Time~\circled{4}.

Finally, note that oscillation does not necessarily arise in a multiple-flow
scenario if the CCA distribution is very unequal, 
e.g., for 9 BBR flows and 1 CUBIC flow in~\cref{fig:dynamic:bbr-cubic}.
In this case, the single CUBIC flow never manages to inflate the 
minimum-RTT estimate of the BBR flows, and hence the rate distribution is 
largely static.

\section{Proof of Theorem~\ref{thm:bbr-cubic:dynamic:oscillation:model:equilibrium}: Short-Term Equilibrium}
\label{prf::bbr-cubic:dynamic:oscillation:model:equilibrium}

In the following, we consistently write~$\tilde{s}$ 
for $\tilde{s}(\alpha)$, as~$\alpha$ is considered fixed
throughout the proof (analogously for~$\tilde{w}^{\max}$ and~$\tilde{x}^{\mathrm{btl}}$)

    The CUBIC maximum window~$\tilde{w}^{\max}$ in equilibrium
    (and by extension also the CUBIC equilibrium rate~$\tilde{x}^{\mathrm{C}}$)
    is directly determined by the CUBIC window-growth duration~$\tilde{s}$ in equilibrium:
    \begin{equation}
        \tilde{w}^{\max} \overset{\text{(\ref{eq:bbr-cubic:system-evolution:cubic:eq:wmax})}}{=} \frac{c}{b} \tilde{s}^3
        \iff \tilde{x}^{\mathrm{C}} = \frac{c\tilde{s}^3}{b\tilde{\tau}_k}.
        \label{eq:bbr-cubic:dynamic:oscillation:model:equilibrium:window}
    \end{equation}
    The short-term equilibrium delay~$\tilde{\tau}_k$ also
    matches the general equilibrium delay~$\overline{\tau}_k$ 
    from~\cref{eq:bbr-cubic:eq:delay} because also the short-term equilibrium
    involves non-zero loss and thus a full buffer.
    To characterize the equilibrium completely, it remains to determine~$\tilde{s}$
    and~$\tilde{x}^{\mathrm{btl}}$, which
    we achieve with a case distinction on~$\alpha$.

\subsection{Low Probing Strength: $\alpha \leq 1$}
\label{prf::bbr-cubic:dynamic:oscillation:model:equilibrium:low}

    We first consider the case~$\alpha \leq 1$. 
    According to~\cref{lmm:bbr-cubic:eq:bbr-starvation}, ~$\tilde{x}_i^{\mathrm{btl}}$
    must equal the lower bound~$\chi$ in this case. For~$\tilde{x}^{\mathrm{btl}}_i = \chi$,
    the CUBIC equilibrium condition in~\cref{lmm:bbr-cubic:eq:cubic-conditions} suggests:
    \begin{align}
        &\frac{b\tilde{\tau}_k}{c \tilde{s}^{4}} \overset{\text{(\ref{eq:bbr-cubic:eq:cubic-conditions})}}{=} 1- \frac{C_{\ell}}{\tilde{y}_{\ell}}
        \overset{\text{(\ref{eq:bbr-cubic:model:basic-model:total-load})}}{=} 
        1 - \frac{C_{\ell}}{\beta\tilde{x}^{\mathrm{btl}} + \tilde{x}^{\mathrm{C}}} 
        \overset{\text{(\ref{eq:bbr-cubic:dynamic:oscillation:model:equilibrium:window})}}{\underset{\beta = \alpha}{=}} 
        1 - \frac{C_{\ell}}{\alpha\chi + \frac{c \tilde{s}^3}{b\tilde{\tau}_k}} \iff\nonumber\\
        &\frac{c^{2}}{b\tilde{\tau}_k}\tilde{s}^{7}-c\left(C_{\ell}-\alpha\chi\right)\tilde{s}^{4}-c\tilde{s}^{3}-\alpha b\tilde{\tau}_k\chi 
        = 0,
        \label{eq:bbr-cubic:dynamic:oscillation:model:equilibrium:S2}
    \end{align} where the polynomial in the last equation 
    corresponds to~$\tilde{S}_2(\tilde{s})$ in~\cref{eq:bbr-cubic:dynamic:oscillation:model:equilibrium:additional:S2}.
    Note that~$\beta = \alpha$ is implied by~$\alpha \leq 1$ 
    (cf.~\cref{eq:bbr-cubic:system-evolution:bbr:beta,eq:bbr-cubic:system-evolution:bbr:alpha}).
    The septic equation~$\tilde{S}_2(s) = 0$ has a
    unique solution for positive~$s$,
    which follows from two arguments:
    \begin{enumerate}
        \item \textit{Negativity at 0:} $\tilde{S}_2(0) = -\alpha b\tilde{\tau}_k\chi < 0$.
        \item \textit{Strict monotonic increase after monotonic decrease:}
        \begin{equation}
            \begin{split}
                \exists s' \geq 0 \quad \text{ s.t. } \quad &\forall s \in [0, s').\ \tilde{S}_2'(s') \leq 0 \quad \text{and}\\ 
                &\forall s>s'.\ \tilde{S}_2'(s) >  0.
            \end{split}
        \end{equation}
        The existence of such a unique turning point~$s'$ is proven in~\cref{prf::bbr-cubic:dynamic:oscillation:model:equilibrium:unique-inflection}. 
        
    \end{enumerate}

\subsection{High Probing Strength: $\alpha > 1$}
\label{prf::bbr-cubic:dynamic:oscillation:model:equilibrium:high}

    Conversely, if~$\alpha > 1$, $\tilde{x}^{\mathrm{btl}}$ may exceed~$\chi$. 
    For this case, we consider the \emph{unrestricted} equilibrium~$\hat{x}^{\mathrm{btl}}$, i.e.,
    the equilibrium bottleneck-bandwidth estimate~$\tilde{x}^{\mathrm{btl}}$
    according to~\cref{lmm:bbr-cubic:eq:bbr-conditions},
    but without restriction to the domain~$[\chi, \infty)$:
    \begin{equation}
        \hat{x}^{\mathrm{btl}} \overset{\text{(\ref{eq:bbr-cubic:eq:bbr-conditions})}}{=} C_{\ell} - \frac{\tilde{x}^{\mathrm{C}}}{\alpha} \overset{\text{(\ref{eq:bbr-cubic:dynamic:oscillation:model:equilibrium:window})}}{=} C_{\ell} - \frac{c\tilde{s}^{3}}{\alpha b \tilde{\tau}_k}.
        \label{eq:bbr-cubic:dynamic:oscillation:model:equilibrium:unrestricted}
    \end{equation}
    We now distinguish the cases~$\hat{x}^{\mathrm{btl}} \geq \chi$
    and~$\hat{x}^{\mathrm{btl}} < \chi$.
    
    \paragraph{$\hat{x}^{\mathrm{btl}} \geq \chi$} 
    In this case, the actual equilibrium $\tilde{x}^{\mathrm{btl}}$
    matches the unrestricted equilibrium $\hat{x}^{\mathrm{btl}}$.
    Plugging $\tilde{x}^{\mathrm{btl}} = \hat{x}^{\mathrm{btl}}$ 
    into the CUBIC equilibrium condition from~\cref{eq:bbr-cubic:eq:cubic-conditions},
    we obtain:
    \begin{equation}
        \begin{split}
            &\frac{b\tilde{\tau}_k}{c \tilde{s}^{4}} \overset{\text{(\ref{eq:bbr-cubic:eq:cubic-conditions})}}{=} 1 - \frac{C_{\ell}}{\beta\tilde{x}^{\mathrm{btl}} + \tilde{x}^{\mathrm{C}}} 
        \overset{\text{(\ref{eq:bbr-cubic:dynamic:oscillation:model:equilibrium:window})}}{\underset{\beta = 1}{=}}  1 - \frac{C_{\ell}}{\tilde{x}^{\mathrm{btl}} + \frac{c\tilde{s}^{3}}{b \tilde{\tau}_k}} \overset{\text{(\ref{eq:bbr-cubic:dynamic:oscillation:model:equilibrium:unrestricted})}}{\iff}\\
            &\frac{(\alpha -1)c^2}{\alpha b\tilde{\tau}_k} \tilde{s}^7
            - \frac{\left(\alpha-1\right)c}{\alpha} \tilde{s}^3
            - b C_{\ell} \tilde{\tau}_k = 0,
        \end{split}
    \end{equation}
    where the polynomial in the second equation corresponds to~$\tilde{S}_1(\tilde{s})$ in~\cref{eq:bbr-cubic:dynamic:oscillation:model:equilibrium:additional:S1}.
    Note that~$\beta = 1$ is implied by~$\alpha > 1$.
    Moreover, the uniqueness of the solution to~$\tilde{S}_1 = 0$ can be shown similarly as before:
    \begin{enumerate}
        \item \textit{Negativity at 0:} $\tilde{S}_1(0) = - b C_{\ell} \tilde{\tau}_k < 0$.
        \item \textit{Strict monotonic increase after monotonic decrease:}
        \begin{align}
             &\tilde{S}_1'(s) > 0 \iff 
             \frac{7(\alpha -1)c^2}{\alpha b\tilde{\tau}_k} s^6
            - 3\frac{\left(\alpha-1\right)c}{\alpha} s^2 > 0\\
            &\overset{/s^2}{\iff} 
            \frac{7(\alpha -1)c^2}{\alpha b\tilde{\tau}_k} s^4
            - \frac{3\left(\alpha-1\right)c}{\alpha} > 0
            \iff 
            s > \sqrt[4]{\frac{ 3b\tilde{\tau}_k}{7 c}}.\nonumber
            \end{align}
    \end{enumerate}

    \paragraph{$\hat{x}^{\mathrm{btl}} < \chi$}
    This case arises if the equilibrium window-growth
    duration~$\tilde{s}$ is sufficiently high:
    \begin{equation}
        \begin{split}
            &\hat{x}^{\mathrm{btl}} = C_{\ell} - \frac{c\tilde{s}^{3}}{\alpha b \tilde{\tau}_k} < \chi
            \iff
            \frac{c\tilde{s}^{3}}{ b \tilde{\tau}_k} > \alpha \left(C_{\ell} - \chi\right)\\
            &\iff
            \tilde{s} > \sqrt[3]{\frac{\alpha b \tilde{\tau}_k}{c} \left(C_{\ell} - \chi\right)} =: \hat{s}(\alpha),
            \label{eq:bbr-cubic:dynamic:oscillation:model:equilibrium:critical-s}
        \end{split}
    \end{equation}
    where~$\hat{s}(\alpha)$ is the window-growth duration
    that leads to
    \begin{equation}
        \hat{x}^{\mathrm{btl}} = C_{\ell} - \frac{c\hat{s}(\alpha)^{3}}{\alpha b \tilde{\tau}_k} = \chi \overset{\geq \chi}{=} \tilde{x}^{\mathrm{btl}},
        \label{eq:bbr-cubic:dynamic:oscillation:model:equilibrium:critical-bbr}
    \end{equation}
    and to
    a CUBIC equilibrium rate
    \begin{equation}
        \tilde{x}^{\mathrm{C}} \overset{\text{(\ref{eq:bbr-cubic:dynamic:oscillation:model:equilibrium:unrestricted})}}{=} \alpha(C_{\ell} - \hat{x}^{\mathrm{btl}})
        \overset{\text{(\ref{eq:bbr-cubic:dynamic:oscillation:model:equilibrium:critical-bbr})}}{=}
        \alpha(C_{\ell} - \chi),
        \label{eq:bbr-cubic:dynamic:oscillation:model:equilibrium:critical-cubic}
    \end{equation}
    given probing strength~$\alpha$.
    In order for~$\hat{s}(\alpha)$ to be an equilibrium 
    (i.e., $\tilde{s} = \hat{s}(\alpha)$),
    $\hat{s}(\alpha)$ has to satisfy the CUBIC equilibrium conditions
    from~\cref{lmm:bbr-cubic:eq:cubic-conditions}:
    \begin{align}
            &\frac{b\tilde{\tau}_k}{c \hat{s}(\alpha)^{4}}
        \overset{\text{(\ref{eq:bbr-cubic:eq:cubic-conditions})}}{=}  1 - \frac{C_{\ell}}{\beta\tilde{x}^{\mathrm{btl}} + \tilde{x}^{\mathrm{C}}} 
        \overset{\text{(\ref{eq:bbr-cubic:dynamic:oscillation:model:equilibrium:critical-bbr})}}{\underset{\text{(\ref{eq:bbr-cubic:dynamic:oscillation:model:equilibrium:critical-cubic})}}{=}} 1 - \frac{C_{\ell}}{\chi + \alpha \left(C_{\ell} - \chi\right)}\nonumber\\
        \overset{\hat{s}(\alpha)}{\underset{\text{(\ref{eq:bbr-cubic:dynamic:oscillation:model:equilibrium:critical-s})}}{\iff}} &\frac{\alpha^4(\alpha-1)^3}{(\chi+\alpha(C_{\ell}-\chi))^3} = \frac{c}{b\tilde{\tau}_k(C_{\ell}-\chi)^7}
        \label{eq:bbr-cubic:dynamic:oscillation:model:equilibrium:critical-alpha}
    \end{align}
    % where
    % \begin{align}
    %     \kappa_7 &= \rho  &\kappa_6 &= -7\rho\chi &\kappa_5 &= 21 \rho\chi^2\\
    %     \kappa_4 &= -35\rho\chi^3 &\kappa_3 &= 35\rho\chi^4 - \alpha^3 &\kappa_2 &= -21 \rho\chi^5 + 3\alpha^{2}\left(\alpha-1\right)\chi\\
    %     \kappa_1 &= 7\rho\chi^{6}-3\alpha\left(\alpha-1\right)^{2}\chi^{2}  &\kappa_0 &= \left(\alpha-1\right)^{3}\chi^{3}-\rho\chi^{7} &
    %     \rho &= \alpha^{4}\left(\alpha-1\right)^{3}\frac{b\tilde{\tau}_k}{c}
    % \end{align}

    Let~$\hat{\alpha}$ be the solution in~$\alpha$ to~\cref{eq:bbr-cubic:dynamic:oscillation:model:equilibrium:critical-alpha},
    which cannot be found analytically in general.
    However, it must hold that~$\hat{\alpha} > 1$,
    as~$\alpha = 1$ yields a zero LHS in~\cref{eq:bbr-cubic:dynamic:oscillation:model:equilibrium:critical-alpha},
    which cannot match the non-zero RHS.

    At probing strength $\hat{\alpha}$, it holds 
    that~$\hat{s}(\hat{\alpha}) = \tilde{s}$.
    Since~$\hat{s}(\alpha)$ is an increasing
    function of~$\alpha$ according to~\cref{eq:bbr-cubic:dynamic:oscillation:model:equilibrium:critical-s},
    any~$\alpha < \hat{\alpha}$ also leads to~$\hat{s}(\alpha) < \tilde{s}$, and thus to
    $\hat{x}^{\mathrm{btl}} < \chi = \tilde{x}^{\mathrm{btl}}$.
    Hence, for $\alpha < \hat{\alpha}$, $\tilde{x}^{\mathrm{C}}$ is found 
    by solving~$\tilde{S}_2$,
    i.e., as for~$\alpha \leq 1$.
    We thus arrive at the condition on~$\alpha$
    in~\cref{thm:bbr-cubic:dynamic:oscillation:model:equilibrium}.
    
    % Let~$\hat{C}_{\ell}$ be the solution in~$C_{\ell}$
    % to~\cref{eq:bbr-cubic:dynamic:oscillation:model:equilibrium:critical-capacity:1}.
    % In general, this solution cannot be found analytically, as a septic equation
    % does not have closed-form solutions in general.
    % However, since~$\chi \approx 0$, we can approximate~$\hat{C}_{\ell}$:
    % \begin{equation}
    %     \sum_{j = 0}^{7} \kappa_i \hat{C}_{\ell}^i \overset{\chi\approx 0}{\approx} \rho \hat{C}_{\ell}^7 - \alpha^3 \hat{C}_{\ell}^3 = 0 \iff \hat{C}_{\ell} \approx \sqrt[4]{\frac{\alpha^3}{\rho}} =
    %     \sqrt[4]{\frac{c}{\alpha\left(\alpha-1\right)^{3}b\tilde{\tau}_k}}
    % \end{equation}
    
    % Hence, $\hat{C}_{\ell}(\alpha, \tilde{\tau}_k)$ is the critical capacity at 
    % which the CUBIC equilibrium
    % rate~$\tilde{x}^{\mathrm{C}}$ is exactly~$\alpha (C_{\ell}-\chi)$
    % given~$\alpha$ and~$\tilde{\tau}_k$.
    % For any capacity below~$\hat{C}_{\ell}(\alpha, \tilde{\tau}_k)$,
    % $\hat{x}^{\mathrm{btl}}$ is below~$\chi$,
    % the actual equilibrium $\tilde{x}^{\mathrm{btl}}$ is identical to~$\chi$,
    % and the equilibrium is therefore found by solving~$\tilde{S}_2$,
    % i.e., as for~$\alpha \leq 1$.
    % Hence, we arrive at the condition on~$C_{\ell}$
    % in~\cref{eq:bbr-cubic:dynamic:oscillation:model:equilibrium:s-xbtl}.

\subsection{Uniqueness of Turning Point for~$\tilde{S}_2$}
\label{prf::bbr-cubic:dynamic:oscillation:model:equilibrium:unique-inflection}

We note that the turning point~$s'$ of~$\tilde{S}_2$ in~\cref{eq:bbr-cubic:dynamic:oscillation:model:equilibrium:S2} should be 
a unique root of the first derivative~$\tilde{S}_2'$, 
where 
\begin{equation}
    \tilde{S}_2'(s) = \frac{7c^{2}}{b\tilde{\tau}_k}s^{6}-4c\left(C_{\ell}-\alpha\chi\right)s^{3}-3cs^{2}.
    \label{eq:bbr-cubic:dynamic:oscillation:model:equilibrium:first-derivative}
\end{equation}

\subsubsection{Strict convexity above critical value~$s'''$}
To find the area in which~$\tilde{S}_2'$ is strictly convex, we solve the following
            inequality for the second derivative of~$\tilde{S}_2'$:
            \begin{align}
                \tilde{S}_2'''(s) > 0 &\iff \frac{210c^{2}}{b\tilde{\tau}_k}s^{4}-24c\left(C_{\ell}-\alpha\chi\right)s > 0\\
                &\overset{/s}{\iff}  \frac{210c^{2}}{b\tilde{\tau}_k}s^{3}-24c\left(C_{\ell}-\alpha\chi\right) > 0
                \label{eq:bbr-cubic:dynamic:oscillation:model:equilibrium:convexity}
            \end{align}
            The division by~$s$ is admissible because we only consider~$s > 0$.
            To identify~$s'''$, we note that the LHS in~\cref{eq:bbr-cubic:dynamic:oscillation:model:equilibrium:convexity}
            increases from non-positive to positive with~$s$ if $C_{\ell} > \alpha\chi$, and is consistently positive for all~$s > 0$
            in the rare case where~$C_{\ell} \leq \alpha\chi$. Hence, we arrive at the following value for~$s'''$,
            marking the start of the convex area of~$\tilde{S}_2'$:
            \begin{equation}
                s''' = \begin{cases}
                    \sqrt[3]{\frac{b\tilde{\tau}_k}{c} \cdot \frac{4}{35} \left(C_{\ell}-\alpha\chi\right)} > 0 & \text{if } C_{\ell} > \alpha\chi\\
                    0 & \text{if } C_{\ell} \leq \alpha\chi.
                \end{cases}
                \label{eq:bbr-cubic:dynamic:oscillation:model:equilibrium:first-derivative:convexity-point}
            \end{equation}

\subsubsection{Non-positivity at critical value~$s'''$}
The function~$\tilde{S}_2'$ yields the following non-positive value 
at the start point~$s'''$ of the convex area:
\begin{equation}
    \tilde{S}_2'(s''') = \begin{cases}
                    -0.11 (C_{\ell} - \alpha\chi)^2 \tilde{\tau}_k - 3c{s'''}^{\frac{2}{3}} < 0 & \text{if } C_{\ell} > \alpha\chi\\
                    0 & \text{if } C_{\ell} \leq \alpha\chi.
                \end{cases}
                \label{eq:bbr-cubic:dynamic:oscillation:model:equilibrium:first-derivative:non-positive}
            \end{equation}

\subsubsection{Non-negativity in convex area}
To find an argument~$s$ at which~$\tilde{S}_2'$ is non-negative,
            we again distinguish the cases $C_{\ell} > \alpha\chi$ and~$C_{\ell} \leq \alpha\chi$.

            \paragraph{$C_{\ell} > \alpha\chi$}
            We consider the following function~$\Psi^-(s)$, which constitutes a lower bound on~$\tilde{S}_2'$,
            i.e., $\forall s > 0.\ \Psi^-(s) \leq \tilde{S}_2'(s)$:
            \begin{equation}
                \Psi^-(s) = \begin{cases}
                    \frac{7c^{2}}{b\tilde{\tau}_k}s^{6}-4c\left(C_{\ell}-\alpha\chi\right)s^{3}-3cs^{\colorbox{gray!30}{3}} & \text{if } s \geq 1\\
                    \frac{7c^{2}}{b\tilde{\tau}_k}s^{6}-4c\left(C_{\ell}-\alpha\chi\right)s^{\colorbox{gray!30}{2}}-3cs^{2} & \text{if } s < 1
                \end{cases}
                \label{eq:bbr-cubic:dynamic:oscillation:model:equilibrium:first-derivative:lower-bound}
            \end{equation}
            The \colorbox{gray!30}{highlights} in \cref{eq:bbr-cubic:dynamic:oscillation:model:equilibrium:first-derivative:lower-bound} mark the differences of~$\Psi^-$ and~$\tilde{S}_{2}'$ from \cref{eq:bbr-cubic:dynamic:oscillation:model:equilibrium:first-derivative}.
            
            The non-zero root of~$\Psi^-$ is~$s^- = \max(\sqrt[3]{s_0}, \sqrt[4]{s_0})$, where
            \begin{equation}
                s_0 = \frac{b\tilde{\tau}_k}{c}\left(\frac{4}{7}\left(C_{\ell}-\alpha\chi\right)+\frac{3}{7}\right).
            \end{equation}
            Since~$\Psi^-$ is a lower bound on~$\tilde{S}_2'$, it holds that
            \begin{equation}
                \tilde{S}_2'(s^{-}) \geq \Psi^-(s^{-}) = 0.
                \label{eq:bbr-cubic:dynamic:oscillation:model:equilibrium:first-derivative:lower-bound:root}
            \end{equation}
            The non-negativity point~$s^{-}$ is in the convex area if~$s''' < s^-$, which demonstrably holds:
            \begin{equation}
                \begin{split}
                    s''' &\overset{\text{(\ref{eq:bbr-cubic:dynamic:oscillation:model:equilibrium:first-derivative:convexity-point})}}{=} \sqrt[3]{\frac{b\tilde{\tau}_k}{c} \cdot \frac{4}{35} \left(C_{\ell}-\alpha\chi\right)} <
                \sqrt[3]{\frac{b\tilde{\tau}_k}{c} \cdot \frac{4}{7} \left(C_{\ell}-\alpha\chi\right)}\\
                &< \sqrt[3]{\frac{b\tilde{\tau}_k}{c}\left(\frac{4}{7}\left(C_{\ell}-\alpha\chi\right)+\frac{3}{7}\right)} \leq s^-.
                \end{split}
            \end{equation}

            \paragraph{$C_{\ell} \leq \alpha\chi$}
            In this rare case, the lower~bound function~$\Psi^-$ and its root~$s^-$ are more simply expressed:
            \begin{align}
                \Psi^-(s) &=  \frac{7c^{2}}{b\tilde{\tau}_k}s^{6} \colorbox{gray!30}{+} 4c\left(C_{\ell}-\alpha\chi\right)s^{\colorbox{gray!30}{2}}-3cs^{\mathbf{2}}\label{eq:bbr-cubic:dynamic:oscillation:model:equilibrium:psi-minus}\\
                s^- &= \sqrt[4]{\frac{b\tilde{\tau}_k}{c}\left(-\frac{4}{7}\left(C_{\ell}-\alpha\chi\right)+\frac{3}{7}\right)}
                \label{eq:bbr-cubic:dynamic:oscillation:model:equilibrium:s-minus}
            \end{align}
            Again, the \colorbox{gray!30}{highlights} in ~\cref{eq:bbr-cubic:dynamic:oscillation:model:equilibrium:psi-minus} mark the differences between~$\Psi^-$ and $\tilde{S}_2'$ from~\cref{eq:bbr-cubic:dynamic:oscillation:model:equilibrium:first-derivative}.
            
            Note that~$\Psi^-$ is a \emph{strict} lower bound (i.e., $\Psi^-(s) < \tilde{S}_2'(s)$) if~$C_{\ell} < \alpha\chi$,
            and equals~$\tilde{S}_2'$ for~$C_{\ell} = \alpha\chi$. 
            Hence, the following property holds on~$s^-$:
            \begin{equation}
                \tilde{S}_2'(s^-) \begin{cases}
                    > \Psi^-(s^-) & \text{if $C_{\ell} < \alpha\chi$}\\
                    = \Psi^-(s^-) & \text{if $C_{\ell} = \alpha\chi$}
                \end{cases} = 0
                \label{eq:bbr-cubic:dynamic:oscillation:model:equilibrium:first-derivative:lower-bound:root:2}
            \end{equation}
            
            The convex-area membership of~$s^- > s'''$ is demonstrated analogously to the previous case:
            \begin{equation}
                s''' \overset{\text{(\ref{eq:bbr-cubic:dynamic:oscillation:model:equilibrium:first-derivative:convexity-point})}}{=} 0 \overset{C_{\ell} \leq \alpha\chi}{<}  \sqrt[4]{\frac{b\tilde{\tau}_k}{c}\left(-\frac{4}{7}\left(C_{\ell}-\alpha\chi\right)+\frac{3}{7}\right)} \overset{\text{(\ref{eq:bbr-cubic:dynamic:oscillation:model:equilibrium:s-minus})}}{=} s^-.
            \end{equation}

\subsubsection{Combination of arguments}
\label{prf::bbr-cubic:dynamic:oscillation:model:equilibrium:unique-inflection:combination}

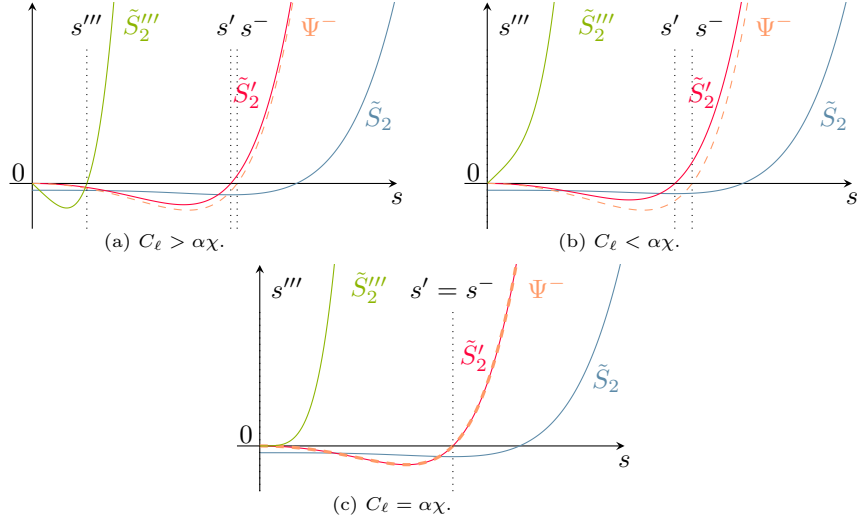
\begin{figure*}
    \centering
    \begin{subfigure}[b]{0.25\linewidth}
        \centering
        \input{figures/turning-point-case1}
        \vspace{-20pt}
        \caption{$C_{\ell} > \alpha\chi$.}
        \label{fig::bbr-cubic:dynamic:oscillation:model:equilibrium:unique-inflection:1}
    \end{subfigure}\hspace{50pt}
    \begin{subfigure}[b]{0.25\linewidth}
        \centering
        \input{figures/turning-point-case2}
        \vspace{-20pt}
        \caption{$C_{\ell} < \alpha\chi$.}
        \label{fig::bbr-cubic:dynamic:oscillation:model:equilibrium:unique-inflection:2}
    \end{subfigure}\hspace{50pt}\\
    \begin{subfigure}[b]{0.25\linewidth}
        \centering
        \input{figures/turning-point-case3}
        \vspace{-20pt}
        \caption{$C_{\ell} = \alpha\chi$.}
        \label{fig::bbr-cubic:dynamic:oscillation:model:equilibrium:unique-inflection:3}
    \end{subfigure}
    \caption{Case distinction for determination of unique turning point in~\cref{prf::bbr-cubic:dynamic:oscillation:model:equilibrium:unique-inflection:combination}.}
    \label{fig::bbr-cubic:dynamic:oscillation:model:equilibrium:unique-inflection}
\end{figure*}

        In summary, the function~$\tilde{S}_2'$ evolves in a strictly convex fashion 
        from~$\tilde{S}_2'(s''') \leq 0$ at~$s'''$ to~$\tilde{S}_2'(s^-) \geq 0$
        at~$s^- > s'''$. To demonstrate that these conditions imply a unique root with
        subsequent increasing behavior, we consider the relevant cases separately
        and visualize them in~\cref{fig::bbr-cubic:dynamic:oscillation:model:equilibrium:unique-inflection}
        to simplify understanding:

        \paragraph{$C_{\ell} > \alpha\chi$ (\cref{fig::bbr-cubic:dynamic:oscillation:model:equilibrium:unique-inflection:1})}
        In this case, we observe that~$\tilde{S}_2'(s''') < 0$ (\cref{eq:bbr-cubic:dynamic:oscillation:model:equilibrium:first-derivative:non-positive}) 
        and $\tilde{S}_2'(s^-) \geq 0$ (\cref{eq:bbr-cubic:dynamic:oscillation:model:equilibrium:first-derivative:lower-bound:root}).
        Given this property, the function~$\tilde{S}_2'$ has at least some increasing part between~$s'''$
        and~$s^-$, and at least the first root is in such an increasing part. 
        Moreover, since~$\tilde{S}_2'$ is strictly convex, the function keeps increasing once it is increasing, 
        proving uniqueness of the root~$s'$ and increasing behavior after~$s'$.

        \paragraph{$C_{\ell} < \alpha\chi$ (\cref{fig::bbr-cubic:dynamic:oscillation:model:equilibrium:unique-inflection:2})}
        In this case, it holds that~$\tilde{S}_2'(s''') = 0$ (\cref{eq:bbr-cubic:dynamic:oscillation:model:equilibrium:first-derivative:non-positive})
        and $\tilde{S}_2'(s^-) > 0$ (\cref{eq:bbr-cubic:dynamic:oscillation:model:equilibrium:first-derivative:lower-bound:root:2}).
        Strict convexity allows either (i)~monotonic increase between~$s''' = 0$ and~$s^- > 0$, 
        which creates a unique root at~$s''' = 0$ such that~$\tilde{S}_2'(s) > 0\ \forall s > 0$, or 
        (ii)~a decrease followed by an increase, causing a unique root~$s' \in (s''', s^-) = (0, s^-)$ above which~$\tilde{S}_2'(s) > 0$.
        
        \paragraph{$C_{\ell} = \alpha\chi$ (\cref{fig::bbr-cubic:dynamic:oscillation:model:equilibrium:unique-inflection:3})}
        In this last case, the function~$\tilde{S}_2'$ has the property~$\tilde{S}_2'(s''') = \tilde{S}_2'(s^-) = 0$ (\cref{eq:bbr-cubic:dynamic:oscillation:model:equilibrium:first-derivative:non-positive,eq:bbr-cubic:dynamic:oscillation:model:equilibrium:first-derivative:lower-bound:root:2}).
        Given strict convexity, the function~$\tilde{S}_2'$ can only return to~$0$ at~$s^-$ if it first
        decreases and then increases forever, causing a unique root~$s' = s^-$ such that~$\tilde{S}_2'(s) > 0\ \forall s > s'$.

\section{Proof of Theorem~\ref{thm:bbr-cubic:dynamic:oscillation:model:convergence}: Stability of the Short-Term Equilibrium}
\label{prf:bbr-cubic:dynamic:oscillation:model:convergence}

 Our proof proceeds in three main steps, namely:
    \begin{enumerate}
        \item Stability investigation of the full linearized dynamics
        \item Dimension reduction by characterization of the center manifold
        \item Stability investigation of the lower-dimensional dynamics
    \end{enumerate}

\subsection{Stability Investigation of the Full Linearized Dynamics}

    \subsubsection{Centering the dynamics}
    Fundamentally, we consider the dynamic system~$\dot{\boldsymbol{\sigma}} = f(\boldsymbol{\sigma})$,
    where the state variables are~$\sigma = \left[\tilde{x}^{\mathrm{btl}}, \tilde{w}^{\max}, \tilde{s}\right]^{\top}$, the adaptation function is~$f = \left[\dot{x}^{\mathrm{btl}}, \dot{w}^{\max}, \dot{s}\right]^{\top}$, and the equilibrium is~$\tilde{\boldsymbol{\sigma}} = \left[\tilde{x}^{\mathrm{btl}}, \tilde{w}^{\max}, \tilde{s}\right]^{\top}$, i.e., $\tilde{\boldsymbol{\sigma}} = f(\tilde{\boldsymbol{\sigma}})$.
    To show the asymptotic stability of the equilibrium~$\tilde{\boldsymbol{\sigma}}$,  
    we first center the dynamic system around the equilibrium:
    We transform the adaptation function~$f$ 
    to~$f^{\circ}$ such that 
    $\tilde{\mathbf{z}} = \boldsymbol{0} = \left[0, 0, 0\right]^{\top}$ 
    is an equilibrium of the dynamic 
    system~$\dot{\mathbf{z}} = f^{\circ}(\mathbf{z})$,
    i.e., $\boldsymbol{0} = \tilde{\mathbf{z}} = f^{\circ}(\tilde{\mathbf{z}})$.
    In our case, the dynamics in the differential
    equations for~$\dot{x}^{\mathrm{btl}}$, 
    $\dot{w}^{\max}$ (\cref{eq:bbr-cubic:system-evolution:cubic:wmax})
    and~$\dot{s}$ (\cref{eq:bbr-cubic:system-evolution:cubic:s}) 
    are centered as follows:
    \begin{align}
        \dot{z}_1 &= \frac{\alpha \left(z_1 + \tilde{x}^{\mathrm{btl}}\right) C}{\alpha \left(z_1 + \tilde{x}^{\mathrm{btl}}\right) + \frac{1}{\tilde{\tau}_k} W^{\circ}(z_2, z_3)} - \left(z_1 + \tilde{x}^{\mathrm{btl}}\right)\\
        \dot{z}_2 &= \left(W^{\circ}(z_2, z_3) - \left(z_2 + \tilde{w}^{\max}\right)\right) \cdot \frac{W^{\circ}(z_2, z_3)}{\tilde{\tau}_k} \cdot p^{\circ}(\mathbf{z})\label{eq:bbr-cubic:stability:center-manifold:z2-dyn}\\
        \dot{z}_3 &= 1 - \left(z_3 + \tilde{s}\right) \cdot \frac{W^{\circ}(z_2, z_3)}{\tilde{\tau}_k} \cdot p^{\circ}(\mathbf{z})
    \end{align}
    where both the CUBIC window-growth function~$W^{\circ}$ and the loss-rate function~$p^{\circ}$ are
    centered as well:
    \begin{equation}
        W^{\circ}(z_2, z_3) \overset{\text{(\ref{eq:bbr-cubic:model:basic-model:cubic-window-growth})}}{=} \left(z_2 + \tilde{w}^{\max}\right) + c\left(\left(z_3 + \tilde{s}\right) - \sqrt[3]{\frac{(z_2 + \tilde{w}^{\max}) b}{c}}\right)^3
    \end{equation}
    \begin{equation}
        p^{\circ}(\mathbf{z}) \overset{\text{(\ref{eq:bbr-cubic:model:basic-model:loss})}}{=} \begin{cases}
            1 - \frac{C_{\ell}}{\beta(z_1 + \tilde{x}^{\mathrm{btl}}) + \frac{1}{\tilde{\tau}_k} W^{\circ}(z_2, z_3)} & \text{if } \parbox[t]{.3\textwidth}{$y_{\ell}(\mathbf{z}) > C_{\ell} \land q_{\ell} = B_{\ell}$}\\
            0 & \text{otherwise}
        \end{cases}
    \end{equation}

    \subsubsection{Linearizing the dynamics}
    We now linearize the adaptation function~$f^{\circ}$ of this centered dynamic system 
    around the equilibrium~$\tilde{\mathbf{z}} = \mathbf{0}$ with the first-order Taylor expansion:
    \begin{equation}
        \mathcal{L}_{\mathbf{0}}[f^{\circ}](\mathbf{z}) = f^{\circ}(\mathbf{0}) + \mathbf{J}_{f^{\circ}}(\mathbf{0}) \mathbf{z} =  \mathbf{J}_{f^{\circ}}(\mathbf{0}) \mathbf{z},
    \end{equation}
    where~$\mathcal{L}_{\mathbf{0}}[f^{\circ}]$ indicates the linearization of the adaptation function~$f^{\circ}$ around
    the new equilibrium~$\mathbf{0}$,~$\mathbf{J}_{f^{\circ}}(\mathbf{0})$ is the Jacobian matrix of~$f^{\circ}$ evaluated at the
    equilibrium~$\mathbf{0}$. Since~$f^{\circ}(\mathbf{0}) = \boldsymbol{0}$, the dynamics around
    the equilibrium are dominated by $\mathbf{J}_{f^{\circ}}(\mathbf{0})\mathbf{z}$. 
    The stability properties of the dynamic system thus depend on~$\mathbf{J}_{f^{\circ}}(\mathbf{0})$,
    especially on its eigenvalues.

    To find these eigenvalues, we need to characterize~$\mathbf{J}_{f^{\circ}}(\mathbf{0})$,
    which is considerably simplified by the following identities:
    \begin{equation}
        \begin{split}
            &W^{\circ}(0,0) = \tilde{w}^{\max}, \hspace{15pt}  \frac{\partial W^{\circ}}{\partial z_2}(0,0) = 1,\\  &\frac{\partial W^{\circ}}{\partial z_3}(0,0) = 0,  \hspace{10pt} \text{and} \hspace{10pt} \frac{\partial p^{\circ}}{\partial z_3}(\mathbf{0}) = 0.
        \end{split}
    \end{equation}
    
    Moreover, we know the following properties of the equilibrium
    from~\cref{sec:new-model}:
    \begin{equation}
        \begin{split}
            &\tilde{s} > 0, \hspace{15pt} 
        \tilde{w}^{\max} = \frac{c}{b}\tilde{s}^3 > 0, \hspace{15pt}
        p^{\circ}(\mathbf{0}) = \frac{b\tilde{\tau}_k}{c\tilde{s}^4} \in (0, 1],\\ 
        &\text{and} \hspace{10pt} 
        \tilde{x}^{\mathrm{btl}} = \max\left(\chi, C_{\ell} - \frac{1}{\alpha} \frac{\tilde{w}^{\max}}{\tilde{\tau}_k}\right) > 0.
        \end{split}
        \label{eq:bbr-cubic:stability:eq-properties}
    \end{equation}
    
    Using these equalities, $\mathbf{J}_{f^{\circ}}(\mathbf{0})$ in our case is:
    \begin{align}
        \mathbf{J}_{f^{\circ}}(\mathbf{0}) = 
        &\begin{bmatrix}
            J_{11} & J_{12} & 0\\
            0 & 0 & 0\\
            J_{31} & J_{32} & J_{33}
            \end{bmatrix}\label{eq:bbr-cubic:stability:J}\\
        J_{11} = &\frac{\alpha C_{\ell} \frac{1}{\tilde{\tau}_k} W^{\circ}(0, 0)}{\left(\alpha \tilde{x}^{\mathrm{btl}} + \frac{1}{\tilde{\tau}_k} W^{\circ}(0, 0)\right)^2} -1 \label{eq:bbr-cubic:stability:j11}\\
        &= \frac{\alpha C_{\ell} \frac{1}{\tilde{\tau}_k} \tilde{w}^{\max}}{\left(\alpha \tilde{x}^{\mathrm{btl}} + \frac{1}{\tilde{\tau}_k} \tilde{w}^{\max}\right)^2} -1 < 0 \nonumber\\
        J_{12} = &-\frac{\alpha C_{\ell} \tilde{x}^{\mathrm{btl}}}{\tilde{\tau}_k \left(\alpha \tilde{x}^{\mathrm{btl}} + \frac{1}{\tilde{\tau}_k} W^{\circ}(0,0)\right)^2} \cdot \frac{\partial W^{\circ}}{\partial z_2}(0,0)  \label{eq:bbr-cubic:stability:j12}\\
        &= -\frac{\alpha C_{\ell} \tilde{x}^{\mathrm{btl}}}{\tilde{\tau}_k \left(\alpha \tilde{x}^{\mathrm{btl}} + \frac{1}{\tilde{\tau}_k} \tilde{w}^{\max}\right)^2}
        < 0 \nonumber\\
        J_{31} = &-\tilde{s} \frac{W^{\circ}(0,0)}{\tilde{\tau}_k} \frac{\beta C}{\left(\beta\tilde{x}^{\mathrm{btl}} + \frac{1}{\tilde{\tau}_k} W^{\circ}(0,0)\right)^2} \label{eq:bbr-cubic:stability:j31}\\
        &= - \frac{\beta C_{\ell} \tilde{s}\tilde{w}^{\max}}{\tilde{\tau}_k\left(\beta\tilde{x}^{\mathrm{btl}} + \frac{1}{\tilde{\tau}_k} \tilde{w}^{\max}\right)^2}  < 0 \nonumber\\
         J_{32} = 
         &-\frac{\tilde{s}}{\tilde{\tau}_k} \cdot \frac{\partial W^{\circ}}{\partial z_2}(0,0) \cdot\left(1 - \frac{C_{\ell}}{ \beta \tilde{x}^{\mathrm{btl}} + \frac{1}{\tilde{\tau}_k} W^{\circ}(0, 0)}\right)\label{eq:bbr-cubic:stability:j32}\\
         &- \tilde{s} \cdot \frac{W^{\circ}(0,0)}{\tilde{\tau}_k} \cdot \frac{C_{\ell}}{\tilde{\tau}_k(\beta \tilde{x}^{\mathrm{btl}} + \frac{1}{\tilde{\tau}_k} W^{\circ}(0,0))^2} \cdot \frac{\partial W^{\circ}}{\partial z_2}(0,0)\nonumber\\
         = &-\frac{\tilde{s}}{\tilde{\tau}_k} \cdot p^{\circ}(\mathbf{0}) - \frac{\tilde{s}\tilde{w}^{\max}}{\tilde{\tau}_k^2} 
         \frac{C_{\ell}}{\left(\beta \tilde{x}^{\mathrm{btl}} + \frac{1}{\tilde{\tau}_k} \tilde{w}^{\max}\right)^2} < 0\nonumber\\
         %&-\frac{1}{\tilde{w}^{\max}} - \frac{\left(\tilde{s}\tilde{w}^{\max} - \tilde{\tau}_k\right)^2}{C_{\ell}\tilde{\tau}_k^2 \tilde{s}\tilde{w}^{\max}} < 0\\
         J_{33} = &-\frac{W^{\circ}(0,0)}{\tilde{\tau}_k} \cdot p^{\circ}(\mathbf{0}) 
         - \frac{\tilde{s}}{\tilde{\tau}_k} \cdot \frac{\partial W^{\circ}}{\partial z_3}(0,0) \cdot p^{\circ}(\mathbf{0}) \label{eq:bbr-cubic:stability:j33}\\
         &- \tilde{s} \cdot \frac{W^{\circ}(0,0)}{\tilde{\tau}_k} \cdot \frac{\partial p^{\circ}}{\partial z_3}(\mathbf{0})\nonumber\\
         = &-\frac{1}{\tilde{s}} < 0\nonumber
    \end{align}
    
    Hence, all entries of~$\mathbf{J}_{f^{\circ}}(\mathbf{0})$ are negative. Among
    the above matrix entries, the bounding of the entry~$J_{11}$ in~\cref{eq:bbr-cubic:stability:j11}
    is not trivial and requires a case distinction on~$\chi$:

    \paragraph{$\chi \leq C_{\ell} - \tilde{w}^{\max}/(\alpha\tilde{\tau}_k)$}
    In this case, the bounding is straightforward:
    \begin{align}
        &\tilde{x}^{\mathrm{btl}} \overset{\text{(\ref{eq:bbr-cubic:stability:eq-properties})}}{=}  \max\left(\chi, C_{\ell} - \frac{\tilde{w}^{\max}}{\alpha \tilde{\tau}_k}\right)
         \overset{\text{Case}}{=} C_{\ell} - \frac{\tilde{w}^{\max}}{\alpha \tilde{\tau}_k}
         \label{eq:bbr-cubic:stability:j11-bounding:xbtl}\\ 
        &\implies \alpha \tilde{x}^{\mathrm{btl}} + \frac{\tilde{w}^{\max}}{\tilde{\tau}_k} = \alpha C_{\ell} \implies \label{eq:bbr-cubic:stability:j11-bounding:alpha-Cl}\\
        &J_{11} = \frac{\alpha C_{\ell} \frac{\tilde{w}^{\max}}{\tilde{\tau}_k}}{\left(\alpha C_{\ell}\right)^2} - 1 = 
        \frac{\frac{\tilde{w}^{\max}}{\tilde{\tau}_k}}{\alpha C_{\ell}} - 1 
        \leq \frac{\alpha(C_{\ell} - \chi)}{\alpha C_{\ell}} - 1 < 0
    \end{align}

    \paragraph{$\chi > C_{\ell} - \tilde{w}^{\max}/(\alpha\tilde{\tau}_k)$}
    For this case, we note that~$\tilde{x}^{\mathrm{btl}}$ is at the minimum~$\chi$
    of its domain:
    \begin{equation}
        \tilde{x}^{\mathrm{btl}} \overset{\text{(\ref{eq:bbr-cubic:stability:eq-properties})}}{=}  \max\left(\chi, C_{\ell} - \frac{\tilde{w}^{\max}}{\alpha \tilde{\tau}_k}\right) = \chi
        \label{eq:bbr-cubic:stability:j11-bounding:xbtl-chi}
    \end{equation}
    Moreover, we obtain a lower bound on the CUBIC equilibrium rate~$\tilde{x}^{\mathrm{C}}$
    for this case:
    \begin{equation}
        \chi > C_{\ell} - \frac{\tilde{w}^{\max}}{\alpha \tilde{\tau}_k} \iff
        \tilde{x}^{\mathrm{C}} = \frac{\tilde{w}^{\max}}{\tilde{\tau}_k} > \alpha\left(C_{\ell}-\chi\right).
        \label{eq:bbr-cubic:stability:j11-bounding:cubic}
    \end{equation}
    
    Moreover, we obtain another lower bound~$\hat{x}^{\mathrm{C}}$ on the CUBIC equilibrium rates
    that lead to a negative Jacobian entry~$J_{11}$:
    \begin{align}
        &J_{11}\left(\tilde{x}^{\mathrm{C}}\right) 
        \overset{\text(\ref{eq:bbr-cubic:stability:j11}){}}{=}
        \frac{\alpha C_{\ell} \tilde{x}^{\mathrm{C}}}{\left(\alpha \chi + \tilde{x}^{\mathrm{C}}\right)^2} -1 < 0\\
        \iff &\tilde{x}^{\mathrm{C}2} + \alpha(2\chi - C_{\ell})\tilde{x}^{\mathrm{C}} + \alpha^{2}\chi^{2} =: \Psi(\tilde{x}^{\mathrm{C}}) > 0
         \label{eq:bbr-cubic:stability:j11-bounding:1}\\
        \iff &\tilde{x}^{\mathrm{C}} > \alpha\frac{\sqrt{C_{\ell}\left(C_{\ell}-4\chi\right)}+\left(C_{\ell}-2\chi\right)}{2} =: \hat{x}^{\mathrm{C}}.
        \label{eq:bbr-cubic:stability:j11-bounding:2}
    \end{align}
    
    We note that the lower bound~$\hat{x}^{\mathrm{C}}$ in~\cref{eq:bbr-cubic:stability:j11-bounding:2} only
    exists if~$4\chi \leq C_{\ell}$. 
    In contrast, the non-existence of~$\hat{x}^{\mathrm{C}}$ means that $\Psi$ from~\cref{eq:bbr-cubic:stability:j11-bounding:1} has no root.
    Since~$\Psi$ is strictly convex, and a strictly convex function with no roots
    is always positive, $\Psi > 0$ from~\cref{eq:bbr-cubic:stability:j11-bounding:1} holds for all~$\tilde{x}^{\mathrm{C}}$.
    % In this case, we thus need to verify that~$\Psi(\tilde{x}^{\mathrm{C}}) > 0$ for \emph{all}~$\tilde{x}^{\mathrm{C}}$.
    % Given~$4\chi > C_{\ell} \iff \chi > C_{\ell}/4$, we can substitute~$C_{\ell}/4$ for~$\chi$
    % in~\cref{eq:bbr-cubic:stability:j11-bounding:1} to obtain
    % a strict lower-bound function 
    % $\Psi^{-}(\tilde{x}^{\mathrm{C}})$ on~$\Psi$, and check whether
    % $\Psi^{-}(\tilde{x}^{\mathrm{C}}) \geq 0$:
    % \begin{equation}
    %     \begin{split}
    %         &\Psi(\tilde{x}^{\mathrm{C}}) = \tilde{x}^{\mathrm{C}2} + \alpha(2\chi - C_{\ell})\tilde{x}^{\mathrm{C}} + \alpha^{2}\chi^{2}
    %         >\\
    %         &\Psi^{-}(\tilde{x}^{\mathrm{C}}) = \tilde{x}^{\mathrm{C}2} + \alpha(\frac{C_{\ell}}{2} - C_{\ell})\tilde{x}^{\mathrm{C}} + \alpha^{2}\left(\frac{C_{\ell}}{4}\right)^{2} \geq 0
    %     \end{split}
    % \end{equation}
    % This condition holds, as the minimum of~$\Psi^-$ for~$\tilde{x}^{\mathrm{C}} > 0$ is exactly 0.

    Conversely, if~$4\chi \leq C_{\ell}$ and~$\hat{x}^{\mathrm{C}}$ therefore exists, 
    we need to verify that~$\hat{x}^{\mathrm{C}} \leq \alpha(C_{\ell} - \chi)$
    such that all~$\tilde{x}^{\mathrm{C}} > \alpha(C_{\ell} - \chi)$
    (i.e., all CUBIC equilibrium rates possible according to~\cref{eq:bbr-cubic:stability:j11-bounding:cubic})
    lead to a negative~$J_{11}$. Given~$\hat{x}^{\mathrm{C}}$ from~\cref{eq:bbr-cubic:stability:j11-bounding:2},
    we obtain:
    \begin{equation}
        \begin{split}
                    &\alpha\frac{\sqrt{C_{\ell}\left(C_{\ell}-4\chi\right)}+\left(C_{\ell}-2\chi\right)}{2} \leq \alpha(C_{\ell} - \chi) \iff\\
                    &- 4\chi C_{\ell} \leq 0 \iff \chi \geq 0 \iff \top.
        \end{split}
    \end{equation}

    In summary, $J_{11} < 0$ thus holds for all~$\tilde{w}^{\max}/\tilde{\tau}_k$
    and all~$\chi$.

    \subsubsection{Finding the eigenpairs}
    Finding the eigenvalues and eigenvectors of~$\mathbf{J}_{f^{\circ}}(\mathbf{0})$
    means
    finding~$(\lambda, \mathbf{v})$ such that $\lambda \in \mathbb{C}$ is an eigenvalue,
    $\mathbf{v} \in \mathbb{C}^3$ is the corresponding eigenvector and must be non-zero ($\mathbf{v} \neq \mathbf{0}$), 
    and~$\mathbf{J}_{f^{\circ}}(\mathbf{0})\mathbf{v} = \lambda\mathbf{v}$.
    Hence, any solution~$(\lambda, \mathbf{v})$ satisfies the following system of equations:
    \begin{align}
        (J_{11}-\lambda) v_1 + J_{12} v_2 &= 0 \label{eq:bbr-cubic:stability:eig-1}\\
        \lambda v_2 &= 0 \label{eq:bbr-cubic:stability:eig-2}\\
        J_{31} v_1 + J_{32} v_2 + (J_{33}-\lambda) v_3 &= 0 \label{eq:bbr-cubic:stability:eig-3}
    \end{align}
    \cref{eq:bbr-cubic:stability:eig-2} implies that~$\lambda$ or~$v_2$ must be zero.

    \paragraph{$\lambda = 0$}
    First, we check whether~$\lambda = 0$ is an eigenvalue of~$\mathbf{J}_{f^{\circ}}(\mathbf{0})$.
    Assuming~$\lambda = 0$, the equation system reduces to two equations:
    \begin{equation}
        J_{11} v_1 + J_{12} v_2 = 0 \hspace{15pt}
        J_{31} v_1 + J_{32} v_2 + J_{33} v_3 = 0
    \end{equation}
    % Depending on the Jacobian entries~$J_{11}$ and~$J_{12}$, we can identify the following eigenvectors for the eigenvalue~$\lambda^{(1)}~=~0$:
    % \begin{itemize}
    %     \item $\boldsymbol{J_{11} < 0,\ J_{12} < 0:}$ The eigenvector~$\mathbf{v}^{(1a)}$
    %     satisfies the following conditions:
    %     \begin{equation}
    %         v_1^{(1a)} \in \mathbb{C} \setminus \{0\} \hspace{15pt}
    %         v_2^{(1a)} = -\frac{J_{11}}{J_{12}}v_1^{(1a)} \hspace{15pt}
    %         v_3^{(1a)} = \frac{J_{11}J_{32} - J_{12}J_{31}}{J_{12}J_{33}} v_1^{(1a)}
    %         \label{eq:bbr-cubic:stability:v_1a}
    %     \end{equation}
    %     In this case, this eigenvector exists because~$J_{12} \neq 0$ and~$J_{12} J_{33} \neq 0$.
    %     Moreover,~$J_{11} \neq 0$ ensures a non-zero eigenvector.
    %     \item $\boldsymbol{J_{11} = 0,\ J_{12} = 0:}$ 
    %     The eigenvector~$\mathbf{v}^{(1b)}$ satisfies the following conditions:
    %     \begin{equation}
    %         J_{31} v_1^{(1b)} + J_{32} v_2^{(1b)} + J_{33} v_3^{(1b)} = 0 \hspace{15pt} \mathbf{v}^{(1b)} \neq \mathbf{0}
    %         \label{eq:bbr-cubic:stability:v_1b}
    %     \end{equation}
    %     \item $\boldsymbol{J_{11} < 0,\ J_{12} = 0:}$
    %     The eigenvector~$\mathbf{v}^{(1c)}$ satisfies the following
    %     conditions:
    %     \begin{equation}
    %         v_1^{(1c)} = 0 \quad v_2^{(1c)} \in \mathbb{C}\setminus\{0\} \quad 
    %         v_3^{(1c)} = -\frac{J_{32}}{J_{33}} v_2^{(1c)}
    %     \end{equation}
    % \end{itemize}
    Given this equation system,
    we can identify the following eigenvector~$v^{(1)}$ for the eigenvalue~$\lambda^{(1)}~=~0$:
    \begin{equation}
            v_1^{(1)} \in \mathbb{C} \setminus \{0\} \hspace{15pt}
            v_2^{(1)} = -\frac{J_{11}}{J_{12}}v_1^{(1)} \hspace{15pt}
            v_3^{(1)} = \frac{J_{11}J_{32} - J_{12}J_{31}}{J_{12}J_{33}} v_1^{(1)}
            \label{eq:bbr-cubic:stability:v_1a}
        \end{equation}
        This eigenvector exists because the denominators are non-zero 
        given~$J_{12} < 0$ and~$J_{12} J_{33} > 0$.
        Moreover,~$J_{11} > 0$ ensures a non-zero eigenvector.
        
    \paragraph{$v_2 = 0$}
    To find further eigenvalues, we now assume~$v_2 = 0$, 
    leading again to a reduced equation system:
    \begin{equation}
          (J_{11}-\lambda) v_1 = 0 \hspace{15pt} J_{31} v_1 + (J_{33} - \lambda) v_3 = 0
    \end{equation}
    For this equation system, we can perform a case distinction on~$v_1$,
    the first entry of the eigenvector:
    \begin{itemize}
        \item $\boldsymbol{v_1 \neq 0:}$ Assuming~$v_1 \neq 0$ yields the eigenpair~$(\lambda^{(2)}, \mathbf{v}^{(2)})$:
    \begin{equation}
        \begin{split}
            \lambda^{(2)} = J_{11} \hspace{15pt} 
        &v^{(2)}_1 \in \mathbb{C} \setminus \{0\}\\
        v^{(2)}_2 = 0 \hspace{15pt}
        &v^{(2)}_3 = \frac{J_{31}}{J_{11} - J_{33}} v^{(2)}_1
        \end{split}
    \end{equation}
    % Note that if~$J_{11} = 0$,~$(\lambda^{(2)}, \mathbf{v}^{(2)})$ is 
    % equivalent to~$(\lambda^{(1b)}, \mathbf{v}^{(1b)})$. Otherwise,~$\lambda^{(2)}$
    % is negative.
        \item $\boldsymbol{v_1 = 0:}$ In this case,
        the equation system collapses to the single equation~$(J_{33}-\lambda) v_3 = 0$,
        where~$v_3 \neq 0$ because not all entries of an eigenvector can be 0.
        Hence, $\mathbf{J}_{f^{\circ}}(\mathbf{0})$ has the following eigenpair~$(\lambda^{(3)}, \mathbf{v}^{(3)})$:
        \begin{equation}
            \begin{split}
                \lambda^{(3)} = J_{33} \overset{\text{(\ref{eq:bbr-cubic:stability:j33})}}{=} -\frac{1}{\tilde{s}} \overset{\text{(\ref{eq:bbr-cubic:stability:eq-properties})}}{<} 0 \hspace{15pt} &v^{(3)}_1 = 0\\
                v^{(3)}_2 = 0 \hspace{15pt}
                &v^{(3)}_3 \in \mathbb{C}\setminus \{0\}
            \end{split}
        \end{equation}
    \end{itemize}

    \paragraph{Summary of eigenpairs}
    The Jacobian matrix~$\mathbf{J}_{f^{\circ}}(\mathbf{0})$ thus has
    zero and negative eigenvalues. For convenience, we categorize the eigenvectors
    into sets corresponding to zero and negative eigenvalues, respectively:
    % \begin{equation}
    %     V^{0} = \begin{cases}
    %         \{\mathbf{v}^{(1a)}\} & \text{if } J_{11} < 0, J_{12} < 0\\
    %         \{\mathbf{v}^{(1b)}\} & \text{if } J_{11} = 0, J_{12} = 0\\
    %         \{\mathbf{v}^{(1c)}\} & \text{if } J_{11} < 0, J_{12} = 0
    %     \end{cases}
    %     \hspace{30pt}
    %     V^{-} = \begin{cases}
    %         \{\mathbf{v}^{(2)}, \mathbf{v}^{(3)}\} & \text{if } J_{11} < 0\\
    %         \{\mathbf{v}^{(3)}\} & \text{if } J_{11} = 0 
    %     \end{cases}
    % \end{equation}
    \begin{equation}
        V^{0} = 
            \{\mathbf{v}^{(1)}\} 
        \hspace{30pt}
        V^{-} = 
            \{\mathbf{v}^{(2)}, \mathbf{v}^{(3)}\}
    \end{equation}
    The presence of both zero and negative eigenvalues means that 
    the stability properties of the original
    nonlinear dynamic system~$\dot{\mathbf{z}}~=~f^{\circ}(\mathbf{z})$ cannot be derived from 
    the linearized system~$\dot{\mathbf{z}}~=~\mathbf{J}_{f^{\circ}}(\mathbf{0})\mathbf{z}$~\cite{vardoyan2021towards}.
    Instead, higher-order terms determine the stability.
    
\subsection{Dimension Reduction via the Center Manifold}
    While the linearized system is inconclusive about the desired stability properties,
    it allows some insight into the dynamics of the nonlinear system when
    using center-manifold theory.
    
    \subsubsection{Center-manifold properties}
    
    To introduce the center manifold, we note that the subspace
    spanned by the eigenvectors in~$V^-$
    is the \emph{stable subspace} of the Jacobian~$\mathbf{J}_{f^{\circ}}(\mathbf{0})$, 
    and the subspace spanned by the eigenvectors in~$V^0$
    is the~\emph{center subspace}~\cite{wiggins2003introduction}.
    Each of these subspaces is associated with a manifold that has the same
    dimension as the corresponding subspace, and is tangential
    to the corresponding subspace at the equilibrium~\cite{guckenheimer2013nonlinear}.
    
    Since our system has only has a stable manifold and a center manifold
    (and no unstable manifold, which would be associated with positive eigenvalues),
    we can use the \emph{center-manifold emergence theorem}~\cite{rard1998topics}.
    This theorem states that given a starting point sufficiently close to
    the center manifold, the dynamics converge exponentially quickly to
    the center manifold, and thus approach a trajectory on the center
    manifold. The overall dynamics of the nonlinear system can thus be
    approximated by the dynamics on the center manifold, which have
    lower dimension and thus allow a more tractable analysis.
    % To analyze these center-manifold dynamics, we separately consider alls 
    % cases regarding~$J_{11}$ and~$J_{12}$.

    % \subsubsection{$J_{11} < 0,\ J_{12} < 0$}
    \subsubsection{Center-manifold dynamics}
    
        To derive the dynamics on the center manifold, we start by decoupling the system state~$\mathbf{z}$
    along subspaces, i.e., we transform it onto a different basis such that every variable
    only effects a change along either the center or the stable subspace.
    We achieve this by a coordinate transformation using the eigenbasis:
    \begin{align}
            &\mathbf{z} = \mathbf{T}\boldsymbol{\zeta} = [\mathbf{v}^{(1)}\ \mathbf{v}^{(2)}\ \mathbf{v}^{(3)}] \boldsymbol{\zeta} = 
        \begin{bmatrix}
            T_{11} & T_{12} & 0\\
            T_{21} & 0 & 0\\
            T_{31} & T_{32} & T_{33}
        \end{bmatrix} \begin{bmatrix}
            \zeta_1\\ \zeta_2\\ \zeta_3 
        \end{bmatrix}\nonumber\\
        \iff &\boldsymbol{\zeta} = \mathbf{T}^{-1} \mathbf{z},
        \label{eq:bbr-cubic:stability:center-manifold:transform}
    \end{align}
    where~$\zeta_1$ is the variable associated with the center subspace, 
    and~$\zeta_2$ and~$\zeta_3$ are the variables associated with the stable subspace.
    From the structure of~$\mathbf{T}$, we see that the center variable~$\zeta_1$
    can be expressed exclusively by~$z_2$ (corresponding to~$w^{\max}$):
    \begin{equation}
        z_2 = T_{21} \zeta_1 \quad \iff \quad \zeta_1 = \frac{z_2}{T_{21}}.
        \label{eq:bbr-cubic:stability:center-manifold:zeta-1}
    \end{equation}

    According to the center-manifold existence theorem, a manifold~$\Gamma_c$
    exists with the following properties around the equilibrium~\cite{wiggins2003introduction}:
    \begin{align}
        \Gamma_c = \{(\zeta_1, \zeta_2, \zeta_3)\ \mid\ &\zeta_2 = h_2(\zeta_1),\ h_2(0) = 0,\ {h_2}'(0) = 0, \label{eq:bbr-cubic:stability:center-manifold:manifold}\\
        &\zeta_3 = h_3(\zeta_1),\ h_3(0) = 0,\ {h_3}'(0) = 0\}\nonumber
    \end{align}
    where both~$h_2$ and~$h_3$ are one-dimensional functions of the form:
    \begin{equation}
        h(\zeta_1) = \sum_{i = 2}^{\infty} a_i\zeta_1^i.
    \end{equation}
    Intuitively, in our case, the center manifold~$\Gamma_c$ corresponds to a curve
    in three-dimensional space, which is tangential to the center subspace at the equilibrium~$\mathbf{0}$
    and is fully describable by the center variable~$\zeta_1$.
    The overall dynamics of the nonlinear system move along that curve, although it is
    not clear yet whether towards or away from the equilibrium.
    To identify this direction, we only need to consider the following one-dimensional dynamics along the center manifold:
    \begin{equation}
        \dot{\zeta}_{1} \overset{\text{(\ref{eq:bbr-cubic:stability:center-manifold:zeta-1})}}{=} \frac{1}{T_{21}}\dot{z}_2(\mathbf{z}) \overset{\text{(\ref{eq:bbr-cubic:stability:center-manifold:transform})}}{=} \frac{1}{T_{21}}\dot{z}_2(\mathbf{T}\boldsymbol{\zeta}')
        \label{eq:bbr-cubic:stability:center-manifold-dyn:1}
    \end{equation}
    where
    \begin{equation}
        \boldsymbol{\zeta}' \overset{\text{(\ref{eq:bbr-cubic:stability:center-manifold:manifold})}}{=} \begin{bmatrix}\zeta_1\\ h_2(\zeta_1)\\ h_3(\zeta_1)\end{bmatrix}
        \overset{\text{(\ref{eq:bbr-cubic:stability:center-manifold:zeta-1})}}{=}
        \begin{bmatrix}z_2/T_{21}\\ h_2(z_2/T_{21})\\ h_3(z_2/T_{21})\end{bmatrix}.
        \label{eq:bbr-cubic:stability:center-manifold-dyn:2}
    \end{equation}

\subsection{Stability Investigation of the Lower-Dimensional Dynamics}
    So far, we have reduced the full dynamics to the lower-dimensional
    center-manifold dynamics, which we can now investigate.
    % We perform this analysis by again distinguishing all
    % cases regarding~$J_{11}$ and~$J_{12}$:

        %\subsubsection{$J_{11} < 0,\ J_{12} < 0$}

        \subsubsection{Taylor expansion of center-manifold dynamics}
        
        The one-dimensional center-manifold dynamics in~\cref{eq:bbr-cubic:stability:center-manifold-dyn:1}
        suggests that the dynamics~$\dot{z}_2$ from~\cref{eq:bbr-cubic:stability:center-manifold:z2-dyn} 
        are solely relevant for stability.
        However, since we investigate these dynamics on the center manifold, 
        the state variables in~$\mathbf{z}$ need to be expressed
        based on the center-manifold functions of~$z_2$:
        \begin{equation}
            \mathbf{\hat{z}}(z_2) = \mathbf{T}\boldsymbol{\zeta}' \overset{\text{(\ref{eq:bbr-cubic:stability:center-manifold:transform})}}{\underset{\text{(\ref{eq:bbr-cubic:stability:center-manifold-dyn:2})}}{=}} \begin{bmatrix}
                T_{11}\frac{z_2}{T_{21}} + T_{12} h_2\left(\frac{z_2}{T_{21}}\right)\\
                z_2\\
                T_{31}\frac{z_2}{T_{21}} + T_{32} h_2\left(\frac{z_2}{T_{21}}\right) + T_{33} h_3\left(\frac{z_2}{T_{21}}\right)
            \end{bmatrix}.
        \end{equation}
        Hence, $\dot{z}_2$ with the variables from~$\mathbf{z}$ substituted by~$\mathbf{\hat{z}}$ is:
        \begin{equation}
            \begin{split}
                 \dot{z}_2 =& \left(W^{\circ}(z_2, \hat{z}_3(z_2)) - \left(z_2 + \tilde{w}^{\max}\right)\right) \cdot\\
                 &\frac{W^{\circ}(z_2, \hat{z}_3(z_2))}{\tilde{\tau}_k} \cdot p^{\circ}(\mathbf{\hat{z}}(z_2)).
            \label{eq:bbr-cubic:stability:reduced-dyn:1}
            \end{split}
        \end{equation}

        Around the equilibrium~$\tilde{z}_2 = 0$, this evolution function~$\dot{z}_2$
        can be approximated via a Taylor expansion of third degree:
        \begin{equation}
            \dot{z}_2(z_2') \approx \dot{z}_2(0) 
            + \frac{\partial \dot{z}_2}{\partial z_2}(0) \cdot z_2' 
            + \frac{\partial^2 \dot{z}_2}{\partial z_2^2}(0) \cdot \frac{{z_2'}^2}{2!}
            + \frac{\partial^3 \dot{z}_2}{\partial z_2^3}(0) \cdot \frac{{z_2'}^3}{3!}.
        \end{equation}
        By using the Symbolic Math Toolbox of Matlab~\cite{symmathmatlab}, this approximated function can be reduced
        to the following expression:
        \begin{equation}
             \dot{z}_2(z_2') \approx \frac{c \tilde{w}^{\max} T_{31}^3 p^{\circ}(\mathbf{0})}{\tilde{\tau}_k T_{21}^3} \cdot {z_2'}^3
             = K{z_2'}^3.
             \label{eq:bbr-cubic:stability:center-manifold:K}
        \end{equation}
        It is easy to see that asymptotic stability of this system requires a negative~$K$:
        If~$z_2' < 0$ (below the equilibrium), $z_2'$ would be positively affected (multiplication of
        negative numbers) and thus drawn closer to the equilibrium at 0, 
        whereas if~$z_2' > 0$, $z_2'$ would be negatively affected,
        and thereby also attracted to the equilibrium at 0.
        To prove asymptotic stability, we thus have to show:
        \begin{equation}
            \begin{split}
                K < 0 &\overset{\text{(\ref{eq:bbr-cubic:stability:center-manifold:K})}}{\iff} \frac{c \tilde{w}^{\max} T_{31}^3 p^{\circ}(\mathbf{0})}{\tilde{\tau}_k T_{21}^3} < 0\\
                &\overset{\text{(\ref{eq:bbr-cubic:stability:center-manifold:transform})}}{\iff} \frac{c \tilde{w}^{\max} {v^{(1)}_{3}}^3 p^{\circ}(\mathbf{0})}{\tilde{\tau}_k {v^{(1)}_{2}}^3} < 0
            \end{split}
            \label{eq:bbr-cubic:stability:case1:taylor:simplified}
        \end{equation} where~$\mathbf{v}^{(1)}$ fulfills the condition in~\cref{eq:bbr-cubic:stability:v_1a}.
        
        Without loss of generality, we set~$v^{(1)}_1 < 0$, which implies (together with~$J_{11} < 0$ and~$J_{12} < 0$):
        \begin{equation}
            v^{(1)}_2 \overset{\text{(\ref{eq:bbr-cubic:stability:v_1a})}}{=} -\frac{J_{11}}{J_{12}} v^{(1)}_1 > 0.
        \end{equation} 
        
        Since~$c$, ~$\tilde{w}^{\max}$,~$p^{\circ}(\mathbf{0})$, and~$\tilde{\tau}_k$ are all known 
        to be positive, asymptotic stability depends on the following condition:
        
        \begin{align}
                &{v^{(1)}_{3}}^3 < 0 \iff v^{(1)}_{3} < 0 \overset{\text{(\ref{eq:bbr-cubic:stability:v_1a})}}{\iff}\\
                & \frac{J_{11}J_{32} - J_{12}J_{31}}{J_{12}J_{33}} v_1^{(1)} < 0
                \iff \frac{J_{11}J_{32} - J_{12}J_{31}}{J_{12}J_{33}} > 0.\nonumber
            \end{align}
        Since~$J_{12} < 0$ and~$J_{33} < 0$, we can simplify this condition even further to:
        \begin{equation}
            J_{11}J_{32} > J_{12}J_{31} \quad \iff \quad J_{11} < \frac{J_{31}}{J_{32}} J_{12}.
            \label{eq:bbr-cubic:stability:ratio-bounding}
        \end{equation}
        To check whether this condition holds, we again perform a case distinction
        regarding~$\chi$, i.e., we distinguish 
        the cases~$\chi \leq C_{\ell} - \tilde{w}^{\max}/(\alpha\tilde{\tau}_k)$
        and~$\chi > C_{\ell} - \tilde{w}^{\max}/(\alpha\tilde{\tau}_k)$.
        
        \subsubsection{$\chi \leq C_{\ell} - \tilde{w}^{\max}/(\alpha\tilde{\tau}_k)$}
        In this case, the expanded form of~\cref{eq:bbr-cubic:stability:ratio-bounding}
        can be simplified by the finding in~\cref{eq:bbr-cubic:stability:j11-bounding:alpha-Cl}:
        \begin{align}
            \begin{split}
                &J_{11} \overset{\text{(\ref{eq:bbr-cubic:stability:j11})}}{=} \frac{\alpha C_{\ell}}{\tilde{\tau}_k\left(\alpha \tilde{x}^{\mathrm{btl}} + \frac{1}{\tilde{\tau}_k} \tilde{w}^{\max}\right)^2}   \cdot \tilde{w}^{\max} - 1 \overset{\text{(\ref{eq:bbr-cubic:stability:ratio-bounding})}}{<}\\
            &- \frac{J_{31}}{J_{32}} \frac{\alpha C_{\ell}}{\tilde{\tau}_k\left(\alpha \tilde{x}^{\mathrm{btl}} + \frac{1}{\tilde{\tau}_k} \tilde{w}^{\max}\right)^2}   \cdot \tilde{x}^{\mathrm{btl}} \overset{\text{(\ref{eq:bbr-cubic:stability:j12})}}{=}  \frac{J_{31}}{J_{32}} J_{12}
            \end{split}
            \\
            \overset{\text{(\ref{eq:bbr-cubic:stability:j11-bounding:alpha-Cl})}}{\iff} 
            &\frac{1}{\tilde{\tau}_k \alpha C_{\ell}}   \cdot \tilde{w}^{\max} - 1 < 
            - \frac{J_{31}}{J_{32}} \frac{1}{\tilde{\tau}_k\alpha C_{\ell}} \cdot \tilde{x}^{\mathrm{btl}}\label{eq:bbr-cubic:stability:ratio-bounding-intermediate}\\
            \overset{\cdot \tilde{\tau}_k \alpha C_{\ell}}{\underset{/\tilde{x}^{\mathrm{btl}}}{\iff}} &\left(\tilde{w}^{\max} - \tilde{\tau}_k \alpha C_{\ell}\right) \cdot \frac{1}{\tilde{x}^{\mathrm{btl}}} < 
            - \frac{J_{31}}{J_{32}}\\
            \overset{\text{(\ref{eq:bbr-cubic:stability:j11-bounding:xbtl})}}{\iff} 
            &\left(\tilde{w}^{\max} - \tilde{\tau}_k \alpha C_{\ell}\right) \cdot \frac{1}{C_{\ell} - \frac{\tilde{w}^{\max}}{\tilde{\tau}_k\alpha}} < 
            - \frac{J_{31}}{J_{32}}\\
            \overset{\text{expand}}{\iff} 
            &\left(\tilde{w}^{\max} - \tilde{\tau}_k \alpha C_{\ell}\right) \cdot \frac{\tilde{\tau}_k \alpha}{\tilde{\tau}_k \alpha C_{\ell}-\tilde{w}^{\max}} < 
            - \frac{J_{31}}{J_{32}}\\
            \overset{\text{cancel}}{\underset{\cdot -1}{\iff}} &\tilde{\tau}_k\alpha > \frac{J_{31}}{J_{32}} \overset{1/}{\iff} \frac{J_{32}}{J_{31}} > \frac{1}{\tilde{\tau}_k \alpha}.
        \end{align} 
        
        In turn, expanding the ratio~$J_{32}/J_{31}$ yields:
        \begin{align}
            \begin{split}
                \frac{J_{32}}{J_{31}} \overset{\text{(\ref{eq:bbr-cubic:stability:j32})}}{\underset{\text{(\ref{eq:bbr-cubic:stability:j31})}}{=}} &-\left(\frac{\tilde{s}}{\tilde{\tau}_k} \cdot p^{\circ}(\mathbf{0}) + \frac{\tilde{s}\tilde{w}^{\max}}{\tilde{\tau}_k^2} 
         \frac{C_{\ell}}{\left(\beta \tilde{x}^{\mathrm{btl}} + \frac{1}{\tilde{\tau}_k} \tilde{w}^{\max}\right)^2}\right)
         \cdot \\
         &\left( - \frac{\tilde{\tau}_k\left(\beta\tilde{x}^{\mathrm{btl}} + \frac{1}{\tilde{\tau}_k} \tilde{w}^{\max}\right)^2}{\beta C_{\ell} \tilde{s}\tilde{w}^{\max}} \right)
            \end{split}\\
         = &\frac{p^{\circ}(\mathbf{0})\left(\beta\tilde{x}^{\mathrm{btl}} + \frac{1}{\tilde{\tau}_k} \tilde{w}^{\max}\right)^2}{\beta C_{\ell} \tilde{w}^{\max}} + \frac{1}{\tilde{\tau}_k \beta} > \frac{1}{\tilde{\tau}_k \alpha}.\label{eq:bbr-cubic:stability:case1:bound}
        \end{align} The bound holds because~$\beta \leq \alpha$ and the first term on the LHS
        in~\cref{eq:bbr-cubic:stability:case1:bound} is strictly positive.
        Hence, also~\cref{eq:bbr-cubic:stability:ratio-bounding} holds.
        
        \subsubsection{$\chi > C_{\ell} - \tilde{w}^{\max}/(\alpha\tilde{\tau}_k)$}
        In this case, $\tilde{x}^{\mathrm{btl}} \overset{\text{(\ref{eq:bbr-cubic:stability:j11-bounding:xbtl-chi})}}{=} \chi$ and the expanded form of~\cref{eq:bbr-cubic:stability:ratio-bounding}
        is thus:
        \begin{align}
            &\frac{\alpha C_{\ell}}{\tilde{\tau}_k\left(\alpha \chi + \frac{1}{\tilde{\tau}_k} \tilde{w}^{\max}\right)^2}   \cdot \tilde{w}^{\max} - 1 \overset{\text{(\ref{eq:bbr-cubic:stability:ratio-bounding})}}{<} 
            - \frac{J_{31}}{J_{32}} \frac{\alpha C_{\ell} \chi}{\tilde{\tau}_k\left(\alpha \chi + \frac{1}{\tilde{\tau}_k} \tilde{w}^{\max}\right)^2}\\
            &\iff \frac{J_{32}}{J_{31}} >
            \frac{\alpha C_{\ell}\tilde{\tau}_k \chi}{\tilde{w}^{\max2} + \alpha(2\chi - C_{\ell})\tilde{\tau}_k \tilde{w}^{\max} + \alpha^2 C_{\ell}^2 \chi^2}
            \label{eq:bbr-cubic:stability:case2:bound}
        \end{align}
        
        Expanding the ratio~$J_{32}/J_{31}$ yields:
        \begin{align}
            \begin{split}
                \frac{J_{32}}{J_{31}} \overset{\text{(\ref{eq:bbr-cubic:stability:j32})}}{\underset{\text{(\ref{eq:bbr-cubic:stability:j31})}}{=}} &-\left(\frac{\tilde{s}}{\tilde{\tau}_k} \cdot p^{\circ}(\mathbf{0}) + \frac{\tilde{s}\tilde{w}^{\max}}{\tilde{\tau}_k^2} 
         \frac{C_{\ell}}{\left(\beta \chi + \frac{1}{\tilde{\tau}_k} \tilde{w}^{\max}\right)^2}\right)
         \cdot\\ &\left( - \frac{\tilde{\tau}_k\left(\beta\chi + \frac{1}{\tilde{\tau}_k} \tilde{w}^{\max}\right)^2}{\beta C_{\ell} \tilde{s}\tilde{w}^{\max}} \right)
            \end{split}\nonumber\\
         = &\frac{\tilde{\tau}_k\left(\beta \chi+\frac{\tilde{w}^{\max}}{\tilde{\tau}_k}\right)^{2}}{\beta C_{\ell}\tilde{s}\tilde{w}^{\max2}}+\frac{1}{\beta \tilde{\tau}_k} > \frac{1}{\beta \tilde{\tau}_k}.
         \label{eq:bbr-cubic:stability:case2:bound-j31j32}
        \end{align}
        Thanks to this neat lower bound of~$J_{32}/J_{31}$,
        \cref{eq:bbr-cubic:stability:case2:bound} is implied by
        the following stronger condition:
        \begin{align}
            \frac{J_{32}}{J_{31}} \overset{\text{(\ref{eq:bbr-cubic:stability:case2:bound-j31j32})}}{>} &\frac{1}{\beta \tilde{\tau}_k} > \frac{\alpha C_{\ell}\tilde{\tau}_k \chi}{\tilde{w}^{\max2} + \alpha(2\chi - C_{\ell})\tilde{\tau}_k \tilde{w}^{\max} + \alpha^2 C_{\ell}^2 \chi^2}\\
            \iff & 
            \frac{1}{\beta \tilde{\tau}_k}\tilde{w}^{\max2}+\frac{\alpha\left(2\chi-C\right)}{\beta}\tilde{w}^{\max}+\frac{\tilde{\tau}_k\alpha^{2}\chi^{2}}{\beta}-\tilde{\tau}_k\alpha C_{\ell}\chi \nonumber\\
            &=: \Psi(\tilde{w}^{\max}) > 0.
            \label{eq:bbr-cubic:stability:case2:bound-stronger}
        \end{align}

        To verify~\cref{eq:bbr-cubic:stability:case2:bound-stronger}
        for all~$\tilde{w}^{\max} > \tilde{\tau}_k \alpha (C_{\ell} - \chi)$,
        we first note that~$\Psi$ is convex.
        Hence, if~$\Psi$ has no roots, 
        the convexity of~$\Psi$ implies
        that~$\Psi(\tilde{w}^{\max}) > 0$ holds for any~$\tilde{w}^{\max}$.
        In contrast, if~$\Psi$ has roots, $\Psi(\tilde{w}^{\max}) > 0$
        holds for any~$\tilde{w}^{\max}$ above the upper root~$\psi$
        of~$\Psi$, which is:
        \begin{equation}
            \psi = \frac{\tilde{\tau}_k}{2}\left(\alpha\left(C_{\ell}-2\chi\right) + \sqrt{\alpha^2 C_{\ell}^{2}+4\alpha C_{\ell}\left(\beta-\alpha\right)\chi}\right)
            \label{eq:bbr-cubic:stability:case2:bound-stronger-roots}
        \end{equation}
        The truth of $\Psi(\tilde{w}^{\max}) > 0$ $\forall \tilde{w}^{\max} > \tilde{\tau}_k \alpha (C_{\ell}-\chi)$
        is confirmed by the fact that~$\psi$ is below the relevant area of argument~$\tilde{w}^{\max}$:
        \begin{equation}
            \begin{split}
                &\psi \leq \tilde{\tau}_k \alpha (C_{\ell} - \chi)
                \overset{\cdot 2/\tilde{\tau}_k}{\iff} \\
                & \alpha\left(C_{\ell}-2\chi\right) + \sqrt{\alpha^2 C_{\ell}^{2}+4\alpha C_{\ell}\left(\beta-\alpha\right)\chi}
                \leq 2 \alpha (C_{\ell} - \chi)\\
                &\overset{-\alpha(C_{\ell} - 2\chi)}{\iff}  
                \sqrt{\alpha^2 C_{\ell}^{2}+4\alpha C_{\ell}\left(\beta-\alpha\right)\chi} \leq \alpha C_{\ell}\\
                &\overset{(\cdot)^2}{\underset{-\alpha C_{\ell}}{\iff}} 4\alpha C_{\ell} (\beta - \alpha) \chi \leq 0 \overset{\beta \leq \alpha}{\underset{\chi > 0}{\iff}} \top.
            \end{split}
        \end{equation}
        
        \subsubsection{Conclusion}
        
        In conclusion, the relation between the Jacobian entries~$J_{11}$,
        $J_{12}$, $J_{31}$, and~$J_{32}$ as given
        in~\cref{eq:bbr-cubic:stability:ratio-bounding} ensures a negative
        third entry~$v^{(1)}_3$ in the center eigenvector, which is required
        for a negative coefficient in the third-order Taylor expansion
        of the center-manifold dynamics in~\cref{eq:bbr-cubic:stability:case1:taylor:simplified}.
        The negativity of this coefficient guarantees asymptotic stability of
        the center-manifold dynamics in~\cref{eq:bbr-cubic:stability:reduced-dyn:1},
        and thus of the full dynamic system.

\section{Proof of Theorem~\ref{thm:bbr-cubic:dynamic:oscillation:model:condition}: BBR-CUBIC Oscillation}
\label{prf::bbr-cubic:dynamic:oscillation:model:condition}

\subsection{Update Functions}
\label{prf::bbr-cubic:dynamic:oscillation:model:condition:update}

We start the proof by characterizing the update function~$w^{\leftarrow}$ and
used in the discrete-time process in~\cref{eq:bbr-cubic:dynamic:oscillation:model:conditions:discrete}.
Conceptually, this function~$w^{\leftarrow}(w)$ yields the CUBIC window size
after an interval of the short-term dynamics given BBR probing strength~$\alpha$,
which in turn results from CUBIC window size~$w$ at the beginning of the interval.
We denote this resulting probing strength by~$\alpha^{\leftarrow}(w)$,
and characterize it in~\cref{prf::bbr-cubic:dynamic:oscillation:model:condition:update:alpha}
before characterizing~$w^{\leftarrow}$ in~\cref{prf::bbr-cubic:dynamic:oscillation:model:condition:update:w}.

\subsubsection{$\alpha$-update function $\alpha^{\leftarrow}$}
\label{prf::bbr-cubic:dynamic:oscillation:model:condition:update:alpha}

From~\cref{eq:bbr-cubic:system-evolution:remaining-queue} in the 
fluid-equilibrium
model in~\cref{sec:new-model}, we know that~$\alpha$
is determined based on the queue length~$q_{\ell}^-$ that remains when
the BBR flow backs off:
\begin{align}
        \alpha^{\leftarrow}(w) = &\min(\nicefrac{5}{4}, \grave{\alpha}^{\leftarrow}(w))\\
        \text{where} \quad \grave{\alpha}^{\leftarrow}(w) = &\frac{2}{\tau_i}\left(\tau^{\mathrm{p}}_i + \frac{q_{\ell}^{-}(w)}{C_{\ell}}\right)\\
    = &\frac{2}{\tau_i}\left(\tau^{\mathrm{p}}_i + \frac{\left[4 + (1-b)w - \tau^{\mathrm{p}}_{\ell}C_{\ell}\right]_{0}^{B_{\ell}}}{C_{\ell}}\right).
\end{align}

This function~$\alpha^{\leftarrow}(w)$ achieves its minimum value~$\alpha_{\min}$
if~$q_{\ell}^- = 0$, which is achieved for all CUBIC window sizes~$w \leq w_0$, where:
\begin{equation}
    w_0 = \frac{1}{1-b}\left(C_{\ell}\tau^{\mathrm{p}}_{\ell} - 4\right) \implies \alpha_{\min} = \alpha^{\leftarrow}(w_0) = \frac{2\tau_i^{\mathrm{p}}}{\tau_i}
    \label{eq:bbr-cubic:dynamic:oscillation:model:condition:update:alpha:w0}
\end{equation}

Conversely, the maximum~$\alpha_{\max}$ is not achieved for~$q_{\ell}^- = B_{\ell}$,
as in this case~$\grave{\alpha}^{\leftarrow} = 2\tau_i/\tau_i= 2$. Instead, the maximum value~$\alpha_{\max}$
is achieved for all window sizes~$w \geq w_1$, where
\begin{align}
        &w_1 = \frac{1}{1-b}\left(C_{\ell}\left(\frac{5}{8} \tau_i + \tau^{\mathrm{p}}_{\ell} - \tau^{\mathrm{p}}_i\right) - 4\right)
    \implies \grave{\alpha}^{\leftarrow}(w_1) = \nicefrac{5}{4}\nonumber\\
    &\implies \alpha^{\leftarrow}(w_1) = \nicefrac{5}{4} = \alpha_{\max}.
    \label{eq:bbr-cubic:dynamic:oscillation:model:condition:update:alpha:w1}
\end{align}

\begin{figure*}
\begin{minipage}{0.45\linewidth}
    \centering
    \includegraphics[width=0.8\linewidth]{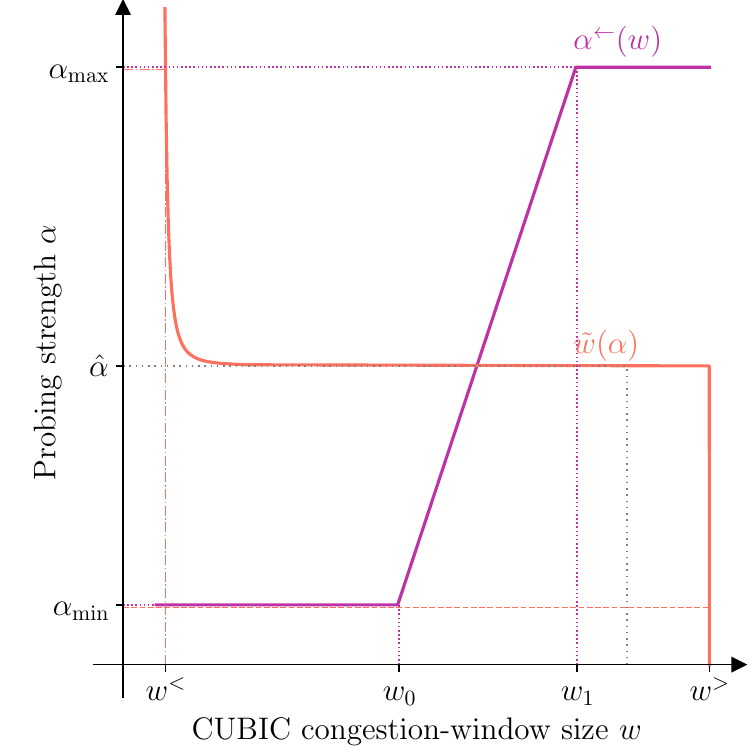}
    \caption{Illustration of functions~$\alpha^{\leftarrow}(w)$ and~$\tilde{w}(\alpha)$.}
    \label{fig:bbr-cubic:dynamic:oscillation:model:condition:update:basic}
\end{minipage}\quad\vrule\quad
\begin{minipage}{0.45\linewidth}
    \centering
    \includegraphics[width=0.8\linewidth]{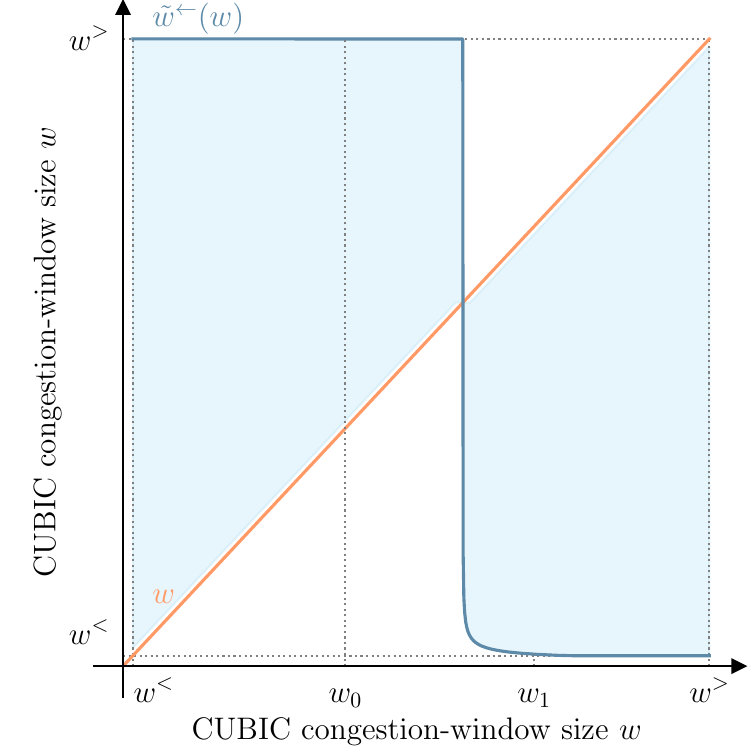}
    \caption{Illustration of function~$\tilde{w}^{\leftarrow}(w)$, and value range
    for function~$w^{\leftarrow}(w)$.}
    \label{fig:bbr-cubic:dynamic:oscillation:model:condition:update:window}
    \end{minipage}
\end{figure*}

Given the above value range, we rewrite the function~$\alpha^{\leftarrow}$ such that its constant and linear pieces
become obvious:
\begin{equation}
    \alpha^{\leftarrow}(w) =
    \begin{cases}
        \alpha_{\min} = \frac{2\tau_i^{\mathrm{p}}}{\tau_i} & \text{if } w \leq w_0,\\
        \grave{\alpha}(w) = \frac{2}{\tau_i} \left(\tau_i^{\mathrm{p}} + \frac{w + 4 - \tau^{\mathrm{p}}_{\ell}C_{\ell}}{C_{\ell}}\right) & \text{if } w \in (w_0, w_1),\\
        \alpha_{\max} = \nicefrac{5}{4} & \text{if } w \geq w_1.
    \end{cases}
    \label{eq:bbr-cubic:dynamic:oscillation:model:condition:alpha-res}
\end{equation}
This function~$\alpha^{\leftarrow}$ is illustrated 
in~\cref{fig:bbr-cubic:dynamic:oscillation:model:condition:update:basic}.

\subsubsection{Window-update function $w^{\leftarrow}$}
\label{prf::bbr-cubic:dynamic:oscillation:model:condition:update:w}

The window-update function~$w^{\leftarrow}(w)$ yields the CUBIC window
size after an interval with initial window size~$w$ and corresponding 
BBR probing strength~$\alpha = \alpha^{\leftarrow}(w)$.
Since this new CUBIC window size is the result of convergence
towards the short-term equilibrium window size~$\tilde{w}(\alpha)$,
we first revisit~$\tilde{w}(\alpha)$ based on~\cref{thm:bbr-cubic:dynamic:oscillation:model:equilibrium}.
While this function does not have a closed-form representation,
we can still conclude that~$\tilde{w}(\alpha)$ has a two-part
structure. In particular, we know that an inflection point~$\hat{\alpha} > 1$ exists
such that~$\tilde{w}(\alpha)$ is found by solving the equation~$\tilde{S}_2(s) = 0$ (cf.~\cref{eq:bbr-cubic:dynamic:oscillation:model:equilibrium:additional:S2})
for all~$\alpha < \hat{\alpha}$,
and and found by solving~$\tilde{S}_1(s) = 0$
(cf.~\cref{eq:bbr-cubic:dynamic:oscillation:model:equilibrium:additional:S1})
for all~$\alpha \geq \hat{\alpha}$
(cf.~\cref{prf::bbr-cubic:dynamic:oscillation:model:equilibrium}).
Crucially, $\tilde{w}(\alpha)$ can be further confirmed
to be strictly monotonically decreasing in~$\alpha$ (cf.~\cref{prf::bbr-cubic:dynamic:oscillation:model:condition:update:w-decr}).

Since~$\tilde{w}(\alpha)$ is thus monotonically decreasing
on its complete domain~$[0, \alpha_{\max}]$,
the value range of~$\tilde{w}$ for that domain
is bounded to~$[w^{<}, w^{>}]$, where
\begin{align}
w^{<} = \tilde{w}(\alpha_{\max}) \overset{\text{(\ref{eq:bbr-cubic:dynamic:oscillation:model:condition:update:alpha:w1})}}{=} \tilde{w}(\alpha^{\leftarrow}(w_1)) = \tilde{w}^{\leftarrow}(w_1)\label{eq:bbr-cubic:dynamic:oscillation:model:condition:w_min}\\
w^{>} = \tilde{w}(\alpha_{\min}) \overset{\text{(\ref{eq:bbr-cubic:dynamic:oscillation:model:condition:update:alpha:w0})}}{=}  \tilde{w}(\alpha^{\leftarrow}(w_0)) = \tilde{w}^{\leftarrow}(w_0)
\label{eq:bbr-cubic:dynamic:oscillation:model:condition:w_max}
\end{align}
The function~$\tilde{w}(\alpha)$ is visualized in~\cref{fig:bbr-cubic:dynamic:oscillation:model:condition:update:basic}.

Note that the above equation also introduces the function~$\tilde{w}^{\leftarrow}(w) = \tilde{w}(\alpha^{\leftarrow}(w))$,
which is of great importance throughout the rest of the proof.
This function~$\tilde{w}^{\leftarrow}$ can be explicitly represented based
on~\cref{eq:bbr-cubic:dynamic:oscillation:model:condition:alpha-res}:
\begin{equation}
    \tilde{w}^{\leftarrow}(w) = \begin{cases}
        w^{>} & \text{if } w \leq w_0,\\
        \tilde{w}(\grave{\alpha}^{\leftarrow}(w)) & \text{if } w \in (w_0, w_1),\\
        w^{<} & \text{if } w \geq w_1,
    \end{cases}
    \label{eq:bbr-cubic:dynamic:oscillation:model:condition:full-update}
\end{equation}
where~$w_0$ and~$w_1$ are such that~$\tilde{w}^{\leftarrow}$
is \emph{strictly} monotonically decreasing for arguments~$w \in [w_0, w_1]$.
% \begin{align}
%     w_0 = &\begin{cases}
%         w_0 & \text{if } \hat{\alpha} < \alpha_{\min}\\
%         \left(\grave{\alpha}^{\leftarrow}\right)^{-1}(\hat{\alpha}) 
%         & \text{if } \hat{\alpha} \in [\alpha_{\min}, \alpha_{\max})\\
%         \infty & \text{if } \hat{\alpha} \geq \alpha_{\max}.
%     \end{cases}\\
%     w_1 = &\begin{cases}
%         w_1 & \text{if } \hat{\alpha} < \alpha_{\max},\\
%         \infty & \text{if } \hat{\alpha} \geq \alpha_{\max}.
%     \end{cases}
% \end{align}
% Notably, if~$\hat{\alpha} < \alpha_{\max}$, the function~$\tilde{w}^{\leftarrow}$
% decreases from value $w^{>}$ at argument~$w_0 \geq w_0$ 
% to value~$w^{<} < w^{>}$ at argument~$w_1 = w_1 > w_0$.
% Conversely, if~$\hat{\alpha} \geq \alpha_{\max}$, $\tilde{w}^{\leftarrow}$
% is never strictly decreasing for any argument~$w$, and thus constant ($w^{>} = w^{<}$).
In summary, the function~$\tilde{w}^{\leftarrow}(w)$
is monotonically decreasing for all~$w$:
\begin{equation}
    \forall w \in \mathbb{R}.\quad \frac{\partial \tilde{w}^{\leftarrow}(w)}{\partial w} \leq 0
    \label{eq::bbr-cubic:dynamic:oscillation:model:condition:update:w-decr}
\end{equation}

Based on~\cref{eq:bbr-cubic:dynamic:oscillation:model:condition:full-update},
we can finally represent~$w^{\leftarrow}(w)$. For every interval, ~$w^{\leftarrow}$
describes the convergence from the interval-start CUBIC window size~$w$ towards its
associated equilibrium~$\tilde{w}(\alpha^{\leftarrow}(w)) = \tilde{w}^{\leftarrow}(w)$,
which follows from the asymptotic stability of any~$\tilde{w}(\alpha)$ proven in~\cref{thm:bbr-cubic:dynamic:oscillation:model:convergence}:
\begin{equation}
    w^{\leftarrow}(w) \in \begin{cases}
        (w, \tilde{w}^{\leftarrow}(w)] & \text{if } w < \tilde{w}^{\leftarrow}(w),\\
        [w] = [\tilde{w}^{\leftarrow}(w)] & \text{if } w = \tilde{w}^{\leftarrow}(w),\\
        [\tilde{w}^{\leftarrow}(w), w) & \text{if } w > \tilde{w}^{\leftarrow}(w).
    \end{cases}
    \label{eq:bbr-cubic:dynamic:oscillation:model:condition:w-arrow}
\end{equation}
The function~$\tilde{w}^{\leftarrow}$ and the value ranges
for~$w^{\leftarrow}(w)$ are illustrated in~\cref{fig:bbr-cubic:dynamic:oscillation:model:condition:update:window}.

\subsubsection{Strict monotonic decrease of $\tilde{w}$}
\label{prf::bbr-cubic:dynamic:oscillation:model:condition:update:w-decr}

To confirm that~$\tilde{w}(\alpha)$ is strictly monotonically decreasing
for all~$\alpha$, we distinguish the cases~$\alpha \geq \hat{\alpha}$
and~$\alpha < \hat{\alpha}$.

\paragraph{$\alpha \geq \hat{\alpha}$}
First, we note that~$\tilde{w} = \tilde{w}^{\max}$
is strictly monotonically increasing in~$\tilde{s}$ (\cref{eq:bbr-cubic:system-evolution:cubic:eq:wmax}).
In turn,~$\tilde{s}(\alpha)$ is strictly monotonically decreasing
in~$\alpha$ if all~$\alpha_0$ and~$\alpha_1$ with~$\hat{\alpha} \leq \alpha_0 < \alpha_1$
satisfy:
\begin{equation}
    \tilde{s}(\alpha_0) > \tilde{s}(\alpha_1) \impliedby \forall s > s_0(\tau_k).\ \tilde{S}_1(s, \alpha_0) < \tilde{S}_1(s, \alpha_1),
    \label{eq:bbr-cubic:dynamic:oscillation:model:condition:s-decreasing}
\end{equation} where~$s_0(\tau_k)$ is the lower bound on all roots of polynomial~$\tilde{S}_1$
from~\cref{thm:bbr-cubic:dynamic:oscillation:model:equilibrium},
given delay~$\tau_k$ for the CUBIC flow. This lower bound is found
by setting~$C_{\ell} = 0$, which yields:
\begin{equation}
    \tilde{S}_1(s_0) = \frac{\alpha -1}{\alpha} \left(\frac{c^2}{b\tau_k} s_0^7 - cs_0^3\right) \overset{!}{=} 0
    \iff s_0 = \sqrt[4]{\frac{b\tau_k}{c}}.
    \label{eq:bbr-cubic:dynamic:oscillation:model:condition:min-root}
\end{equation}
Returning to~\cref{eq:bbr-cubic:dynamic:oscillation:model:condition:s-decreasing},
we note that
\begin{equation}
    \begin{split}
        &\tilde{S}_1(s, \alpha_0) < \tilde{S}_1(s, \alpha_1)\\
        \iff &\frac{\alpha_0 -1}{\alpha_0} \left(\frac{c^2}{b\tau_k} s^7 - cs^3\right) < \frac{\alpha_1 -1}{\alpha_1} \left(\frac{c^2}{b\tau_k} s^7 - cs^3\right)
    \end{split}
\end{equation} holds if the $s$-polynomial in the parentheses
is strictly positive, which is again true for all~$s > s_0(\tau_k)$.
This insight implies~$\tilde{s}(\alpha_0) > \tilde{s}(\alpha_1)$
and also $\tilde{w}(\alpha_0) > \tilde{w}(\alpha_1)$, i.e.,
$\tilde{w}(\alpha)$ is strictly monotonically decreasing for all~$\alpha \geq \hat{\alpha}$.

\paragraph{$\alpha < \hat{\alpha}$}
The demonstration of strict negative monotonicity
for~$\alpha < \hat{\alpha}$ works analogously to the
previous case, but is based on~$\tilde{S}_2$
instead of~$\tilde{S}_1$.
Intriguingly, we again find~$s_0(\tau_k)$
as in~\cref{eq:bbr-cubic:dynamic:oscillation:model:condition:min-root},
marking both the lower bound on possible roots
and the lower bound on roots whose position
is a strictly monotonically decreasing function of~$\alpha$.
Hence,~$\tilde{w}$ is a strictly monotonically decreasing
function for all~$\alpha$.

\subsection{CUBIC Long-Term Equilibrium Window Size~$\overline{w}$}
\label{prf::bbr-cubic:dynamic:oscillation:model:condition:equilibrium}

Given the update functions from~\cref{prf::bbr-cubic:dynamic:oscillation:model:condition:update},
we can characterize the equilibrium of the
discrete dynamic process in~\cref{eq:bbr-cubic:dynamic:oscillation:model:conditions:discrete}.
In particular, we will prove the existence and
the uniqueness of an equilibrium~$\overline{w}$
such that~$\overline{w} = w^{\leftarrow}(\overline{w})$.
From~\cref{eq:bbr-cubic:dynamic:oscillation:model:condition:w-arrow},
we learn that~$w^{\leftarrow}(w)$ only returns the argument~$w$ as a value
if~$w = \tilde{w}^{\leftarrow}(w)$.
Hence, the equilibrium condition can only hold if
\begin{equation}
    \overline{w} = \tilde{w}^{\leftarrow}(\overline{w}).
    \label{eq:bbr-cubic:dynamic:oscillation:model:condition:eq-condition:2}
\end{equation}

Since~$\tilde{w}^{\leftarrow}$ is monotonically decreasing (\cref{eq::bbr-cubic:dynamic:oscillation:model:condition:update:w-decr}),
we know that its value range~$\mathcal{R}(\tilde{w}^{\leftarrow})$ 
given argument range $[w^{<}, w^{>}]$ is
\begin{equation}
    \mathcal{R}(\tilde{w}^{\leftarrow}) = [\tilde{w}^{\leftarrow}(w^{>}), \tilde{w}^{\leftarrow}(w^{<})]
    \subseteq [w^{<}, w^{>}]
    \label{eq:bbr-cubic:dynamic:oscillation:model:condition:update-val-range}
\end{equation}
where the subset relation holds for the following reason (symmetric for~$w^{<}$ and~$w_0$):
\begin{align}
        \text{If } w^{>} < w_1: \quad &\tilde{w}^{\leftarrow}(w^{>}) \overset{\text{\cref{eq::bbr-cubic:dynamic:oscillation:model:condition:update:w-decr}}}{\geq} \tilde{w}^{\leftarrow}(w_1) \overset{\text{\cref{eq:bbr-cubic:dynamic:oscillation:model:condition:w_min}}}{=} w^{<}\nonumber\\
        \text{If } w^{>} \geq w_1: \quad &\tilde{w}^{\leftarrow}(w^{>}) \overset{\text{\cref{eq:bbr-cubic:dynamic:oscillation:model:condition:full-update}}}{=} w^{<}
\end{align}

In a next step, we consider the fixed-point function~$w^{\ast}(w) = w - \tilde{w}^{\leftarrow}(w)$.
Since any equilibrium~$\overline{w}$ must satisfy~$\overline{w} = \tilde{w}^{\leftarrow}(\overline{w})$,
it must satisfy~$w^{\ast}(\overline{w}) = 0$. Note that~$w^{\ast}(w)$ is
now \emph{strictly monotonically increasing}.
The value range~$\mathcal{R}(w^{\ast})$ for argument range~$[w^{<}, w^{>}]$ is therefore:
\begin{equation}
    \mathcal{R}(w^{\ast}) = [w^{<} - \tilde{w}^{\leftarrow}(w^{<}), w^{>} - \tilde{w}^{\leftarrow}(w^{>})].
\end{equation}

\cref{eq:bbr-cubic:dynamic:oscillation:model:condition:update-val-range} suggests 
that~$w^{<} \leq \tilde{w}^{\leftarrow}(w^{>}) \leq \tilde{w}^{\leftarrow}(w^{<}) $
and~$w^{>} \geq  \tilde{w}^{\leftarrow}(w^{<}) \geq \tilde{w}^{\leftarrow}(w^{>})$.
As a result, we find that~$w^{<} - \tilde{w}^{\leftarrow}(w^{<}) \leq 0$
and~$w^{>} - \tilde{w}^{\leftarrow}(w^{<}) \geq 0$.
This finding, together with the continuousness and strict monotonicity of~$w^{\ast}$,
imply (by the intermediate-value theorem) that 
a unique~$\overline{w} \in [w^{<}, w^{>}]$ exists
such that~$w^{\ast}(\overline{w}) = 0 \iff \overline{w} = \tilde{w}^{\leftarrow}(\overline{w})$.
This~$\overline{w}$ is thus a unique equilibrium which is guaranteed
to exist.

\subsection{Convergence Trajectories}
\label{prf::bbr-cubic:dynamic:oscillation:model:condition:convergence}

Given the unique equilibrium~$\overline{w}$, we now investigate
the stability properties of this equilibrium.
In other words, we characterize the convergence trajectories determined
by the discrete process in~\cref{eq:bbr-cubic:dynamic:oscillation:model:conditions:discrete},
and elicit a condition under which these trajectories may or may
not lead to the equilibrium.
This stability property depends on the neighborhood
of the equilibrium~$\overline{w}$ with respect to the
update function~$w^{\leftarrow}$.
In particular, we now show that
the equilibrium~$\overline{w}$ is unstable if an 
equilibrium neighborhood~$\Omega$ exists such that
the window-update function~$w$ has a slope of less than $-1$
in the complete neighborhood:
\begin{equation}
    \exists \Omega = (\omega_0, \omega_1),\ \omega_0 < \overline{w} < \omega_1.
    \quad \forall \omega \in \Omega.\ \frac{\partial{w^{\leftarrow}}}{\partial w} < -1
    \label{eq:bbr-cubic:dynamic:oscillation:model:condition:neighborhood}
\end{equation}
In other words, we will show that all evolution trajectories from a window size~$\omega \in \Omega$
in this neighborhood lead out of the neighborhood. In that case,
even if the dynamics outside the neighborhood converge into the neighborhood,
the trajectory then leaves the neighborhood again and the equilibrium is not converged upon.

In particular, we consider a state~$w(t)$
with~$w(t) \in \Omega$ and $w(t) < \overline{w}$ (The proof for~$w(t) > \overline{w}$ is symmetric).
Since both~$w(t)$ and~$\overline{w}$ are in~$\Omega$,~$w^{\leftarrow}$ is strictly
monotonically decreasing between~$w(t)$ and~$\overline{w}$, and therefore 
\begin{equation}
    w(t) < \overline{w} \implies w(t+1) = w^{\leftarrow}(w(t)) > w^{\leftarrow}(\overline{w}) \overset{\text{(\ref{eq:bbr-cubic:dynamic:oscillation:model:condition:eq-condition:2})}}{=} \overline{w}.
    \label{eq:bbr-cubic:dynamic:oscillation:model:condition:first-step}
\end{equation}

At this point,~$w(t+1)$ might already be outside of~$\Omega$ if~$w(t+1) > \omega_1$.
Otherwise, if~$w(t+1) \in \Omega$, we know by the same argument 
that~$w(t+2) = w^{\leftarrow}(w(t+1)) < \overline{w}$.
Again,~$w(t+2)$ might be out of the neighborhood~$\Omega$ if~$w(t+2) < \omega_0$.

Importantly,~$w$ will \emph{eventually} leave the neighborhood if for all any~$t$,
$w(t+2) < w(t)$, i.e., the window state moves away
from the equilibrium~$\overline{w}$ and will eventually fall below~$\omega_0$. 
This condition can be reformulated:
\begin{align}
    &w(t+2) < w(t) \overset{\text{Expand}}{\iff} w^{\leftarrow}(w^{\leftarrow}(w(t))) < w(t)\nonumber
    \overset{-w^{\leftarrow}(w(t))}{\iff}\\  &w^{\leftarrow}(w^{\leftarrow}(w(t))) - w^{\leftarrow}(w(t)) < w(t) - w^{\leftarrow}(w(t))\nonumber\\
    &\overset{/\text{RHS}}{\underset{\text{RHS} \overset{\text{(\ref{eq:bbr-cubic:dynamic:oscillation:model:condition:first-step})}}{<} 0}{\iff}} \frac{w^{\leftarrow}(w^{\leftarrow}(w(t))) - w^{\leftarrow}(w(t))}{w(t) - w^{\leftarrow}(w(t))} > 1\nonumber\\
    &\overset{\cdot -1}{\iff} \frac{w^{\leftarrow}(w^{\leftarrow}(w(t))) - w^{\leftarrow}(w(t))}{w^{\leftarrow}(w(t)) - w(t)} < -1.\label{eq:bbr-cubic:dynamic:oscillation:model:condition:average-slope}
\end{align}
The last inequality in \cref{eq:bbr-cubic:dynamic:oscillation:model:condition:average-slope} can be understood
as a condition on the average slope of~$w^{\leftarrow}$ in~$[w(t), w^{\leftarrow}(w(t))]$.
Since we are considering the case where both~$w(t)$ and~$w^{\leftarrow}(w(t))$ are in the neighborhood~$\Omega$,
the derivative condition in \cref{eq:bbr-cubic:dynamic:oscillation:model:condition:neighborhood}
implies this average-slope condition.

\color{black}

\section{Proof of Theorem~\ref{thm:bbr-cubic:dynamic:oscillation:model:maximal}: Maximally intensive oscillation}
\label{prf:bbr-cubic:dynamic:oscillation:model:maximal}

\subsection{Maximum Oscillation Amplitude}
\label{prf:bbr-cubic:dynamic:oscillation:model:maximal:assumptions}

In order to find plausible bounds of the flow-size
distribution in BBR-CUBIC competition, 
we consider the worst case in terms of oscillation amplitude.
To characterize this worst-case oscillation amplitude,
we make two observations.

\paragraph{Convergence speed} We note that the oscillation amplitude is proportional
to the convergence speed of the short-term dynamics in each update interval:
In interval~$[t, t+1]$, these short-term dynamics involve 
a CUBIC window-size change from the initial CUBIC window size~$w(t)$
to window size~$w(t+1) = w^{\leftarrow}(w(t))$,
where~$w^{\leftarrow}(w(t))$ is between the initial window size~$w(t)$
and the short-term equilibrium window size~$\tilde{w}^{\leftarrow}(w(t))$.
Hence, a high similarity between~$w^{\leftarrow}$ and~$\tilde{w}^{\leftarrow}$
indicates a high convergence speed, quick window-size changes, and
thus a large amplitude of the oscillation. In the following proof,
we therefore assume~$w^{\leftarrow} = \tilde{w}^{\leftarrow}$
to maximize the oscillation amplitude.

\paragraph{Neighborhood size}
We note that the oscillation amplitude is proportional
to the size of the unstable neighborhood~$\Omega = [\omega_0, \omega_1]$ 
of the long-term equilibrium window size~$\overline{w}$. 
Since the process~$\{w(t)\}_{t\in \mathbb{N}, t\geq 0}$
evolves away from this equilibrium~$\overline{w}$ when in neighborhood~$\Omega$,
the oscillation mostly involves window sizes~$w \neq \Omega$
left and right of the neighborhood. Moreover, many steps of the oscillation
also need to cross~$\Omega$, e.g., change the window size from ~$w(t) < \omega_0$
to~$w(t+1) > \omega_1$. Hence, $\omega_1 - \omega_0$ is proportional to
the window-size changes within the update periods,
and thus to the oscillation amplitude.
Since the update function~$w^{\leftarrow}$ (in our case: $\tilde{w}^{\leftarrow}$)
must be strictly decreasing in~$\Omega$, the maximally large 
neighborhood~$\Omega  = (\omega_0, \omega_1)$
corresponds to~$(w_0, w_1)$, i.e., the decreasing part of~$\tilde{w}^{\leftarrow}$ 
(cf.~\cref{eq:bbr-cubic:dynamic:oscillation:model:condition:full-update}). 

\subsection{Existence of Limit Cycle}
\label{prf:bbr-cubic:dynamic:oscillation:model:maximal:stability}

With the two assumptions from~\cref{prf:bbr-cubic:dynamic:oscillation:model:maximal:assumptions},
we now prove that the BBR-CUBIC oscillation has a limit cycle
of CUBIC window sizes~$\tilde{w}^{\leftarrow}(w^{<})$ and~$\tilde{w}^{\leftarrow}(w^{>})$,
i.e., the CUBIC window size persistently alternates between these two values.

Clearly, such a limit cycle only exists given an unstable long-term equilibrium~$\overline{w}$.
As noted in~\cref{prf::bbr-cubic:dynamic:oscillation:model:condition:convergence},
an equilibrium~$\overline{w}$ is only unstable if~$\overline{w} \in (w_0, w_1)$ with~$w_0 < w_1$,
i.e., $\overline{w} = \tilde{w}^{\leftarrow}(\overline{w})$ is a value of~$\tilde{w}^{\leftarrow}$ 
in the decreasing part of~$\tilde{w}^{\leftarrow}$.
Since the values of~$\tilde{w}^{\leftarrow}$ are restricted to~$[w^{<}, w^{>}]$ 
(cf.~\cref{eq:bbr-cubic:dynamic:oscillation:model:condition:w_max,eq:bbr-cubic:dynamic:oscillation:model:condition:w_min}),
it holds that~$\overline{w} \in ([w^{<}, w^{>}] \cap (w_0, w_1)) =: W_{\cap}$.
Since~$\overline{w}$ is guaranteed to exist, we know that~$W_{\cap} \neq \varnothing$, and
thus
\begin{equation}
    w_0 < w^{>} \quad \text{ and } \quad w_1 > w^{<}.
    \label{prf:bbr-cubic:dynamic:oscillation:model:maximal:stability:intersection}
\end{equation}

In the following, we consider all possible cases for~$W_{\cap}$,
and verify whether the limit cycle is sound, i.e.,:
\begin{equation}
    \begin{split}
        \tilde{w}^{\leftarrow}(w^{<}) &= \tilde{w}^{\leftarrow}(\tilde{w}^{\leftarrow}(w^{>}))\\
        \tilde{w}^{\leftarrow}(w^{>}) &= \tilde{w}^{\leftarrow}(\tilde{w}^{\leftarrow}(w^{<}))
    \end{split}
    \label{eq:bbr-cubic:dynamic:oscillation:model:maximal:stability:soundness}
\end{equation}

\subsubsection{$W_{\cap} = (w_0, w_1)$}
Given the intersection range, we know that~$w^{<} \leq w_0 < w_1 \leq w^{>}$.
This interleaving implies that
\begin{equation}
    \tilde{w}^{\leftarrow}(w^{<}) \overset{\text{(\ref{eq:bbr-cubic:dynamic:oscillation:model:condition:full-update})}}{=} w^{>} \quad 
\text{and} \quad \tilde{w}^{\leftarrow}(w^{>}) \overset{\text{(\ref{eq:bbr-cubic:dynamic:oscillation:model:condition:full-update})}}{=} w^{<}.
\label{eq:bbr-cubic:dynamic:oscillation:model:maximal:stability:case-1}
\end{equation}
Clearly, the limit cycle is sound in this case:
\begin{equation}
    \begin{split}
        \tilde{w}^{\leftarrow}(w^{<}) &\overset{\text{(\ref{eq:bbr-cubic:dynamic:oscillation:model:maximal:stability:case-1})}}{=} \tilde{w}^{\leftarrow}(\tilde{w}^{\leftarrow}(w^{>}))\\
        \tilde{w}^{\leftarrow}(w^{>}) &\overset{\text{(\ref{eq:bbr-cubic:dynamic:oscillation:model:maximal:stability:case-1})}}{=} \tilde{w}^{\leftarrow}(\tilde{w}^{\leftarrow}(w^{<}))\\
    \end{split}
\end{equation}

\subsubsection{$W_{\cap} = [w^{<}, w_1)$}
Given the intersection range, we know that~$w_0 < w^{<} < w_1 \leq w^{>}$.
Hence, we know that
\begin{equation}
    \tilde{w}^{\leftarrow}(w^{<}) \overset{\text{(\ref{eq:bbr-cubic:dynamic:oscillation:model:condition:full-update})}}{<} w^{>} \quad 
\text{and} \quad \tilde{w}^{\leftarrow}(w^{>}) \overset{\text{(\ref{eq:bbr-cubic:dynamic:oscillation:model:condition:full-update})}}{=} w^{<}.
\label{eq:bbr-cubic:dynamic:oscillation:model:maximal:stability:case-2}
\end{equation}
With this knowledge, the first soundness condition in~\cref{eq:bbr-cubic:dynamic:oscillation:model:maximal:stability:soundness} holds:
\begin{align}
    \tilde{w}^{\leftarrow}(w^{<}) &\overset{\text{(\ref{eq:bbr-cubic:dynamic:oscillation:model:maximal:stability:case-2})}}{=} \tilde{w}^{\leftarrow}(\tilde{w}^{\leftarrow}(w^{>}))
\end{align}
The second condition, however, is not directly satisfied,
but can be converted to:
\begin{align}
        &\tilde{w}^{\leftarrow}(w^{>}) = \tilde{w}^{\leftarrow}(\tilde{w}^{\leftarrow}(w^{<}))\overset{\text{(\ref{eq:bbr-cubic:dynamic:oscillation:model:maximal:stability:case-2})}}{\iff}\\ 
        &w^{<} = \tilde{w}^{\leftarrow}(\tilde{w}^{\leftarrow}(w^{<})) \overset{\text{(\ref{eq:bbr-cubic:dynamic:oscillation:model:condition:full-update})}}{\iff}\\
        &\tilde{w}^{\leftarrow}(w^{<}) \geq w_1
        \overset{\text{(\ref{eq:bbr-cubic:dynamic:oscillation:model:condition:w_min})}}{\iff} \tilde{w}^{\leftarrow}(\tilde{w}^{\leftarrow}(w_1)) \geq w_1\\
        &\overset{-\tilde{w}^{\leftarrow}(w_1)}{\iff}
        \tilde{w}^{\leftarrow}(\tilde{w}^{\leftarrow}(w_1)) - \tilde{w}^{\leftarrow}(w_1) \geq w_1 - \tilde{w}^{\leftarrow}(w_1)\nonumber\\
        &\overset{\text{(\ref{eq:bbr-cubic:dynamic:oscillation:model:condition:full-update})}}{=} w_1 - w^{<} \overset{\text{(\ref{prf:bbr-cubic:dynamic:oscillation:model:maximal:stability:intersection})}}{>} 0 \label{prf:bbr-cubic:dynamic:oscillation:model:maximal:stability:w_min_w_1}  \\
        &\overset{/\text{RHS}}{\underset{\cdot -1}{\iff}} \frac{\tilde{w}^{\leftarrow}(\tilde{w}^{\leftarrow}(w_1)) - \tilde{w}^{\leftarrow}(w_1)}{\tilde{w}^{\leftarrow}(w_1) - w_1} \leq -1
\end{align}
In turn, the last condition on~$w_1$ is ensured by the derivative condition in 
neighborhood $\Omega = [w_0, w_1]$, i.e., by~\cref{eq:bbr-cubic:dynamic:oscillation:model:condition:neighborhood}.

\subsubsection{$W_{\cap} = (w_0, w^{>}]$}
This case is symmetric to the case for~$W_{\cap} = [w^{<}, w_1)$
since~$w^{<} \leq w_0 < w^{>} < w_1$.

\subsubsection{$W_{\cap} = [w^{<}, w^{>}]$}
From the intersection range, we derive~$w^{<} > w_0$
and~$w^{>} < w_1$. These conditions can be shown
to be incompatible with an unstable equilibrium (again
via the derivative condition):
\begin{equation}
    w^{>} - w^{<} < w_1 - w_0 \overset{/\text{RHS}}{\underset{\cdot-1}{\iff}} \frac{\tilde{w}^{\leftarrow}(w_0) - \tilde{w}^{\leftarrow}(w_1)}{w_0 - w_1} > -1.
\end{equation}
Hence, this case is outside the scope of this proof.

\subsection{Convergence to Limit Cycle}
\label{prf:bbr-cubic:dynamic:oscillation:model:maximal:convergence}

In the following, we show that the long-term dynamics of the CUBIC
window size~$w$ converge to the limit cycle constituted by
$\tilde{w}^{\leftarrow}(w^{<})$
and~$\tilde{w}^{\leftarrow}(w^{>})$.
For this purpose, we revisit all relevant cases from~\cref{prf:bbr-cubic:dynamic:oscillation:model:maximal:stability}.

\subsubsection{$W_{\cap} = (w_0, w_1)$}
Without loss of generality, we assume an initial state~$w(t) < \overline{w}$,
 which allows the following case distinction:
 \begin{itemize}
     \item $\boldsymbol{w(t) \leq w_0}$: We know that~$w(t+1) = \tilde{w}^{\leftarrow}(w(t)) \overset{\text{(\ref{eq:bbr-cubic:dynamic:oscillation:model:condition:full-update})}}{=} w^{>}$.
    Hence, the limit cycle is entered in the first step, as~$w^{>}$ is
part of the limit cycle by~\cref{eq:bbr-cubic:dynamic:oscillation:model:maximal:stability:case-1}.
    \item $\boldsymbol{w(t) \in (w_0, \overline{w}):}$ We note that~$\tilde{w}^{\leftarrow}$
    is strictly monotonically decreasing between~$w(t) > w_0$ and~$\overline{w} < w_1$ 
    by~\cref{eq:bbr-cubic:dynamic:oscillation:model:condition:full-update},
    and thus
    \begin{equation}
        w(t+1) = \tilde{w}^{\leftarrow}(w(t)) > \tilde{w}^{\leftarrow}(\overline{w}) \overset{\text{(\ref{eq:bbr-cubic:dynamic:oscillation:model:condition:eq-condition:2})}}{=} \overline{w}.
        \label{eq:bbr-cubic:dynamic:oscillation:model:maximal:stability:eq-comparison}
    \end{equation}
    \begin{itemize}
        \item $\boldsymbol{w(t+1) \geq w_1}$: The limit cycle is entered at~$w(t+2) = \tilde{w}^{\leftarrow}(w(t+1)) = w^{<}$, as~$w^{<}$ is
part of the limit cycle by~\cref{eq:bbr-cubic:dynamic:oscillation:model:maximal:stability:case-1}.
        \item $\boldsymbol{w(t+1) \in (\overline{w}, w_1)}$: 
        We again note that~$\tilde{w}^{\leftarrow}$ is strictly monotonically
        decreasing between~$\overline{w} > w_0$ and $w(t+1) < w_1$, and
        thus~$w(t+2) = \tilde{w}^{\leftarrow}(w(t+1)) < \overline{w}$. 
        Moreover, the window size makes progress towards
the limit cycle if~$w(t+2) < w(t)$, which can once more be guaranteed
via the derivative condition.
    \end{itemize}
 \end{itemize}

\subsubsection{$W_{\cap} = [w^{<}, w_1)$}
\label{prf:bbr-cubic:dynamic:oscillation:model:maximal:convergence:case-2}
In this case, the limit cycle~$L$ is composed of~$w^{<}$
and~$\tilde{w}^{\leftarrow}(w^{<}) < w^{>}$ (\cref{eq:bbr-cubic:dynamic:oscillation:model:maximal:stability:case-2}).
Without loss of generality, we assume an initial state~$w(t) < \overline{w}$,
which allows the following case distinction:

\begin{itemize}
    \item $\boldsymbol{w(t) \leq w_0}$: In that case, $w(t+1) = \tilde{w}^{\leftarrow}(w(t)) = w^{>}$
    by~\cref{eq:bbr-cubic:dynamic:oscillation:model:condition:full-update}. 
    Since~$w^{>} \geq w_1$ in the current case,
    the limit cycle is entered at~$w(t+2) = \tilde{w}^{\leftarrow}(w^{>}) = w^{<}$.
    \item $\boldsymbol{w(t) \in (w_0, w^{<})}$: We note that $\tilde{w}^{\leftarrow}$ 
    is strictly monotonically decreasing between~$w(t) > w_0$ 
    and~$w^{<} \leq w_1$ (\cref{eq:bbr-cubic:dynamic:oscillation:model:condition:full-update},
    where~$w(t) < w^{<}$. 
    This property implies that
    \begin{equation}
        \quad w(t+1) = \tilde{w}^{\leftarrow}(w(t)) > \tilde{w}^{\leftarrow}(w^{<})
        \overset{\text{(\ref{prf:bbr-cubic:dynamic:oscillation:model:maximal:stability:w_min_w_1})}} \geq
        w_1.
    \end{equation}
    Since~$w(t+1) \geq w_1$, the limit cycle is then entered at~$w(t+2) = \tilde{w}^{\leftarrow}(w(t+1)) = w^{<}$.
    \item $\boldsymbol{w(t) \in [w^{<}, \overline{w})}$:
    This case only arises if~$w^{<} <\overline{w}$.
    Hence, $\tilde{w}^{\leftarrow}$ is again strictly monotonically decreasing
    between~$w^{<} > w_0$ and~$\overline{w} < w_1$, and therefore~$w(t+1) > \overline{w}$
    from~\cref{eq:bbr-cubic:dynamic:oscillation:model:maximal:stability:eq-comparison} holds again.
    \begin{itemize}
        \item $\boldsymbol{w(t+1) \geq w_1}$: The limit cycle is entered at~$w(t+2) = \tilde{w}^{\leftarrow}(w(t+1)) = w^{<}$ by~\cref{eq:bbr-cubic:dynamic:oscillation:model:condition:full-update}.
        \item $\boldsymbol{w(t+1) \in (\overline{w}, w_1)}$: Based on arguments symmetric to the above
        analysis, plus the derivative condition from~\cref{eq:bbr-cubic:dynamic:oscillation:model:condition:neighborhood},
        we find that~$w(t+2) \in [w^{<}, w(t))$, which implies that the limit cycle is eventually entered.
    \end{itemize}
\end{itemize}

\subsubsection{$W_{\cap} = (w_0, w^{>}]$}
This case is symmetric to the case 
in~\cref{prf:bbr-cubic:dynamic:oscillation:model:maximal:convergence:case-2} (directly above).

\section{Derivation of Non-Pessimal Fairness Bounds}
\label{sec:non-pessimal}

\begin{figure*}
    \centering
    \input{figures/bbr_vs_cubic__remedies_annotated}
    \caption{Evaluation of two strategies for oscillation suppression.
        `Random Probing' desynchronizes RTT-probing steps, and
        `Oscillation Detection' freezes min-RTT estimate
        at high estimate variance.}
        \label{fig:bbr-cubic:dynamic:oscillation:remedies:stabilizing}
\end{figure*}
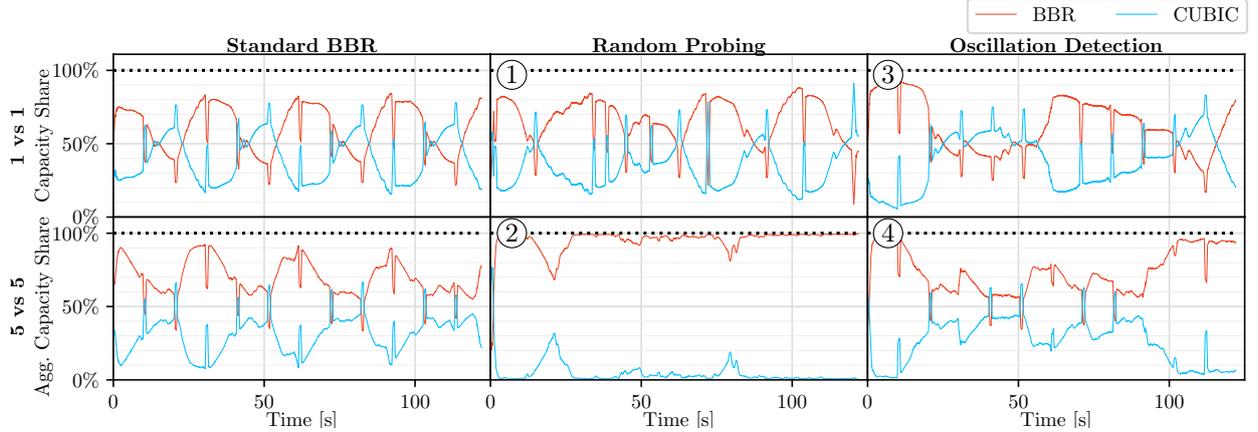

Fundamentally, we note that the evolution of the
CUBIC congestion-window size~$w$ 
is controlled by the CUBIC window-growth
function~$W(w^{\max}, s)$ 
from~\cref{eq:bbr-cubic:model:basic-model:cubic-window-growth}.
% :
% \begin{equation}
%     W(w^{\max}, s) = w^{\max} + c \cdot \left(s - \sqrt[3]{\frac{b}{c} w^{\max}}\right)^3
% \end{equation}
% As this window-growth function forms the basis of
% our non-pessimal bounds, we refer to these bounds as
% \emph{non-pessimal}.

The first argument of~$W$, i.e., the recorded maximum window~$w^{\max}$, converges sub-exponentially, i.e., slowly, to its current short-term
equilibrium value~$\tilde{w}^{\max}$, 
as both the proof 
of~\cref{thm:bbr-cubic:dynamic:oscillation:model:convergence}
and the visualization in~\cref{fig:bbr-cubic:dynamic:oscillation:model:center-manifold}
demonstrate.
The short-term equilibrium values~$\tilde{w}^{\max}$ themselves
alternate between values above and below the long-term
equilibrium~$\overline{w}^{\max} = \overline{w}$,
which is implied by the structure of the 
function~$\tilde{w}^{\leftarrow}$ (\cref{fig:bbr-cubic:dynamic:oscillation:model:oscillation})
and the fact~$\tilde{w}^{\max} = \tilde{w}$ for
any short-term equilibrium (\cref{eq:bbr-cubic:system-evolution:cubic:eq:wmax}). 
Given slow convergence in alternating directions, it is 
plausible that~$w^{\max}$ is usually close to the 
intermediate value~$\overline{w}^{\max} = \overline{w}$, 
i.e., the long-term equilibrium. 
Hence, we assume~$w^{\max} = \overline{w}$ for the
non-pessimal bound.

Given this fixed first argument of~$W$, we derive
the non-pessimal bounds from the variance in the
second argument, the window-growth duration~$s$.
Clearly, the minimum possible value of~$s$ is 0,
leading to the maximum non-pessimal bound~$\phi^{\mathrm{B}}(\hat{W}_0)$
of the BBR capacity share, where
\begin{equation}
    \hat{W}_0 = W(\overline{w}, 0) =  \overline{w} + c \cdot \left(- \sqrt[3]{\frac{b}{c} \overline{w}}\right)^3 = (1-b) \overline{w}.
\end{equation}

Conversely, the window-growth duration $s$ 
(and thus also the CUBIC window size~$w$) 
attains its highest value if
it grows without interruption 
for the entire 10 seconds 
between RTT-probing steps.
In fact, $s$ rarely grows beyond a total of 10 seconds
in realistic settings, as $s$ causes rapid
growth in the CUBIC congestion-window size~$w$,
which quickly leads to loss and thus resets~$s$.
More formally, the maximum possible short-term 
equilibrium value~$\tilde{s}(\alpha_{\min}) = \tilde{s}(2\tau^{\mathrm{p}}/\overline{\tau})$
is below 10 seconds for all settings in~\cref{fig:bbr-cubic:dynamic:oscillation:model:prediction:experiments}.
While~$s$ can temporarily 
exceed this equilibrium value during its evolution 
(\cref{fig:bbr-cubic:dynamic:oscillation:model:center-manifold}), 
the proof of~\cref{thm:bbr-cubic:dynamic:oscillation:model:convergence}
confirms that~$s$ converges exponentially fast to its short-term 
equilibrium value~$\tilde{s}$, suggesting that~$s$
never grows much beyond~$\tilde{s}$.
In summary, assuming a maximum~$s$ of 10 seconds
leads to the minimum non-pessimal bound~$\phi^{\mathrm{B}}(\hat{W}_1)$
for the BBR capacity share, where
\begin{equation}
    \hat{W}_1 = W(\overline{w}, 10) = \overline{w} + c \cdot \left(10 - \sqrt[3]{\frac{b}{c} \overline{w}}\right)^3.
\end{equation}

\section{Evaluation of Further BBR Modifications}
\label{sec:bbr-modification:evaluation}

As explained in~\cref{sec:bbr-cubic:dynamic},
BBR/CUBIC oscillation happens because the BBR flows simultaneously 
perform RTT probing by briefly, but sharply reducing their rate. 
Therefore, the BBR flows regularly estimate a relatively low minimum RTT, 
especially if the CUBIC flows are small at the time of probing.
This low minimum-RTT estimate then decreases the BBR 
sending rate (via the BBR congestion window), increases
the CUBIC sending rate, and thus leads to a higher
minimum-RTT estimate at the next RTT probing.
If that next minimum-RTT estimate is high enough,
the evolution of the sending rates is reversed,
causing oscillation.

This causal chain may be disrupted by
a number of possible modifications to the BBR algorithm,
which we discuss and experimentally evaluate in the following.

\paragraph{Randomize RTT probing} As described above, 
oscillation requires that all BBR flows simultaneously
perform RTT probing, namely every 10 seconds.
If these RTT-probing periods were randomized,
each flow would probe the RTT when other BBR flows
still contribute to the queue. Hence, the high minimum-RTT estimates 
would be consistently high rather than varying over time. 
While such randomization might avoid
oscillation, it is undesirable for three reasons.

First, the synchronization of the RTT probing among BBR flows
is a conscious feature of BBR, enabling BBR flows to 
discover the path propagation delay in a pure BBR scenario;
this discovery would be prevented by randomization.
Second, randomization does not prevent oscillation
given a single BBR flow, as visible in~\cref{fig:bbr-cubic:dynamic:oscillation:remedies:stabilizing} \circled{1}.
Given a single BBR flow, no other BBR flows exist that could inflate
the minimum-RTT estimate. Hence, the rate of the randomized single BBR flow
still oscillates, although not in 10-second steps anymore,
but in intervals with varying duration.
Third, randomization can suppress oscillation only at the cost
of even lower fairness, as the experiment in~\cref{fig:bbr-cubic:dynamic:oscillation:remedies:stabilizing} \circled{2}
demonstrates: Since randomization causes consistently high minimum-RTT estimates,
it also causes consistently high BBR rates and near-starvation of the CUBIC flows.

\paragraph{Detect oscillation and freeze} To suppress oscillation
in a more targeted fashion,
we envision that a BBR flow (i)~keeps a recent history of
its minimum-RTT estimates, (ii)~maintains the mean~$\mu$
and the standard deviation~$\sigma$ of these estimates, 
and (iii)~concludes that oscillation is ongoing
if the standard deviation~$\sigma$ exceeds a configured share~$\kappa$ 
of the mean~$\mu$. In case of oscillation, the BBR flow then
considers~$\mu$ its minimum-RTT estimate.

Unfortunately, this oscillation-suppression strategy is self-defeating:
When oscillation is suppressed for a sufficiently long time,
the variance of minimum-RTT estimates decreases,
the oscillation-suppression mechanism is deactivated,
and oscillation resumes.
Hence, this mechanism does not eliminate the oscillation,
but only prolongs the oscillation period
(see experiments~\circled{3} and~\circled{4} 
in~\cref{fig:bbr-cubic:dynamic:oscillation:remedies:stabilizing}).
Moreover, the mechanism relies on a suitable value~$\kappa$ 
to distinguish structural oscillation from acceptable fluctuation of
the minimum-RTT estimates, which might be difficult to find in practice.

\end{appendix}

\end{document}

%% file: setup.tex
\usepackage{hyperref}

\usepackage{booktabs,tabularx}

\usepackage{amsmath}

\usepackage[mode=text, math-rm=\text, detect-all, binary-units=true, per-mode=symbol, range-phrase=\,--\,, range-units=single, detect-weight=true, detect-family=true]{siunitx}
\DeclareSIUnit{\bit}{bit}
\DeclareSIUnit{\bps}{\bit\per\s}

\usepackage{tikz}
\usepackage{pgfplots}
\pgfplotsset{compat=1.17}
\usetikzlibrary{calc,arrows.meta,shapes.geometric,shapes.misc}
\usepackage{subcaption}
\usepackage{dblfloatfix} 

\usepackage{csquotes}

\graphicspath{{figures}}

\usepackage[nice]{nicefrac} % Display ½ nicely
\usepackage{xcolor}         % Colors!
\usepackage{xifthen}
\usepackage{makecell}
\usepackage{mathtools}
\usepackage{multicol}
\usepackage{multirow}
\usepackage{placeins}
\usepackage{paralist}

\usepackage{amsmath} % not needed for acmart
\usepackage{amssymb} % not needed for acmart
\usepackage{amsthm} % not needed for acmart

\usepackage[capitalize]{cleveref}

\crefname{section}{\S}{Sections}
\Crefname{section}{Section}{Sections}
\crefformat{section}{\S#2#1#3}
\Crefformat{section}{Section~#2#1#3}
\crefformat{footnote}{#2\footnotemark[#1]#3}

\usepackage{enumitem}       % Allows customizing enumerate counters
\newcommand{\enumfont}{\bfseries\sffamily}
\SetEnumitemKey{myenum}{
}
\newlist{benenum}{enumerate}{1}
\setlist[benenum]{myenum, label={\enumfont B\arabic*}, ref=B\arabic*}
\crefname{benenumi}{benefit}{benefits}

\usepackage[normalem]{ulem}
\newcommand{\gennote}[5][blue]{{\color{#1}%
		$\rule{8pt}{8pt}_\textsf{\bfseries #2}^\textsf{\bfseries #3}$
		\textcolor{gray}{\emph{\sout{#4}}}#5}%
}

\newcommand{\authorcomment}[3]{%
	\expandafter\newcommand\csname#1\endcsname[1]{\gennote[#3]{#2}{}{}{##1}}%
	\expandafter\newcommand\csname#1S\endcsname[2][]{\gennote[#3]{#2}{}{##1}{##2}}%
	\expandafter\newcommand\csname#1Q\endcsname[1]{\gennote[#3]{#2}{Question}{}{##1}}%
	\expandafter\newcommand\csname#1N\endcsname[1]{\gennote[#3]{#2}{Note}{}{##1}}%
	\expandafter\newcommand\csname#1C\endcsname[1]{\gennote[#3]{#2}{Comment}{}{##1}}%
	\expandafter\newcommand\csname#1T\endcsname[1]{\gennote[#3]{#2}{TODO}{}{##1}}%
}
\authorcomment{stefan}{StS}{blue}
\authorcomment{simon}{SiS}{orange}
\authorcomment{adrian}{AP}{red}
\authorcomment{ml}{ML}{purple}

\newenvironment{proofsketch}{%
  \proof}{\endproof}

\newcommand*\circled[1]{\tikz[baseline=(char.base)]{\node[shape=circle,draw,inner sep=1pt,fill=white] (char) {#1};}}

\definecolor{bittersweet}{rgb}{1.0, 0.44, 0.37}
\definecolor{lightgray}{rgb}{0.96, 0.96, 0.96}
\definecolor{internationalkleinblue}{rgb}{0.0, 0.18, 0.65}
\definecolor{mediumslateblue}{rgb}{0.48, 0.41, 0.93}
\definecolor{lava}{rgb}{0.81, 0.06, 0.13}

\crefname{appendix}{}{}

%% file: figures/bbr_explainer.tex
\def\coordwidth{6}
\def\coordheight{2.7}
\newcommand{\coord}[2]{
    ($(anchor_point) + #1*(\coordwidth, 0) + #2*(0, \coordheight)$)
}
\begin{tikzpicture}

    \node (anchor_point) at (0, 0) {};

    % Modes
    \fill[fill=lightgray] \coord{0}{0} -- \coord{0.8}{0} -- \coord{0.8}{1.0} -- \coord{0}{1.0};
    \fill[fill=lightgray] \coord{0.9}{0} -- \coord{0.95}{0} -- \coord{0.95}{1.0} -- \coord{0.9}{1.0};
    \node[align=center,anchor=north] at \coord{0.4}{1.02} {\scriptsize Bandwidth\\[-2mm]\scriptsize probing};
    
    % Coordinate system
    \draw[-latex] \coord{0}{-0.028} -- \coord{0}{1};
    \draw[-latex] \coord{-0.037}{0} -- \coord{1}{0};
    \node at \coord{0.5}{-0.05} {\scriptsize Time};
    \node[rotate=90,anchor=south] at \coord{0}{0.5} {\scriptsize BBR Flow Rate};
    
    % Phases
    \def\phasewidth{0.05}
    \foreach \i in {1, 2, 3, 4, 5, 6, 7, 9, 10, 11, 12, 13, 14, 15, 19} {
        \node (ph\i) at ($(anchor_point) + \i*\phasewidth*(\coordwidth, 0)$) {};
        \draw[dotted] (ph\i.center) -- ($(ph\i.center) + 0.8*(0, \coordheight)$);
    }
    \foreach \i in {8, 16, 18} {
        \node (ph\i) at ($(anchor_point) + \i*\phasewidth*(\coordwidth, 0)$) {};
        \draw[dotted,line width=0.3mm] (ph\i.center) -- ($(ph\i.center) + 0.8*(0, \coordheight)$);
    }
    \node[anchor=south,rotate=90,inner sep=0.3mm] (rttprobing) at \coord{0.89}{0.35} {\scriptsize RTT probing};
    \draw[{latex[length=1.6mm,width=1mm]}-{latex[length=1.6mm,width=1mm]},draw=mediumslateblue] \coord{0.8}{0.77} -- \coord{0.9}{0.77};
    \node[inner sep=0.2mm,anchor=north,fill=white,text=mediumslateblue] at \coord{0.85}{0.75} {\scriptsize 200ms}; 

    % BBR Rate
    \def\btllevel{0.4}
    \draw[internationalkleinblue,line width=0.3mm] 
                \coord{0}{\btllevel} -- 
                \coord{0.4}{\btllevel} --
                \coord{0.4}{1.25*\btllevel} --
                \coord{0.8}{1.25*\btllevel} --
                \coord{0.8}{1.25*1.25*\btllevel} --
                \coord{0.95}{1.25*1.25*\btllevel};
    \draw[bittersweet,line width=0.2mm] 
                \coord{0}{\btllevel} -- 
                \coord{0.05}{\btllevel} -- 
                \coord{0.05}{1.25*\btllevel} -- 
                \coord{0.1}{1.25*\btllevel} -- 
                \coord{0.1}{0.75*\btllevel} --
                \coord{0.15}{0.75*\btllevel} --
                \coord{0.15}{\btllevel} -- 
                \coord{0.4}{\btllevel} --
                \coord{0.4}{1.25*\btllevel} --
                \coord{0.45}{1.25*\btllevel} --
                \coord{0.45}{1.25*1.25*\btllevel} --
                \coord{0.5}{1.25*1.25*\btllevel} --
                \coord{0.5}{0.75*1.25*\btllevel} --
                \coord{0.55}{0.75*1.25*\btllevel} --
                \coord{0.55}{1.25*\btllevel} --
                \coord{0.8}{1.25*\btllevel} --
                \coord{0.8}{0.05} -- 
                \coord{0.9}{0.05} --
                \coord{0.9}{1.25*1.25*\btllevel} --
                \coord{0.95}{1.25*1.25*\btllevel};

    % MinRTT
    \draw[{<[length=0.5mm]}-{>[length=0.5mm]},draw=mediumslateblue] \coord{0.15}{0.1*\btllevel} -- \coord{0.2}{0.1*\btllevel};
    \node[fill=white,anchor=south,fill=lightgray,inner sep=0.1mm,text=mediumslateblue] (tmin) at \coord{0.175}{0.2*\btllevel} {\scriptsize $\tau^{\min}$};
    \draw[-{latex[length=1.6mm,width=1mm]},densely dashed] (rttprobing.west) to[out=210,in=0] (tmin.east);
    \node[fill=lightgray,inner sep=0.1mm] at \coord{0.6}{0.15} {\scriptsize update};

    % BtlBw
    \node[fill=lightgray,text=internationalkleinblue,inner sep=0.3mm] (xbtl) at \coord{0.65}{0.68} {\scriptsize $x^{\mathrm{btl}}$};
    \node (xbtlline) at \coord{0.8}{0.55} {};
    \draw[-{latex[length=1.6mm,width=1mm]},draw=internationalkleinblue] ($(xbtl.south) - (0.1, 0.05)$) to[out=270,in=180] (xbtlline.center);

    % Pacing rate
    \node[fill=lightgray,text=bittersweet,inner sep=0.3mm,align=center] (pacing) at \coord{0.2}{0.6} {\scriptsize Sending\\[-2mm]\scriptsize rate};
    \node (pacingline) at \coord{0.1}{0.45} {};
    \draw[-{latex[length=1.6mm,width=1mm]},draw=bittersweet] ($(pacing.south) - (0, 0.05)$) to[out=270,in=0] (pacingline.center);

    % Phase
    \draw [decorate,decoration={brace,amplitude=0.75mm}] ($(ph14.center) +  + 0.8*(0, \coordheight)$) -- ($(ph15.center) +  + 0.8*(0, \coordheight)$) node[midway,yshift=0.5mm,anchor=south]  {\scriptsize \textit{Phase}};
    
\end{tikzpicture}

%% file: figures/cubic_explainer.tex
\def\coordwidth{6}
\def\coordheight{2.7}
\newcommand{\coord}[2]{
    ($(anchor_point.center) + #1*(\coordwidth, 0) + #2*(0, \coordheight)$)
}
\begin{tikzpicture}

    \node (anchor_point) at (0, 0) {};
    
    % Coordinate system
    \draw[-latex] \coord{0}{-0.028} -- \coord{0}{1};
    \draw[-latex] \coord{-0.037}{0} -- \coord{1}{0};
    \node at \coord{0.5}{-0.05} {\scriptsize Time};
    \node[rotate=90,anchor=south] at \coord{0}{0.5} {\scriptsize Congestion-Window Size};

    % CUBIC rate
    % \draw[domain=0:1, smooth, variable=\x, blue] plot ({\x},{cubic(\x,0,1)});
    \def\wmax{0.3}
    \def\loss{0}
    \def\lossnext{0.4}
    \pgfmathsetmacro{\wmaxnext}{\wmax+(3*\lossnext-\loss-(0.3*\wmax)^(0.333))^3}
    \draw[draw=internationalkleinblue,line width=0.2mm] \coord{0}{\wmax} -- \coord{\lossnext}{\wmax} -- \coord{\lossnext}{\wmaxnext} -- \coord{0.95}{\wmaxnext};
    \foreach \i in {1, 2, ..., 40} {
        \pgfmathsetmacro{\t}{\i*0.01}
        \pgfmathsetmacro{\tlast}{\t-0.01}
        \pgfmathsetmacro{\cubicval}{\wmax+(3*\t-\loss-(0.3*\wmax)^(0.333))^3}
        \pgfmathsetmacro{\cubicvallast}{\wmax+(3*\tlast-\loss-(0.3*\wmax)^(0.333))^3}
        \draw[draw=bittersweet] \coord{\tlast}{\cubicvallast} -- \coord{\t}{\cubicval};
    }
    \draw[draw=bittersweet] \coord{\lossnext}{\wmaxnext} -- \coord{\lossnext}{0.7*\wmaxnext};
    \foreach \i in {1, 2, ..., 41} {
        \pgfmathsetmacro{\t}{\i*0.01}
        \pgfmathsetmacro{\tlast}{\t-0.01}
        \pgfmathsetmacro{\tactual}{\t+\lossnext}
        \pgfmathsetmacro{\tlastactual}{\t+\lossnext-0.01}
        \pgfmathsetmacro{\cubicval}{\wmaxnext+(3*\t-(0.3*\wmaxnext)^(0.333))^3}
        \pgfmathsetmacro{\cubicvallast}{\wmaxnext+(3*\tlast-(0.3*\wmaxnext)^(0.333))^3}
        \draw[draw=bittersweet] \coord{\tlastactual}{\cubicvallast} -- \coord{\tactual}{\cubicval};
    }
    
    \node[text=bittersweet,align=center,inner sep=0.2mm] (cwnd) at \coord{0.45}{0.9} {\scriptsize Congestion-window\\[-2mm]\scriptsize size $w$};
    \def\cwndlinex{0.7}
    \pgfmathsetmacro{\cubicval}{\wmaxnext+(3*(\cwndlinex-\lossnext)-(0.3*\wmaxnext)^(0.333))^3}
    \node (cwndline) at \coord{\cwndlinex}{\cubicval} {};
    \draw[-{latex[length=0.5mm]},draw=bittersweet] (cwnd.south) to[out=300,in=150] (cwndline.center);

    \node[text=internationalkleinblue,inner sep=0.2mm] (wmax) at \coord{0.7}{0.2} {\scriptsize $w^{\max}$};
    \node (wmaxline) at \coord{\lossnext}{\wmax+0.1} {};
    \draw[-{latex[length=0.5mm]},draw=internationalkleinblue] (wmax.north) to[out=90,in=0] (wmaxline.center);

    \draw[{latex[length=0.5mm]}-{latex[length=0.5mm]},draw=mediumslateblue] \coord{\lossnext}{\wmaxnext-0.1} -- \coord{\lossnext+0.42}{\wmaxnext-0.1};
    \node[text=mediumslateblue,anchor=north,inner sep=0.5mm] at \coord{0.6}{\wmaxnext-0.1} {\scriptsize $s$};
    
    \node[inner sep=0] (loss) at \coord{\lossnext}{\wmaxnext} {\includegraphics[width=3mm]{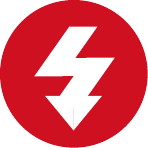}};
    \node[text=lava,align=center,anchor=north,inner sep=0.2mm] (loss_text) at \coord{0.15}{0.65} {\scriptsize Packet\\[-2mm]\scriptsize loss};
    \draw[-{latex[length=0.5mm]},draw=lava] (loss_text.north) to[out=90,in=180] (loss.west);

\end{tikzpicture}

%% file: figures/oscillation-mechanism.tex
\definecolor{blue(munsell)}{rgb}{0.0, 0.5, 0.69}
\begin{tikzpicture}[
longterm/.style={draw=black!60, shading=radial,outer color=brown(web)light,inner color=white, thick,minimum width=61mm,minimum height=20mm,align=left,rounded corners=0.2cm},
shortterm/.style={draw=black!60, shading=radial,outer color=amberlight,inner color=white, thick,minimum width=61mm,minimum height=20mm,align=left,rounded corners=0.2cm},
shorttermline/.style={shape=circle,minimum width=1mm,minimum height=3.5mm,text width=-2mm,fill=amber},
update/.style={shape=diamond,minimum width=1mm,minimum height=3.5mm,text width=-2mm,fill=brown(web)},
 every node/.style={scale=0.8}
]
    
    \node[longterm] (longterm) at (0, 2) {
        $*$\\
        CUBIC congestion-window size~$w$\\
        $\rightarrow$ back-off queue length~$q_{\ell}^-$\\
        $\rightarrow$ BBR minimum-RTT estimate~$\tau^{\min}$ %\\
        % $\rightarrow$ BBR probing strength~$\alpha$
    };

    \node[fill=black,rounded corners=0.2cm,text=white,minimum width=61mm,align=center,draw=black] (longterm_title) at ([yshift=0.8cm]longterm) {\textbf{BBR RTT Probing}\\(every 10 seconds)};
    \node[update] at ([xshift=-1.75cm]longterm_title) {};
    \node[update] at ([xshift=1.75cm]longterm_title) {};

    \node[shortterm] (shortterm) at ([yshift=-2.5cm]longterm) {
        $*$\\
        % BBR probing strength~$\alpha$ (fixed)\\
        BBR minimum-RTT estimate~$\tau^{\min}$\\
        $\rightarrow$ BBR sending rate~$x^{\mathrm{B}}$\\
        $\rightarrow$ Packet loss~$p_{\ell}$ \\
        $\rightarrow$ CUBIC congestion-window size~$w$
    };

    \node[fill=black,rounded corners=0.2cm,text=white,minimum width=61mm,align=center,draw=black] (shortterm_title) at ([yshift=1cm]shortterm) {\textbf{Short-Term Dynamics}\\(during 10 seconds)};
    \node[shorttermline] at ([xshift=-1.75cm]shortterm_title.center) {};
    \node[shorttermline] at ([xshift=1.75cm]shortterm_title.center) {};

    \draw[-latex,line width=1.5mm,color=blue(munsell)] ([xshift=-1.4cm]longterm_title.north) -- ([xshift=2.3cm]longterm_title.north) to[out=0,in=90] ([xshift=2.5cm,yshift=-0.3cm]longterm_title.north) -- ([xshift=2.5cm,yshift=0.3cm]shortterm.south) to[out=270,in=0] ([xshift=2.3cm]shortterm.south)
    -- ([xshift=-2.3cm]shortterm.south) to[out=180,in=270] ([xshift=-2.5cm,yshift=0.3cm]shortterm.south) -- ([xshift=-2.5cm,yshift=-0.3cm]longterm_title.north) to[out=90,in=180] ([xshift=-2.3cm]longterm_title.north) -- ([xshift=-1.5cm]longterm_title.north);
    \node[text=blue(munsell)] at ([xshift=0.6cm,yshift=0.3cm]longterm_title.north) {\textbf{Long-Term Dynamics}};

\end{tikzpicture}

%% file: figures/oscillation-example.tex
\begin{tikzpicture}[
    update/.style={shape=diamond,minimum width=1mm,minimum height=3.5mm,text width=-2mm,fill=brown(web)},
    shorttermline/.style={shape=circle,minimum width=1mm,minimum height=3.5mm,text width=-2mm,fill=amber},
]

    \node at (0, 0) {\includegraphics[width=0.99\linewidth,trim=5 10 5 0]{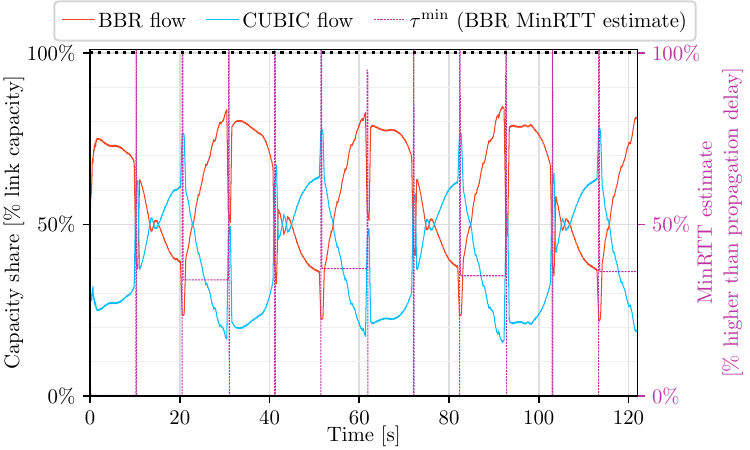}};

    \def\lineypos{-1.75}
    \def\updatelinexstart{-3.1}
    \def\updatelinexend{2.85}
    \def\dist{0.595}
    \draw[] (\updatelinexstart, \lineypos) -- (\updatelinexend, \lineypos);

    % \foreach \i\offset in {0/0.02, 1/0.0, 2/0, 3/-0.02, 4/-0.04, 5/-0.04, 6/-0.07, 7/-0.08, 8/-0.1, 9/-0.12} {
    \foreach \i/\offset in {0/0.0, 1/0.0, 2/0.0, 3/0.0, 4/0.0, 5/0.0, 6/0.01, 7/0.01, 8/0.02, 9/0.02} {
        \node[shorttermline] (st\i) at ($(\updatelinexstart, \lineypos) + \i*(\dist, 0) + 0*\offset*(1, 0) + 0.5*(\dist, 0)$) {};
    }
    
    % \foreach \i\offset in {0/0.02, 1/0.0, 2/0, 3/-0.02, 4/-0.04, 5/-0.04, 6/-0.07, 7/-0.08, 8/-0.08, 9/-0.1, 10/-0.12} {
    \foreach \i/\offset in {0/0.0, 1/0.0, 2/0.0, 3/0.0, 4/0.0, 5/0.0, 6/0.01, 7/0.01, 8/0.02, 9/0.02, 10/0.02} {
        \node[update] (update\i) at ($(\updatelinexstart, \lineypos) + \i*(\dist, 0) + \offset*(1, 0)$) {};
    }

    \foreach \i\offset in {1/0.0, 2/0, 3/-0.02, 4/-0.02} {
        \node[] at ($(\updatelinexstart, 1.5) + \i*(\dist, 0) + \offset*(1, 0)$) {\circled{\i}};
    }

% \node[update] at (-1.8,-1.85) {};
% \node[update] at (-1.4227272727272728,-1.85) {};
% \node[update] at (-1.0454545454545454,-1.85) {};
% \node[update] at (-0.6681818181818182,-1.85) {};
% \node[update] at (-0.2909090909090908,-1.85) {};
% \node[update] at (0.08636363636363642,-1.85) {};
% \node[update] at (0.4636363636363636,-1.85) {};
% \node[update] at (0.840909090909091,-1.85) {};
% \node[update] at (1.21,-1.85) {};
% \node[update] at (1.585,-1.85) {};
% \node[update] at (1.96,-1.85) {};
% \node[update] at (2.335,-1.85) {};

    % \node[] at (-1.0454545454545454, 1.8) {\circled{1}};
    % \node[] at (-0.6681818181818182, 1.8) {\circled{2}};
    % \node[] at (-0.2909090909090908, 1.8) {\circled{3}};
    % \node[] at (0.08636363636363642, 1.8) {\circled{4}};

    \node[fill=asparagus,fill opacity=0.4,minimum width=5.95mm,anchor=west] (inflation) at ($(\updatelinexstart, -0.85) + (\dist, 0)$) {}; %(-2.1, -0.7) {};
    \node (infl_y) at ($(inflation.east) + 9*(\dist, 0)$) {};
    \draw[densely dotted,draw=asparagus,line width=0.3mm] (inflation.east) -- (infl_y.center);
    
    \node[anchor=west,inner sep=0.5mm,text=asparagus] at (infl_y) {\footnotesize 34\%};
    
    \node[align=center,fill=asparagus, fill opacity=0.6, text opacity=1.0,inner sep=0.4mm] (inflation_text) at ($(inflation) + (3, 2.5)$) {\scriptsize Inflated\\[-2mm]\scriptsize MinRTT\\[-2mm]\scriptsize estimate};
    
    \draw[-latex,draw=asparagus,line width=0.3mm] (inflation_text.south) to[out=270,in=90] (inflation.north);
        
\end{tikzpicture}

%% file: figures/dynamics_arrows.tex
\begin{tikzpicture}[overlay]

    \node[] at (0, 0) {};

    \draw[-latex,densely dashed,brown(web),line width=0.3mm] plot [smooth] coordinates {(-6, 5) (-4.7, 5.5) (-3.8, 5.2) (-3.5, 3.4) (-3, 1.9)};

    \draw[-latex,densely dashed,amber,line width=0.3mm] plot [smooth] coordinates 
    {(-6, 2.6) (-4.5, 0.7) (-3.2, 0.6) (-2.7, 1.5)};
    
\end{tikzpicture}

%% file: figures/oscillation-synchronization.tex
\begin{tikzpicture}[
    update/.style={draw=brown(web),thin,minimum width=-2mm,minimum height=5mm},
]

    \node at (0, 0) {\includegraphics[width=0.5\linewidth,trim=15 10 5 0]{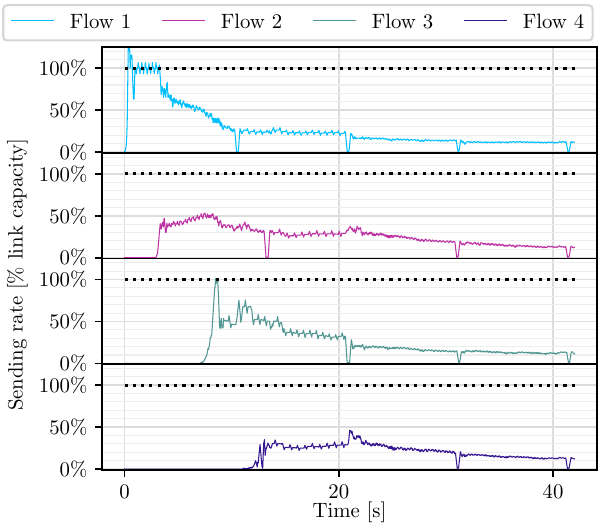}};

    \def\labely{2.7}
    \def\flowonedrop{1.7}
    \def\flowtwodrop{0.1}
    \def\flowthreedrop{-1.4}
    \def\flowfourdrop{-3}
    \def\timelineend{-3.2}

    \def\rttprobing{-1.1}
    \node (uplabel) at (\rttprobing, \labely) {\circled{1}};
    \node[update] (up1) at (\rttprobing, \flowonedrop) {};
    \draw[densely dashed,thin] (uplabel.center) -- (up1.north);
    \draw[densely dashed,thin] (up1.south) -- (\rttprobing, \timelineend);

    \def\rttprobing{-0.65}
    \node (uplabel_pos) at (\rttprobing, \labely) {};
    \node[update] (up1) at (\rttprobing, \flowtwodrop) {};
    \draw[densely dashed,thin] (uplabel_pos.center) -- (up1.north);
    \draw[densely dashed,thin] (up1.south) -- (\rttprobing, \timelineend);
    \node (uplabel) at (uplabel_pos) {\circled{2}};

    \def\rttprobing{0.55}
    \node (uplabel_pos) at (\rttprobing, \labely) {};
    \node[update] (up1) at (\rttprobing, \flowonedrop) {};
    \node[update] (up2) at (\rttprobing, \flowthreedrop) {};
    \draw[densely dashed,thin] (uplabel_pos.center) -- (up1.north);
    \draw[densely dashed,thin] (up1.south) -- (up2.north);
    \draw[densely dashed,thin] (up2.south) -- (\rttprobing, \timelineend);
    \node (uplabel) at (uplabel_pos) {\circled{3}};

    \def\rttprobing{2.2}
    \node (uplabel_pos) at (\rttprobing, \labely) {};
    \node[update] (up5) at (\rttprobing, \flowonedrop) {};
    \node[update] (up6) at (\rttprobing, \flowtwodrop) {};
    \node[update] (up7) at (\rttprobing, \flowthreedrop) {};
    \node[update] (up8) at (\rttprobing, \flowfourdrop) {};
    \draw[densely dashed,thin] (uplabel_pos.center) -- (up5.north);
    \draw[densely dashed,thin] (up5.south) -- (up6.north);
    \draw[densely dashed,thin] (up6.south) -- (up7.north);
    \draw[densely dashed,thin] (up7.south) -- (up8.north);
    \node (uplabel) at (uplabel_pos) {\circled{4}};

    % \node[update] (up3) at (0.35, 1.1) {};
    % \node[update] (up4) at (0.35, -1) {};
    % \draw[densely dashed,thin] (0.35, 1.9) -- (up3.north);
    % \draw[densely dashed,thin] (up3.south) -- (up4.north);
    % \draw[densely dashed,thin] (up4.south) -- (0.35, -2.2);
    % \node at (0.35, 1.9) {\circled{3}};

    % \node[update] (up5) at (1.5, 1.1) {};
    % \node[update] (up6) at (1.5, 0.1) {};
    % \node[update] (up7) at (1.5, -1) {};
    % \node[update] (up8) at (1.5, -2) {};
    % \draw[densely dashed,thin] (1.5, 1.9) -- (up5.north);
    % \draw[densely dashed,thin] (up5.south) -- (up6.north);
    % \draw[densely dashed,thin] (up6.south) -- (up7.north);
    % \draw[densely dashed,thin] (up7.south) -- (up8.north);
    % \node at (1.5, 1.9) {\circled{4}};

\end{tikzpicture}

%% file: figures/turning-point-case1.tex
\def\cubicb{0.3}
\def\cubicc{0.4}
\def\cubict{0.1}
\def\cubicchi{1}
\def\cubicC{1.2}
\definecolor{airforceblue}{rgb}{0.36, 0.54, 0.66}
\definecolor{americanrose}{rgb}{1.0, 0.01, 0.24}
\definecolor{applegreen}{rgb}{0.55, 0.71, 0.0}
\definecolor{atomictangerine}{rgb}{1.0, 0.6, 0.4}
\begin{tikzpicture}[declare function={
    S(\x) = \cubicc^2/(\cubicb*\cubict) * \x^7 - \cubicc * (\cubicC - \cubicchi) * \x^4 - \cubicc*\x^3 - \cubicb*\cubict;
    S1(\x) = 7 * \cubicc^2 / (\cubicb*\cubict) * \x^6 - 4* \cubicc * (\cubicC - \cubicchi) * \x^3 - 3*\cubicc*\x^2;
    S3(\x) = 210 * \cubicc^2 / (\cubicb*\cubict) * \x^4 - 24 * \cubicc * (\cubicC - \cubicchi) * \x;
    Psi(\x) =  7 * \cubicc^2 / (\cubicb*\cubict) * \x^6 - 4* \cubicc * (\cubicC - \cubicchi) * \x^2 - 3*\cubicc*\x^2;
    s3 = pow(\cubicb*\cubict/\cubicc * 4/35 * (\cubicC-\cubicchi), 1/3);
    sm = pow(\cubicb*\cubict/\cubicc * (4/7 * (\cubicC-\cubicchi) + 3/7), 1/4);
    s1 = 0.4352;
}]

    \begin{axis}%
    [
        xmin=-0.05,
        xmax=0.8,
        xtick={},
        xticklabels={},
        ytick={},
        yticklabels={},
        xtick style={draw=none},
        ytick style={draw=none},
        ymin=-0.2,
        ymax=0.8,
        x=6cm,
        y=3cm,
        axis y line=left,
        axis y line shift=-0.05,
        axis x line shift=-0.2,
        samples=200,
        domain=0:0.8,
        xlabel={$s$},
        ylabel={},
        x label style={at={(axis description cs:1,0.2)}},
        y label style={at={(axis description cs:0,1)}},
        axis x line=bottom,
        restrict y to domain=-10:10
    ]

        \node at (-0.03, 0.05) {0};
        
        \addplot[mark=None,color=airforceblue] (x,{S(\x)});
        \node[color=airforceblue] at (0.76, 0.3) {$\tilde{S}_2$};
        
        \addplot[mark=None,color=americanrose] (x,{S1(\x)});
        \node[color=americanrose] at (0.47, 0.4) {$\tilde{S}_2'$};

        \addplot[mark=None,color=atomictangerine,dashed] (x,{Psi(\x)});
        \node[color=atomictangerine] at (0.63, 0.7) {$\Psi^-$};
        
        \addplot[mark=None,color=applegreen] (x,{S3(\x)});
        \node[color=applegreen] at (0.24, 0.7) {$\tilde{S}_2'''$};

        \node at (s3, 0.7) {$s'''$};
        \draw[dotted] (s3,-0.2) -- (s3, 0.6);

        \node at ({sm+0.04}, 0.7) {$s^-$};
        \draw[dotted] (sm,-0.2) -- (sm, 0.6);

        \node at ({s1-0.01}, 0.7) {$s'$};
        \draw[dotted] (s1,-0.2) -- (s1, 0.6);

    \end{axis}
    
\end{tikzpicture}

%% file: figures/turning-point-case2.tex
\def\cubicb{0.3}
\def\cubicc{0.4}
\def\cubict{0.1}
\def\cubicchi{1}
\def\cubicC{0.8}
\definecolor{airforceblue}{rgb}{0.36, 0.54, 0.66}
\definecolor{americanrose}{rgb}{1.0, 0.01, 0.24}
\definecolor{applegreen}{rgb}{0.55, 0.71, 0.0}
\definecolor{atomictangerine}{rgb}{1.0, 0.6, 0.4}
\begin{tikzpicture}[declare function={
    S(\x) = \cubicc^2/(\cubicb*\cubict) * \x^7 - \cubicc * (\cubicC - \cubicchi) * \x^4 - \cubicc*\x^3 - \cubicb*\cubict;
    S1(\x) = 7 * \cubicc^2 / (\cubicb*\cubict) * \x^6 - 4* \cubicc * (\cubicC - \cubicchi) * \x^3 - 3*\cubicc*\x^2;
    S3(\x) = 210 * \cubicc^2 / (\cubicb*\cubict) * \x^4 - 24 * \cubicc * (\cubicC - \cubicchi) * \x;
    Psi(\x) =  7 * \cubicc^2 / (\cubicb*\cubict) * \x^6 + 4* \cubicc * (\cubicC - \cubicchi) * \x^2 - 3*\cubicc*\x^2;
    s3 = 0;
    sm = pow(\cubicb*\cubict/\cubicc * (-4/7 * (\cubicC-\cubicchi) + 3/7), 1/4);
    s1 = 0.4113;
}]

    \begin{axis}%
    [
        xmin=-0.05,
        xmax=0.8,
        xtick={},
        xticklabels={},
        ytick={},
        yticklabels={},
        xtick style={draw=none},
        ytick style={draw=none},
        ymin=-0.2,
        ymax=0.8,
        x=6cm,
        y=3cm,
        axis y line=left,
        axis y line shift=-0.05,
        axis x line shift=-0.2,
        samples=200,
        domain=0:0.8,
        xlabel={$s$},
        ylabel={},
        x label style={at={(axis description cs:1,0.2)}},
        y label style={at={(axis description cs:0,1)}},
        axis x line=bottom,
        restrict y to domain=-10:10
    ]

        \node at (-0.03, 0.05) {0};

        \addplot[mark=None,color=airforceblue] (x,{S(\x)});
        \node[color=airforceblue] at (0.76, 0.3) {$\tilde{S}_2$};
        
        \addplot[mark=None,color=americanrose] (x,{S1(\x)});
        \node[color=americanrose] at (0.47, 0.4) {$\tilde{S}_2'$};

        \addplot[mark=None,color=atomictangerine,dashed] (x,{Psi(\x)});
        \node[color=atomictangerine] at (0.63, 0.7) {$\Psi^-$};
        
        \addplot[mark=None,color=applegreen] (x,{S3(\x)});
        \node[color=applegreen] at (0.24, 0.7) {$\tilde{S}_2'''$};

        \node[align=center] at ({s3+0.06}, 0.7) {$s'''$};
        \draw[dotted] (s3,-0.2) -- (s3, 0.6);

        \node at ({sm+0.04}, 0.7) {$s^-$};
        \draw[dotted] (sm,-0.2) -- (sm, 0.6);

        \node at ({s1-0.01}, 0.7) {$s'$};
        \draw[dotted] (s1,-0.2) -- (s1, 0.6);

    \end{axis}
    
\end{tikzpicture}

%% file: figures/turning-point-case3.tex
\def\cubicb{0.3}
\def\cubicc{0.4}
\def\cubict{0.1}
\def\cubicchi{1}
\def\cubicC{1}
\definecolor{airforceblue}{rgb}{0.36, 0.54, 0.66}
\definecolor{americanrose}{rgb}{1.0, 0.01, 0.24}
\definecolor{applegreen}{rgb}{0.55, 0.71, 0.0}
\definecolor{atomictangerine}{rgb}{1.0, 0.6, 0.4}
\begin{tikzpicture}[declare function={
    S(\x) = \cubicc^2/(\cubicb*\cubict) * \x^7 - \cubicc * (\cubicC - \cubicchi) * \x^4 - \cubicc*\x^3 - \cubicb*\cubict;
    S1(\x) = 7 * \cubicc^2 / (\cubicb*\cubict) * \x^6 - 4* \cubicc * (\cubicC - \cubicchi) * \x^3 - 3*\cubicc*\x^2;
    S3(\x) = 210 * \cubicc^2 / (\cubicb*\cubict) * \x^4 - 24 * \cubicc * (\cubicC - \cubicchi) * \x;
    Psi(\x) =  7 * \cubicc^2 / (\cubicb*\cubict) * \x^6 + 4* \cubicc * (\cubicC - \cubicchi) * \x^2 - 3*\cubicc*\x^2;
    s3 = 0;
    sm = pow(\cubicb*\cubict/\cubicc * (-4/7 * (\cubicC-\cubicchi) + 3/7), 1/4);
    s1 = sm;
}]

    \begin{axis}%
    [
        xmin=-0.05,
        xmax=0.8,
        xtick={},
        xticklabels={},
        ytick={},
        yticklabels={},
        xtick style={draw=none},
        ytick style={draw=none},
        ymin=-0.2,
        ymax=0.8,
        x=6cm,
        y=3cm,
        axis y line=left,
        axis y line shift=-0.05,
        axis x line shift=-0.2,
        samples=200,
        domain=0:0.8,
        xlabel={$s$},
        ylabel={},
        x label style={at={(axis description cs:1,0.2)}},
        y label style={at={(axis description cs:0,1)}},
        axis x line=bottom,
        restrict y to domain=-10:10
    ]

        \node at (-0.03, 0.05) {0};

        \addplot[mark=None,color=airforceblue] (x,{S(\x)});
        \node[color=airforceblue] at (0.76, 0.3) {$\tilde{S}_2$};
        
        \addplot[mark=None,color=americanrose] (x,{S1(\x)});
        \node[color=americanrose] at (0.47, 0.4) {$\tilde{S}_2'$};

        \addplot[mark=None,color=atomictangerine,dashed,very thick] (x,{Psi(\x)});
        \node[color=atomictangerine] at (0.63, 0.7) {$\Psi^-$};
        
        \addplot[mark=None,color=applegreen] (x,{S3(\x)});
        \node[color=applegreen] at (0.24, 0.7) {$\tilde{S}_2'''$};

        \node[align=center] at ({s3+0.06}, 0.7) {$s'''$};
        \draw[dotted] (s3,-0.2) -- (s3, 0.6);

        \node at (sm, 0.7) {$s' = s^-$};
        \draw[dotted] (sm,-0.2) -- (sm, 0.6);

    \end{axis}
    
\end{tikzpicture}

%% file: figures/bbr_vs_cubic__remedies_annotated.tex
\begin{tikzpicture}

    \node at (0, 0) {\includegraphics[width=\linewidth,trim=0 30 0 0]{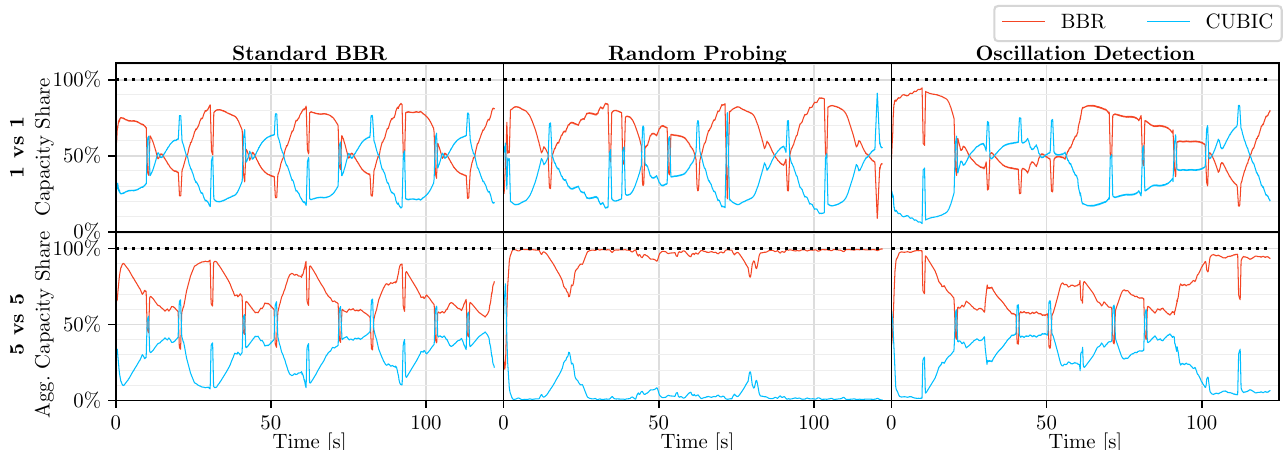}};

    \def\rowone{1.4}
    \def\rowtwo{-0.7}

    \def\colone{-1.5}
    \def\coltwo{3.45}

    \node at (\colone, \rowone) {\circled{1}};
    \node at (\colone, \rowtwo) {\circled{2}};

    \node at (\coltwo, \rowone) {\circled{3}};
    \node at (\coltwo, \rowtwo) {\circled{4}};

    % \node at (\colthree, \rowone) {\circled{5}};
    % \node at (\colthree, \rowtwo) {\circled{6}};
    
\end{tikzpicture}

%% file: sigmetrics2025.bbl
\begin{thebibliography}{10}

\bibitem{akella2002selfish}
Aditya Akella, Srinivasan Seshan, Richard Karp, Scott Shenker, and Christos
  Papadimitriou.
\newblock {Selfish behavior and stability of the Internet: A game-theoretic
  analysis of TCP}.
\newblock {\em ACM SIGCOMM Computer Communication Review}, 32(4):117--130,
  2002.

\bibitem{arun2022starvation}
Venkat Arun, Mohammad Alizadeh, and Hari Balakrishnan.
\newblock Starvation in end-to-end congestion control.
\newblock In {\em Proceedings of the ACM SIGCOMM 2022 Conference}, pages
  177--192, 2022.

\bibitem{arun2021toward}
Venkat Arun, Mina~Tahmasbi Arashloo, Ahmed Saeed, Mohammad Alizadeh, and Hari
  Balakrishnan.
\newblock Toward formally verifying congestion control behavior.
\newblock In {\em Proceedings of the ACM SIGCOMM 2021 Conference}, pages 1--16,
  2021.

\bibitem{bao2010model}
Wei Bao, Vincent~WS Wong, and Victor~CM Leung.
\newblock {A model for steady state throughput of TCP CUBIC}.
\newblock In {\em IEEE Global Telecommunications Conference GLOBECOM 2010},
  pages 1--6. IEEE, 2010.

\bibitem{bonald1998comparison}
Thomas Bonald.
\newblock {\em {Comparison of TCP Reno and TCP Vegas via fluid approximation}}.
\newblock PhD thesis, INRIA, 1998.

\bibitem{brakmo1994tcp}
Lawrence~S Brakmo, Sean~W O'Malley, and Larry~L Peterson.
\newblock {TCP Vegas: New techniques for congestion detection and avoidance}.
\newblock In {\em Proceedings of the conference on Communications
  architectures, protocols and applications}, pages 24--35, 1994.

\bibitem{briscoe2007flow}
Bob Briscoe.
\newblock Flow rate fairness: Dismantling a religion.
\newblock {\em ACM SIGCOMM Computer Communication Review}, 37(2):63--74, 2007.

\bibitem{cardwell2017bbr}
Neal Cardwell, Yuchung Cheng, C~Stephen Gunn, Soheil~Hassas Yeganeh, and Van
  Jacobson.
\newblock {BBR}: congestion-based congestion control.
\newblock {\em Communications of the ACM}, 60(2):58--66, 2017.

\bibitem{cardwell2017bbr-ietf}
Neal Cardwell, Yuchung Cheng, C~Stephen Gunn, Soheil~Hassas Yeganeh, Ian Swett,
  Jana Iyengar, Victor Vasiliev, and Van Jacobson.
\newblock {BBR} congestion control: {IETF} 99 update.
\newblock In {\em Presentation in ICCRG at IETF 99}, 2017.
\newblock URL:
  \url{https://datatracker.ietf.org/meeting/99/materials/slides-99-iccrg-iccrg-presentation-2-00.pdf}.

\bibitem{cardwell2023bbrv3}
Neal Cardwell, Yuchung Cheng, Kevin Yang, David Morley, Soheil~Hassas Yeganeh,
  Priyaranjan Jha, Yousuk Seung, and Van Jacobson.
\newblock {BBRv3: Algorithm Bug Fixes and Public Internet Deployment}.
\newblock
  \url{https://datatracker.ietf.org/meeting/117/materials/slides-117-ccwg-bbrv3-algorithm-bug-fixes-and-public-internet-deployment-00},
  2023.

\bibitem{cardwell2019bbrv2}
Neal Cardwell, Yuchung Cheng, S~Hassas Yeganeh, Ian Swett, Victor Vasiliev,
  Priyaranjan Jha, Yousuk Seung, Matt Mathis, and Van Jacobson.
\newblock {BBRv2: A model-based congestion control}.
\newblock In {\em Presentation in ICCRG at IETF 104th meeting}, 2019.

\bibitem{cardwell2000modeling}
Neal Cardwell, Stefan Savage, and Thomas Anderson.
\newblock {Modeling TCP latency}.
\newblock In {\em Proceedings of IEEE INFOCOM 2000}, volume~3, pages
  1742--1751. IEEE, 2000.

\bibitem{dong2018pcc}
Mo~Dong, Tong Meng, Doron Zarchy, Engin Arslan, Yossi Gilad, Brighten Godfrey,
  and Michael Schapira.
\newblock {PCC Vivace}: Online-learning congestion control.
\newblock pages 343--356, 2018.

\bibitem{dukkipati2006flow}
Nandita Dukkipati and Nick McKeown.
\newblock Why flow-completion time is the right metric for congestion control.
\newblock {\em ACM SIGCOMM Computer Communication Review}, 36(1):59--62, 2006.

\bibitem{fall1996simulation}
Kevin Fall and Sally Floyd.
\newblock {Simulation-based comparisons of Tahoe, Reno and SACK TCP}.
\newblock {\em ACM SIGCOMM Computer Communication Review}, 26(3):5--21, 1996.

\bibitem{gomez2020performance}
Jose Gomez, Elie Kfoury, Jorge Crichigno, Elias Bou-Harb, and Gautam
  Srivastava.
\newblock {A performance evaluation of TCP BBRv2 alpha}.
\newblock In {\em 2020 43rd International Conference on Telecommunications and
  Signal Processing (TSP)}, pages 309--312. IEEE, 2020.

\bibitem{guckenheimer2013nonlinear}
John Guckenheimer and Philip Holmes.
\newblock {\em Nonlinear oscillations, dynamical systems, and bifurcations of
  vector fields ({S}ection 3.2)}, volume~42.
\newblock Springer Science \& Business Media, 2013.

\bibitem{ha2008cubic}
Sangtae Ha, Injong Rhee, and Lisong Xu.
\newblock {CUBIC: a new TCP-friendly high-speed TCP variant}.
\newblock {\em ACM SIGOPS operating systems review}, 42(5):64--74, 2008.

\bibitem{hock2017experimental}
Mario Hock, Roland Bless, and Martina Zitterbart.
\newblock Experimental evaluation of {BBR} congestion control.
\newblock pages 1--10. IEEE, 2017.

\bibitem{hossfeld2016definition}
Tobias Ho{\ss}feld, Lea Skorin-Kapov, Poul~E Heegaard, and Martin Varela.
\newblock Definition of {QoE} fairness in shared systems.
\newblock {\em IEEE Communications Letters}, 21(1):184--187, 2016.

\bibitem{rard1998topics}
Gerard Iooss and Moritz Adelmeyer.
\newblock {\em Topics in bifurcation theory and applications}, volume~3.
\newblock World Scientific, 1998.

\bibitem{bbrv1-source}
Van Jacobson, Neal Cardwell, Yuchung Cheng, and Soheil~Hassas Yeganeh.
\newblock {BBRv1} {S}ource {C}ode.
\newblock
  \url{https://github.com/torvalds/linux/blob/master/net/ipv4/tcp_bbr.c}, 2023.
\newblock Accessed on 2023-03-28.

\bibitem{jain1999throughput}
Raj Jain, Arjan Durresi, and Gojko Babic.
\newblock Throughput fairness index: An explanation.
\newblock In {\em ATM Forum contribution}, volume~99, 1999.

\bibitem{johari2001end}
Ramesh Johari and David Kim~Hong Tan.
\newblock End-to-end congestion control for the internet: Delays and stability.
\newblock {\em IEEE/ACM Transactions on networking}, 9(6):818--832, 2001.

\bibitem{kelly2003fairness}
Frank Kelly.
\newblock Fairness and stability of end-to-end congestion control.
\newblock {\em European journal of control}, 9(2-3):159--176, 2003.

\bibitem{kelly1998rate}
Frank~P Kelly, Aman~K Maulloo, and David Kim~Hong Tan.
\newblock Rate control for communication networks: shadow prices, proportional
  fairness and stability.
\newblock {\em Journal of the Operational Research society}, 49(3):237--252,
  1998.

\bibitem{kfoury2020emulation}
Elie~F Kfoury, Jose Gomez, Jorge Crichigno, and Elias Bou-Harb.
\newblock {An emulation-based evaluation of TCP BBRv2 alpha for wired
  broadband}.
\newblock {\em Computer Communications}, 161:212--224, 2020.

\bibitem{kumar1998comparative}
Anurag Kumar.
\newblock {Comparative performance analysis of versions of TCP in a local
  network with a lossy link}.
\newblock {\em Ieee/Acm Transactions on Networking}, 6(4):485--498, 1998.

\bibitem{mininet_intro}
Bob Lantz, Nikhil Handigol, Brandon Heller, and Vimal Jeyakumar.
\newblock Introduction to {M}ininet.
\newblock
  \url{https://github.com/mininet/mininet/wiki/Introduction-to-Mininet}, 2021.

\bibitem{liu2003fluid}
Yong Liu, Francesco Lo~Presti, Vishal Misra, Don Towsley, and Yu~Gu.
\newblock Fluid models and solutions for large-scale {IP} networks.
\newblock pages 91--101, 2003.

\bibitem{low2002internet}
Steven~H Low, Fernando Paganini, and John~C Doyle.
\newblock Internet congestion control.
\newblock {\em IEEE control systems magazine}, 22(1):28--43, 2002.

\bibitem{mathis1997macroscopic}
Matthew Mathis, Jeffrey Semke, Jamshid Mahdavi, and Teunis Ott.
\newblock {The macroscopic behavior of the TCP congestion avoidance algorithm}.
\newblock {\em ACM SIGCOMM Computer Communication Review}, 27(3):67--82, 1997.

\bibitem{symmathmatlab}
{MathWorks}.
\newblock {Symbolic Math Toolbox}.
\newblock \url{https://ch.mathworks.com/products/symbolic.html}, 2023.

\bibitem{iperf}
Robert McMahon.
\newblock i{P}erf.
\newblock \url{https://sourceforge.net/projects/iperf2/}, 2021.

\bibitem{mishra2019great}
Ayush Mishra, Xiangpeng Sun, Atishya Jain, Sameer Pande, Raj Joshi, and Ben
  Leong.
\newblock {The great internet TCP congestion control census}.
\newblock {\em Proceedings of the ACM on Measurement and Analysis of Computing
  Systems}, 3(3):1--24, 2019.

\bibitem{mishra2022we}
Ayush Mishra, Wee~Han Tiu, and Ben Leong.
\newblock Are we heading towards a {BBR}-dominant {I}nternet?
\newblock In {\em Proceedings of the 22nd ACM Internet Measurement Conference
  (IMC)}, pages 538--550, 2022.

\bibitem{misra2000fluid}
Vishal Misra, Wei-Bo Gong, and Don Towsley.
\newblock {Fluid-based analysis of a network of {AQM} routers supporting {TCP}
  flows with an application to {RED}}.
\newblock In {\em Proceedings of the ACM Conference on Applications,
  Technologies, Architectures, and Protocols for Computer Communication
  (SIGCOMM)}, pages 151--160, 2000.

\bibitem{nandagiri2020bbrvl}
Aarti Nandagiri, Mohit~P Tahiliani, Vishal Misra, and KK~Ramakrishnan.
\newblock {BBRv1 vs BBRv2: Examining performance differences through
  experimental evaluation}.
\newblock In {\em 2020 IEEE International Symposium on Local and Metropolitan
  Area Networks (LANMAN}, pages 1--6. IEEE, 2020.

\bibitem{olsen2003stochastic}
J{\"o}rgen Ols{\'e}n.
\newblock {\em {Stochastic modeling and simulation of the TCP protocol}}.
\newblock PhD thesis, {Department of Mathematics, Uppsala University}, 2003.

\bibitem{padhye1998modeling}
Jitendra Padhye, Victor Firoiu, Don Towsley, and Jim Kurose.
\newblock {Modeling TCP throughput: A simple model and its empirical
  validation}.
\newblock In {\em Proceedings of the ACM SIGCOMM'98 conference on Applications,
  technologies, architectures, and protocols for computer communication}, pages
  303--314, 1998.

\bibitem{poojary2011analytical}
Sudheer Poojary and Vinod Sharma.
\newblock {Analytical model for congestion control and throughput with TCP
  CUBIC connections}.
\newblock In {\em 2011 IEEE Global Telecommunications Conference-GLOBECOM
  2011}, pages 1--6. IEEE, 2011.

\bibitem{rhee2018rfc}
I~Rhee, L~Xu, S~Ha, A~Zimmermann, L~Eggert, and R~Scheffenegger.
\newblock {RFC 8312: CUBIC for Fast Long-Distance Networks}, 2018.

\bibitem{scherrer2022model}
Simon Scherrer, Markus Legner, Adrian Perrig, and Stefan Schmid.
\newblock Model-based insights on the performance, fairness, and stability of
  {BBR}.
\newblock In {\em Proceedings of the 22nd ACM Internet Measurement Conference
  (IMC)}, pages 519--537, 2022.

\bibitem{scholz2018towards}
Dominik Scholz, Benedikt Jaeger, Lukas Schwaighofer, Daniel Raumer, Fabien
  Geyer, and Georg Carle.
\newblock Towards a deeper understanding of {TCP} {BBR} congestion control.
\newblock In {\em Proceedings of the IFIP Networking Conference (IFIP
  Networking) and Workshops}, pages 1--9. IEEE, 2018.

\bibitem{song2021understanding}
Yeong-Jun Song, Geon-Hwan Kim, Imtiaz Mahmud, Won-Kyeong Seo, and You-Ze Cho.
\newblock Understanding of {BBR}v2: {E}valuation and comparison with {BBR}v1
  congestion control algorithm.
\newblock {\em IEEE Access}, 9:37131--37145, 2021.

\bibitem{srikant2004mathematics}
Rayadurgam Srikant.
\newblock {\em The mathematics of Internet congestion control}.
\newblock Springer Science \& Business Media, 2004.
\newblock \href {https://doi.org/10.1007/978-0-8176-8216-3}
  {\path{doi:10.1007/978-0-8176-8216-3}}.

\bibitem{turkovic2019fifty}
Belma Turkovic, Fernando~A Kuipers, and Steve Uhlig.
\newblock Fifty shades of congestion control: A performance and interactions
  evaluation.
\newblock {\em arXiv preprint arXiv:1903.03852}, 2019.

\bibitem{vardoyan2021towards}
Gayane Vardoyan, CV~Hollot, and Don Towsley.
\newblock Towards stability analysis of data transport mechanisms: A fluid
  model and its applications.
\newblock {\em IEEE/ACM Transactions on Networking}, 29(4):1730--1744, 2021.

\bibitem{ware2019beyond}
Ranysha Ware, Matthew~K Mukerjee, Srinivasan Seshan, and Justine Sherry.
\newblock Beyond jain's fairness index: Setting the bar for the deployment of
  congestion control algorithms.
\newblock In {\em Proceedings of the 18th ACM Workshop on Hot Topics in
  Networks}, pages 17--24, 2019.

\bibitem{ware2019modeling}
Ranysha Ware, Matthew~K Mukerjee, Srinivasan Seshan, and Justine Sherry.
\newblock {Modeling BBR's interactions with loss-based congestion control}.
\newblock In {\em Proceedings of the {I}nternet {M}easurement {C}onference},
  pages 137--143, 2019.

\bibitem{wiggins2003introduction}
Stephen Wiggins and Martin Golubitsky.
\newblock {\em Introduction to applied nonlinear dynamical systems and chaos
  ({C}hapter 18)}, volume~2.
\newblock Springer, 2003.

\bibitem{xu2022measurement}
Xiaokun Xu and Mark Claypool.
\newblock {Measurement of cloud-based game streaming system response to
  competing TCP CUBIC or TCP BBR flows}.
\newblock In {\em Proceedings of the 22nd ACM Internet Measurement Conference
  (IMC)}, pages 305--316, 2022.

\bibitem{yang2022bbrv2+}
Furong Yang, Qinghua Wu, Zhenyu Li, Yanmei Liu, Giovanni Pau, and Gaogang Xie.
\newblock {BBRv2+: Towards balancing aggressiveness and fairness with
  delay-based bandwidth probing}.
\newblock {\em Computer Networks}, 206:108789, 2022.

\bibitem{zarchy2019axiomatizing}
Doron Zarchy, Radhika Mittal, Michael Schapira, and Scott Shenker.
\newblock Axiomatizing congestion control.
\newblock {\em Proceedings of the ACM on Measurement and Analysis of Computing
  Systems}, 3(2):1--33, 2019.

\end{thebibliography}
